\documentclass[preprint,aps,12pt,showpacs,nofootinbib,tightenlines,amsmath,amssymb]{revtex4}
\usepackage{amsmath}
\usepackage{graphicx}
\usepackage{amssymb}
\usepackage{color}

\newcommand{\btd}{\bigtriangledown}
\newcommand{\be}{\begin{eqnarray}}
\newcommand{\ee}{\end{eqnarray}}

\newcommand{\Tr}{\mathop{\rm Tr}\nolimits}

\newcommand{\ob}{\bar{\omega}}
\newcommand{\ot}{\tilde{\omega}}

\newcommand{\Gmt}{\tilde{\mathsf \Gamma}}
\newcommand{\Gm}{\mathsf \Gamma}

\newcommand{\Gmc}{\mathsf \varGamma}
\newcommand{\Gmct}{\tilde{\mathsf \varGamma}}

\newcommand{\Rc}{\mathcal R}
\newcommand{\Rct}{\tilde{\mathcal R}}
\newcommand{\Lm}{\mathsf L}

\newcommand{\Sm}{\mathsf S}
\newcommand{\Rm}{\mathsf R}
\newcommand{\Rmt}{\tilde{\mathsf R}}
\newcommand{\wb}{\bar{w}}
\newcommand{\pb}{\bar{\varphi}}
\newcommand{\ab}{\bar{a}}
\newcommand{\achi}{{a}_{\chi}}
\newcommand{\gb}{\bar{g}}
\newcommand{\chib}{\bar{\chi}}
\newcommand{\Ob}{\bar{\Omega}}
\newcommand{\Om}{\mathsf \Omega }
\newcommand{\Omt}{\tilde{\mathsf \Omega }}
\newcommand{\td}{\bigtriangledown}

\newcommand{\chih}{\hat{\chi}}
\newcommand{\hh}{\hat{h}}

\newcommand{\Hh}{\hat{H}}
\newcommand{\Nh}{\hat{N}}
\newcommand{\mk}{m_{\kappa}}
\newcommand{\lk}{l_{\kappa}}
\newcommand{\ek}{\eta_{\kappa}}
\newcommand{\Mk}{H_{\kappa}}
\newcommand{\Lk}{L_{\kappa}}
\newcommand{\kh}{\hat{\kappa}}
\newcommand{\bvt}{\bar{\vartheta}}
\newcommand{\vt}{\vartheta}

\begin{document}
\def\intdk{\int\frac{d^4k}{(2\pi)^4}}
\def\sla{\hspace{-0.17cm}\slash}
\hfill


\title{Quantum Field Theory of Gravity with Spin and Scaling Gauge Invariance  and Spacetime Dynamics with Quantum Inflation}

\author{Yue-Liang Wu}\email{ylwu@itp.ac.cn; ylwu@ucas.ac.cn}
\affiliation{State Key Laboratory of Theoretical Physics(SKLTP)\\
Kavli Institute for Theoretical Physics China (KITPC) \\
Institute of Theoretical Physics, Chinese Academy of Sciences, Beijing, 100190, China, \\
and International Centre for Theoretical Physics Asia-Pacific (ICTP-AP) \\
University of Chinese Academy of Sciences (UCAS), Beijing 100049, China }


\begin{abstract}
Treating the gravitational force on the same footing as the electroweak and strong forces, we present a quantum field theory of gravity based on spin and scaling gauge symmetries. A biframe spacetime is initiated to describe such a quantum gravity theory. The gravifield sided on both locally flat noncoordinate spacetime and globally flat Minkowski spacetime is an essential ingredient for gauging global spin and scaling symmetries. The locally flat gravifield spacetime spanned by the gravifield is associated with a non-commutative geometry characterized by a gauge-type field strength of the gravifield. A coordinate-independent and gauge-invariant action for the quantum gravity is built in the gravifield basis. In the coordinate basis, we derive equations of motion for all quantum fields including the gravitational effect and obtain basic conservation laws for all symmetries. The equation of motion for gravifield tensor is deduced in connection directly with the total energy-momentum tensor. When the spin and scaling gauge symmetries are broken down to a background structure that possesses the global Lorentz and scaling symmetries, we obtain exact solutions by solving equations of motion for the background fields in a unitary basis. The massless graviton and massive spinon result as physical quantum degrees of freedom.  The resulting Lorentz-invariant and conformally flat background gravifield spacetime is characterized by a cosmic vector with a nonzero cosmological mass scale. The evolving Universe is, in general, not isotropic in terms of conformal proper time. The conformal size of the Universe becomes singular at the cosmological horizon and turns out to be inflationary in light of cosmic proper time. A mechanism for quantum scalinon inflation is demonstrated such that it is the quantum effect that causes the breaking of global scaling symmetry and generates the inflation of the early Universe, which is ended when the evolving vacuum expectation value of the scalar potential gets a minimal.  Regarding the gravifield as a Goldstone-like field that transmutes the local spin gauge symmetry into the global Lorentz symmetry with a hidden general coordinate invariance, a spacetime gauge field is constructed from the spin gauge field that becomes a hidden gauge field. The bosonic gravitational interactions are described by the Goldstone-like gravimetric field and spacetime gauge field. Two types of gravity equation result; one is as the extension to Einstein's equation of general relativity, and the other is a new one that characterizes spinon dynamics. The Einstein theory of general relativity is considered to be an effective low-energy theory.  
\end{abstract}

\maketitle

\begin{widetext}
\tableofcontents
\end{widetext}

\section{Introduction}

The gravitational force as a basic force of nature has well been characterized in the macroscopic world by the theory of general relativity formulated by Einstein\cite{GR} as the dynamics of a metric in Riemannian spacetime. It remains unsatisfactory that the other three basic forces, i.e., electromagnetic, weak and strong interactions, have successfully been described by the so-called standard model (SM) of gauge interactions within the framework of relativistic quantum field theory (QFT) \cite{QED1, QED2, QED3, QED4,EW1,EW2,EW3,QCD1,QCD2,QCD3} in the flat Minkowski spacetime, which is in contrast to the gravitational force that is formulated by a dynamic curved spacetime. Such an odd dichotomy causes an obstacle for a unified description of gravity with three basic forces and a difficulty for the quantization of gravitational force. On the other hand, it becomes more and more clear that a QFT description for the gravitational force will play a significant role in understanding and exploring the origin of Universe, such as the singularity and inflation of the early Universe\cite{IU1,IU2,IU3,IU4}.

The relativistic QFT as a successful theory that unifies quantum mechanics and special relativity provides a remarkable theoretical framework to describe the microscopic world.  The three basic forces have well been characterized by the SM based on the gauge symmetries of quarks and leptons, which have been tested by more and more precise experiments. The advent of the SM-like Higgs boson at the LHC \cite{higgs2012a,higgs2012c} motivates us to study a more fundamental theory including quantum gravity. A consistent description for the gravitational force has not yet  been established successfully within the framework of QFT. The foundation of Einstein's general relativity is based on the postulate: {\it The laws of physics must be of such a nature that they apply to a system of reference in any kind of motion}. Such a postulate appears to be a natural extension of the one of special relativity. The principle of special relativity concerns two postulates. One is the postulate of relativity, namely, {\it if a system of coordinates $K$ is chosen so that, in relation to it, physical laws hold good in their simplest form, the same laws also hold good in relation to any other system of coordinates $K'$ moving in uniform translation relatively to $K$}. The other is the postulate that {\it the velocity of light in vacuum is a constant}. The first postulate of relativity is actually satisfied by the classical mechanics of Galileo and Newton. It is the second postulate of the  constancy of the velocity of light in vacuum that leads to the relativity of simultaneity and the related laws for the behavior of moving bodies and clocks. Consequently, it leads space and time to be a four-dimensional flat Minkowski spacetime and the physical laws to be invariant under the global Lorentz transformations of SO(1,3) symmetry.  In Einstein's general relativity, the postulate of relativity is extended to be: {\it the physical laws of nature are to be expressed by equations which hold good for all systems of coordinates}, which leads the gravitational force to be characterized by a Riemann geometry in a curved spacetime. Namely, the gravitational force lies in the dynamic Riemannian geometry of curved spacetime. Thus, the physical laws, in the theory of general relativity, are invariant under the general linear transformations of local GL($4,R$) symmetry, which indicates that space and time cannot be well defined in such a way that the differences of the spatial coordinates or time coordinates can be directly measured by the standard ways proposed in special relativity. 

As gauge theories have been shown to be consistently described by QFT, it motivates one to make efforts for finding out a gauge theory description on the gravitational force. In analogue to the Yang-Mills gauge theory\cite{YM}, early works on gravity gauge theories were proposed by many pioneers\cite{GGT1,GGT2,GGT3,GGT4,GGT5}. As gauge theories have successfully been achieved in describing electromagnetic, weak and strong interactions of quarks and leptons,  numerous efforts on gravity gauge theories have been made over the past half century, and more literatures may be found in some review articles\cite{GGTR1,GGTR2,GGTR3}. It is noticed that most gravity gauge theories were built based on Riemannian or non-Riemannian geometry on curved spacetime manifolds. Thus the main issues on the definition of space and time and the quantization of gravity gauge theories remain open questions. Also, the basic structure of Lagrangian and dynamic properties of gravity gauge theories are still not well understood.  

In this paper, we shall present an alternative gauge theory of gravity within the framework of QFT in the flat Minkowski spacetime and treat the gravitational force on the same footing as the electroweak and strong forces. Namely, all basic interactions of elementary particles will be governed by gauge symmetries and characterized by quantum fields defined in the flat Minkowski spacetime of coordinates. As all known symmetry groups are related to the intrinsic quantum numbers of the basic building blocks of nature, i.e., quarks and leptons, the key point for the gauge theory of gravity in our present consideration is that we will distinguish the spin symmetry group SO(1,3)$\equiv$ SP(1,3) of fermion fields from the Lorentz symmetry group SO(1,3) of coordinates in the Minkowski spacetime, and treat the spin symmetry SP(1,3) as an internal symmetry. The notations SP(1,3) and SO(1,3) are used to mark explicitly such two different kinds of symmetries. Since special relativity and quantum mechanics have a strong foundation based on the well-established principles, we shall make postulates for the QFT description of gravity based on the principles of relativistic QFT. The main postulates are  (i) a biframe spacetime is proposed to describe the quantum field theory of gravity. One frame spacetime is the globally flat coordinate Minkowski spacetime that acts as an inertial reference frame for the motions of fields, and the other is the locally flat noncoordinate gravifield spacetime that functions as an interaction representation frame for the degrees of freedom of fields. (ii) The kinematics of all quantum fields obeys the principles of special relativity and quantum mechanics. (iii) The dynamics of all quantum fields is characterized by basic interactions governed by the gauge symmetries. (iv)  The action of quantum gravity is to be expressed in the so-called gravifield spacetime to be {\it coordinate independent and gauge invariant}. (v) The theory is invariant not only under the local spin and scaling gauge transformations of quantum fields defined in the gravifield spacetime, but also under the global Lorentz and scaling as well as translational transformations of coordinates in the flat Minkowski spacetime.  We shall show that a theory of quantum gravity based on the spin and scaling gauge symmetries can, in general, be constructed in a coordinate-independent formalism and naturally transmuted into the coordinate basis through a basic gravitational field. 

The success of a relativistic QFT description on the SM is based on a global symmetry of the inhomogeneous Lorentz group or Poincar$\acute{\mbox{e}}$  group that contains both the Lorentz group SO(1,3) and translational group $T^{1,3}$ in a globally flat Minkowski spacetime of coordinates. Namely, a fundamental theory built within the framework of relativistic QFT is invariant under a global symmetry of Poincar$\acute{\mbox{e}}$  group P(1,3) = SO(1,3)$\times T^{1,3}$.  To build a gauge theory of gravity within the framework of relativistic QFT, we suppose that it remains essential to keep a global symmetry of Poincar$\acute{\mbox{e}}$ group P(1,3) in a flat Minkowski spacetime and in the meantime introduce a locally flat noncoordinate spacetime through gauging an internal Poincar$\acute{\mbox{e}}$  group PG(1,3)  to characterize fermionic degrees of freedom, such as quarks and leptons that are regarded as the basic building blocks of matter in the SM. Such a conceptual idea on establishing a gauge theory of gravity is distinguished from the original thought of Einstein, who extended the special theory of relativity to a general theory of relativity with a straightforward speculation that the physical laws that are invariant under a global symmetry of Poincar$\acute{\mbox{e}}$  group P(1,3) in a flat Minkowski spacetime are to be expressed by more general formalisms that are invariant under a local symmetry of general linear group GL($4,R$) in a curved Riemannian spacetime of coordinates. It is known that the GL($4,R$) group contains no symmetry of translational group $T^{1,3}$, though it includes a symmetry of Lorentz group SO(1,3). We will show that a gauge theory of gravity constructed based on the postulate of gauge invariance and coordinate independence possesses a gauge symmetry of the spin group or gauged Lorentz group SP(1,3) that is a subgroup of internal Poincar$\acute{\mbox{e}}$  gauge group PG(1,3) in a locally flat noncoordinate spacetime. Here the notation PG(1,3) is used to distinguish from the global symmetry of Poincar$\acute{\mbox{e}}$  group P(1,3) in the Minkowski spacetime of coordinates. In general, the resulting gauge theory of gravity will be demonstrated to have a basic local and global symmetry SP(1,3)$\times$ P(1,3) in either the gravitational fermionic interactions or bosonic interactions. A new {\it gravifield} defined on both a globally flat coordinate Minkowski spacetime and a locally flat noncoordinate spacetime is necessarily introduced as a gauge-type field that transforms homogeneously under the spin gauge group SP(1,3) and forms a basis for a locally flat noncoordinate spacetime. Such a locally flat noncoordinate spacetime spanned by the gravifield basis forms a coordinate-independence gravifield spacetime. Thus, the gravifield is thought to be associated with a gauge field in the coset $T_G^{1,3}$ = PG(1,3)/SP(1,3) of Poincar$\acute{\mbox{e}}$  gauge group PG(1,3) in the spinor representation of the locally flat noncoordinate gravifield spacetime. We shall show that the gauge-type gravifield is a basic field characterizing the gravitational interactions and the spin gauge field of spin group SP(1,3) reflects a torsional interaction. When transmuting the spin gauge symmetry into a hidden gauge symmetry via the gravifield, we arrive at a new formalism of the gauge theory of gravity with a global Poincar$\acute{\mbox{e}}$  symmetry in the flat Minkowski spacetime of coordinates. As a consequence, the general theory of relativity will be found to result as an effective theory of the gauge theory of gravity in the low-energy limit, and a general coordinate invariance of general linear group GL($4,R$) appears as a hidden symmetry embedded in the global Poincar$\acute{\mbox{e}}$  group P(1,3) via a gauge-fixing coordinate transformation of the general linear group. 

Our paper is organized as follows: After a brief introduction in this section, we are going to extend in Sec.II the global spin and scaling symmetries of basic fermion fields to be local spin and scaling gauge symmetries within the framework of QFT in the flat Minkowski spacetime, which enables us to treat the gravitational force on the same footing as other three basic forces. As a consequence,  it is inevitable to introduce simultaneously a bicovariant vector field $ \chi_{\mu}^{\;\; a} (x) $ defined on the globally flat Minkowski spacetime and valued in the locally flat noncoordinate spacetime. In Sec. III,  it is shown that the bicovariant vector field $\chi^{\;\;a}_{\mu}(x)$ is essential in characterizing the gravitational interactions and will be referred as {\it gravifield} for short. The so-called {\it gravifield} $\chi^{\;\; a}_{\mu}(x)$ can form a {\it gravifield basis} $\{\chi^{a}\}$. Such a  gravifield basis describes the locally flat noncoordinate spacetime that is named as a {\it gravifield spacetime}. Based on the locally flat gravifield spacetime,  we are able to construct a coordinate-independent and gauge-invariant action for a gravitational gauge theory,  where the fermion fields and gauge fields belong to the spinor representations and vector representations of the spin symmetry group SP(1,3), respectively. In Sec. IV,  we naturally express the action of gravity gauge theory in the globally flat Minkowski spacetime by simply converting the gravifield basis into the coordinate basis. Taking the globally flat Minkowski spacetime as an inertial frame for a reference to describe the motions of all quantum fields, we obtain explicitly their equations of motion in the existence of gravitational interactions, which enables us to study in principle the dynamics of all quantum fields including the gravitational effects. As the action provides a unified description for all basic forces based on the gauge symmetries within the framework of QFT, it makes a meaningful definition for momentum and energy and allows us to discuss the basic conservation laws of the theory under the local gauge invariance and global transformation invariance  in Sec. V.  We demonstrate that, alternative to Einstein's equation for the theory of general relativity, the equation of motion for the gravifield tensor is obtained in connection directly to the energy-momentum tensor, which leads to a basic conservation law for a gravifield tensor current in light of the energy-momentum conservation. In Sec. VI, we discuss the gravitational gauge symmetry breaking based on the fact that the spin and scaling gauge fields are not observed experimentally. It is postulated that the spin and scaling gauge symmetries are broken down to the global Lorentz and scaling symmetries, which results in a background structure described by the background fields. From the equations of motion for the background fields, we obtain a set of exact Lorentz-invariant solutions characterized by a cosmic vector with a nonzero cosmological mass scale. In Sec. VII, a gauge-invariant line element in the locally flat gravifield spacetime is defined with the gravifield basis,  the scaling scalar field ensures the line element to be invariant under both the local and global scaling transformations.  It is demonstrated that the background structure after gravitational gauge symmetry breaking generates a background gravifield spacetime which coincides with a conformally flat Minkowski spacetime governed by a background conformal scale field. Such a conformal scale field becomes singular at the cosmological horizon in terms of the conformal proper time. We then show that, in light of the cosmic proper time, the evolving Universe based on the background gravifield spacetime becomes conformally inflationary or deflationary.  Such a resulting Universe is in general not isotropic; only in a comoving reference frame where the cosmic proper time is directly correlated to the comoving Lorentz time does the Universe look like homogeneous and isotropic.  In Sec. VIII, we describe the quantization of the gravitational gauge theory based on the background structure of the gravifield spacetime. The physical quantum degrees of freedom are analyzed by applying for the usual counting rule of the gauge theory. In the path integral approach, we present the gauge-fixing contributions to the quantization of gravity theory and the Faddeev-Popov ghost term. The leading effective action is explicitly yielded for writing down the Feynman rules of the quantum gravity theory. The renormalizability of the theory is discussed. Then we demonstrate how the quantum effect causes the inflation of the early Universe and leads the inflationary Universe to end via spontaneous scaling symmetry breaking in the effective background scalar potential. In Sec. IX, an alternative spacetime gauge field defined in the flat Minkowski spacetime is constructed through the spin gauge field and the gravifield, where it is shown that the spin gauge symmetry becomes a hidden symmetry and the gravifield appears as a Goldstone-like field that transmutes the local spin gauge symmetry into the global Lorentz symmetry. Then we present an alternative formalism for the action of quantum gravity and show that the bosonic gravitational interactions can be described by the Goldstone-like gravimetric field and spacetime gauge field. In Sec. X, we explicitly show that the gravity equation can lead to two types of equation; one type of equation is the extension to Einstein's equation of general relativity, and the other is a new type of equation that characterizes the twisting and torsional effects. It indicates that the action of gravity gauge theory has a hidden symmetry for a local linear coordinate transformation GL($4,R$). The Einstein theory of general relativity is thought to be an effective low-energy theory. Our conclusions and remarks are presented in the last section.

\section{Gauge Symmetries and Gravitational Fields in QFT}

In the SM, the basic building blocks of matter are quarks and leptons which are all fermionic quantum fields with a half integer spin. Three basic forces electromagnetic, weak and strong interactions among quarks and leptons are all governed by the gauge symmetries U(1)$_Y \times$ SU(2)$_L \times$ SU(3)$_c$, respectively. The symmetry groups are related to the intrinsic quantum numbers of quarks and leptons. For a general consideration, let us begin with an action of Dirac fermion with gauge interactions,
\begin{eqnarray}
\label{action1}
S & = &  \int d^{4}x\, \{ \bar{\Psi}(x)  \frac{1}{2}\gamma^{a}  \delta_{a}^{\;\;\mu} i D_{\mu} \Psi(x) + H.c.\} 
+\frac{1}{2g_A^2}   \eta^{\mu\mu'}\eta^{\nu\nu'} Tr {\cal F}_{\mu\nu} (x) {\cal F}_{\mu'\nu'} (x) 
\end{eqnarray}
where we have used the definitions
\begin{eqnarray}
& &  i D_{\mu}\equiv  i\partial_{\mu}  + {\cal A}_{\mu}(x)\; ,\quad {\cal A}_{\mu}(x) = g_A {\cal A}_{\mu}^I(x) T^I\, , \quad  \eta^{\mu\nu} = \delta_{a}^{\;\;\mu} \delta_{b}^{\;\;\nu} \eta^{ab}\, , \nonumber \\
& & {\cal F}_{\mu\nu} (x) \equiv i[ D_{\mu}, D_{\nu}] = \partial_{\mu} {\cal A}_{\nu}(x) - \partial_{\nu} {\cal A}_{\mu}(x) - i [{\cal A}_{\mu}(x) ,  {\cal A}_{\nu}(x) ]  = g_A\; {\cal F}_{\mu\nu}(x)^I \; T^I \, ,
\end{eqnarray}
with $\delta_{a}^{\;\;\mu} $ the Kronecker symbol and $\gamma^a$ the Dirac $\gamma$ matrices. The quantum fields $\Psi_n(x)$ $(n=1,2,\dots)$ denote Dirac fermions which can be quarks and leptons in the SM, and ${\cal A}_{\mu}(x) = g_A {\cal A}_{\mu}^I(x) T^I$ ($I=1,2,\dots$) represent gauge fields with $T^I$ the generators of symmetry group $G$ and $g_A$ the coupling constant of gauge interactions. The generators satisfy the group algebra $[T^I, T^J] = i f^{IJK} T^K$  with the trace normalization $tr T^I T^J = \frac{1}{2} \delta^{IJ}$. ${\cal A}_{\mu}^I$ can be taken as the quantum gauge fields in the adjoint representations of gauge group G=U(1)$_Y \times$ SU(2)$_L \times$ SU(3)$_c$ in the SM. 
The Greek alphabet ($\mu,\nu = 0,1,2,3$) and  Latin alphabet ($a,b,=0,1,2,3$)  are used to distinguish four-vector indices in coordinate spacetime and noncoordinate spacetime, respectively.  As the theory discussed in this paper is based on the globally flat and locally flat spacetime, both the Greek and Latin indices are raised and lowered by the constant metric matrices, i.e., $\eta^{\mu\nu} $ and $\eta^{ab}$ with the signature $-2$, $\eta^{\mu\nu} $ or $\eta_{\mu\nu} =$ diag.$(1,-1,-1,-1)$ and $\eta^{ab}$ or $\eta_{ab} = $diag. $(1,-1,-1,-1)$.  Thus the scalar product of vectors and tensors is obtained via the contraction with the constant metric matrices $\eta_{\mu\nu}$ and $\eta_{ab}$, i.e., ${\cal A}^{\mu} {\cal A}_{\mu} = \eta^{\mu\nu} {\cal A}_{\mu}  {\cal A}_{\nu}$ and $\gamma^{a}\gamma_{a} = \eta_{ab} \gamma^a \gamma^{b}$. The system of units is chosen such that $c = \hbar = 1$.

The action is invariant under the gauge transformation $g(x) \in G$:
\begin{eqnarray}
& & {\cal A}_{\mu}(x) \to {\cal A}'_{\mu}(x) = g(x) {\cal A}_{\mu}(x) g^{\dagger}(x) +   g(x) i\partial_{\mu} g^{\dagger}(x) \; , \nonumber \\
& & \Psi(x) \to \Psi'(x) = g(x) \Psi(x)
\end{eqnarray}
Without gravitational interaction, the action is invariant under the global Lorentz transformations, where the gauge fields and Dirac fermions transform in the vector and spinor representations of Lorentz group, respectively, in the flat Minkowski spacetime:
\begin{eqnarray}
& & x^{\mu} \to x^{'\mu} = L^{\mu}_{\; \;\; \nu}\; x^{\nu}, \qquad \qquad   A_{\mu} (x) \to A'_{\mu}(x') = L_{\mu}^{\; \;\; \nu}\; A_{\nu}(x)  \; , \nonumber \\
& &\Psi(x) \to \Psi'(x') = S(L) \Psi(x), \quad S(L) \gamma^{a} S^{-1}(L) = L^a_{\;\;\; b}\; \gamma^b\; , 
\ee
with
\be
& &  S(L) = e^{i\alpha_{ab} \Sigma^{ab}/2}  \in \mbox{SP}(1,3), \quad  \Sigma^{ab} = \frac{i}{4}[\gamma^a, \gamma^b] 
\end{eqnarray}   
where $\Sigma^{ab}$ are the generators of spin group SP(1,3)$\sim$ SO(1,3) in the spinor representation,
\begin{eqnarray}
& & [\Sigma^{ab}, \Sigma^{cd}] = i (\Sigma^{ad}\eta^{bc} -\Sigma^{bd}  \eta^{ac} - \Sigma^{ac} \eta^{bd} + \Sigma^{bc} \eta^{ad}) , \nonumber \\
& & [ \Sigma^{ab}, \gamma^c] = i ( \gamma^a \eta^{bc} - \gamma^b\eta^{ac} ).
\end{eqnarray}
The action is invariant under the parallel translation for coordinates
\begin{eqnarray}
 x^{\mu} \to x^{'\mu} = x^{\mu} + a^{\mu}
 \end{eqnarray}
with $a^{\mu}$ the constant vector, and also under the global scaling transformation for coordinates and quantum fields,
\begin{eqnarray}
x^{\mu} \to x^{'\mu} = \lambda^{-1}\; x^{\mu}, \quad {\cal A}_{\mu} (x) \to {\cal A}'_{\mu}(x') = \lambda\; {\cal A}_{\mu}(x), \quad
\Psi(x) \to \Psi'(x') = \lambda^{3/2}\; \Psi(x), 
\end{eqnarray}
with $\lambda$ the constant scaling factor. 

Before discussing gravitational force, let us briefly analyze how the SM introduces the basic forces of electromagnetic, weak and strong interactions which are governed by the corresponding internal gauge symmetries U(1)$_Y \times$ SU(2)$_L \times$ SU(3)$_c$. Such gauge symmetries actually reflect the correlations of quantum numbers which characterize the intrinsic properties of quarks and leptons.  The U(1)$_Y$ gauge group reflects the charge quantum number of the particle and antiparticle, the SU(2)$_L$ gauge group characterizes the symmetry between two isospin quantum numbers of quarks and leptons, and the SU(3)$_c$ gauge group is introduced to describe the symmetry among three color quantum numbers for each flavor quark. Quarks and leptons as Dirac fermions all carry spin and chirality quantum numbers, which is known to be characterized by the spin symmetry group SP(1,3) $\sim$ SO(1,3). in the SM, such a symmetry of quarks and leptons must be a global symmetry, so that it coincides with the global Lorentz symmetry SO(1,3) of coordinates in the flat Minkowski spacetime to ensure the Lorentz invariance and the covariance of action in special relativity. 

Analogous to the introduction of internal gauge symmetries U(1)$_Y \times$ SU(2)$_L \times$ SU(3)$_c$ for the basic forces of electromagnetic, weak and strong interactions among quarks and leptons, it is natural to take the spin symmetry group SP(1,3) of quarks and leptons as an internal gauge symmetry which governs a basic force of {\it spin gauge interaction}. The corresponding spin gauge field and field strength are defined as follows:
\begin{eqnarray}
& & \Om_{\mu}(x) = g_s\; \Om_{\mu}^{ab}(x)\,  \frac{1}{2} \Sigma_{ab} \; , \nonumber \\
& & {\cal R}_{\mu\nu}(x) =  \partial_{\mu} \Om_{\nu}(x) - \partial_{\nu} \Om_{\mu}(x) - i [\Om_{\mu}(x) ,  \Om_{\nu}(x) ] = 
g_s {\cal R}_{\mu\nu}^{ab} (x)\,  \frac{1}{2} \Sigma_{ab}\; , 
\end{eqnarray}
with $g_s$ the coupling constant. It is required that the action is SP(1,3) gauge invariance under the spin gauge transformation $S(x) = e^{i\alpha_{ab}(x) \Sigma^{ab}/2}\in $ SP(1,3) for gauge fields and Dirac fermions,
\begin{eqnarray}
& & \Om_{\mu}(x) \to \Om'_{\mu}(x) = S(x) \Om_{\mu}(x) S^{-1}(x) + S(x)i \partial_{\mu} S^{-1}(x), \nonumber \\
& &\Psi(x) \to \Psi'(x) = S(x) \Psi(x) \; , \quad S^{-1}(x) \gamma^{a} S(x) = \Lambda^a_{\;\;\; b}(x)\; \gamma^b\, .
\end{eqnarray}

With the same consideration, we shall take the global scaling symmetry of Dirac fermions to be a local scaling gauge symmetry. Namely, the Dirac fermions transform under the local scaling gauge transformation as follows 
\begin{eqnarray}
& & \Psi(x) \to \Psi'(x) = \xi^{3/2} (x) \Psi(x) \; ,
\end{eqnarray}
while the internal gauge fields $ {\cal A}_{\mu} (x)$ and spin gauge field $\Om_{\mu}(x)$ are unchanged in the local scaling gauge transformation.

Analogously,  when extending the global scaling symmetry of quantum fields to be a local scaling gauge symmetry, but keeping a global scaling symmetry of coordinates, we  shall introduce the Weyl gauge field\cite{WG} $W_{\mu}(x) $, which governs a basic force of {\it scaling gauge interaction} and transforms under the local scaling gauge transformation as follows: 
\begin{eqnarray}
 W_{\mu}(x) \to W'_{\mu}(x) = W_{\mu}(x) + g_w^{-1} \partial_{\mu} \ln \xi(x)\, .
\end{eqnarray}
As the local scaling transformation is an Abelian gauge symmetry, the field strength is simply given by 
\be
{\cal W}_{\mu\nu} = \partial_{\mu}W_{\nu} - \partial_{\nu}W_{\mu} \, .
\ee

In demanding the action to be invariant under both the spin gauge transformation for the quantum fields and the global Lorentz transformation for the coordinates, namely, keeping the action to have a covariant form of special relativity in the flat Minkowski spacetime, it is necessary to introduce a bicovariant vector field defined on a locally flat noncoordinate spacetime and valued in a vector representation of Lorentz group SO(1,3) in the flat Minkowski spacetime, i.e., 
\be
\hat{\chi}_a = \hat{\chi}_a^{\;\;\mu}(x)\partial_{\mu} = \eta_{\mu\nu} \hat{\chi}^{\;\;\mu}_a(x) \partial^{\nu}\; ,
\ee
which transforms 
\begin{eqnarray}
& & \hat{\chi}_a^{\;\; \mu}(x) \to  \hat{\chi}_a^{'\;\mu}(x) =  \Lambda_a^{\;\; b}(x) \hat{\chi}_b^{\;\; \mu}(x) \; , \nonumber \\
& & \hat{\chi}_a^{\;\; \mu}(x) \to  \hat{\chi}_a^{'\; \mu}(x) = \xi(x) \hat{\chi}_a^{\;\; \mu}(x) \, ,
\end{eqnarray}
under the local spin and scaling gauge transformations, respectively, and
\begin{eqnarray}
& &  \hat{\chi}_a^{\;\; \mu}(x) \to  \hat{\chi}_a^{'\;\mu}(x') = L^{\mu}_{\; \;\; \nu}\; \hat{\chi}_a^{\;\; \nu}(x)\, , \quad 
 x^{\mu} \to x^{'\mu} = L^{\mu}_{\; \;\; \nu}\; x^{\nu}\; , \nonumber \\
& & \hat{\chi}_a^{\;\; \mu}(x) \to  \hat{\chi}_a^{'\; \mu}(x') = \hat{\chi}_a^{\;\;\mu}(x)\, , \quad 
 x^{\mu} \to x^{'\mu} = \lambda^{-1}\; x^{\mu}\; ,
\end{eqnarray}
under the global Lorentz and scaling transformations, respectively, where the transformations satisfy the following properties 
\be
& & \Lambda_a^{\;\; b}(x) \Lambda_c^{\;\; d}(x) \eta^{ac} = \eta^{bd}\; , \quad \Lambda_a^{\;\; b}(x) \in \mbox{SP}(1,3)\; , \nonumber \\
& & L^{\mu}_{\; \;\; \nu} L^{\rho}_{\; \;\; \sigma} \eta_{\mu\rho} = \eta_{\nu\sigma}\; , \qquad \quad  L^{\mu}_{\; \;\; \nu} \in \mbox{SO}(1,3)\; .
\ee

Regarding the bicovariant vector field $ \hat{\chi}^{\;\;\mu}_a(x)$ as a kind of matrix field and defining an inverse of $ \hat{\chi}^{\;\;\mu}_a(x)$, we obtain a dual bicovariant vector field $\chi_{\mu}^{\;\; a} (x)$ which satisfies the following orthonormal conditions  
\begin{eqnarray}
\chi_{\mu}^{\;\; a}(x) \hat{\chi}^{\;\;\nu}_a(x) = \chi_{\mu\, a} (x) \hat{\chi}_b^{\;\; \nu}(x)  \eta^{ab} = \eta_{\mu}^{\;\;\nu} \; ,\quad 
\hat{\chi}^{\;\; \mu}_a(x) \chi_{\mu}^{\;\; b} (x)  =   \hat{\chi}_{a\, \mu}(x)\chi_{\nu}^{\;\; b} (x) \eta^{\mu\nu} = \eta_{a}^{\;\;b} \; .
\end{eqnarray}
Such a bicovariant vector  field $\chi_{\mu}^{\;\, a}(x)$ exists once the determinant of $\hat{\chi}_{a}^{\;\; \mu}(x)$ is nonzero, namely,
\begin{eqnarray}
\det \chi_{\mu}^{\;\; a}(x)\equiv \chi(x) = \frac{1}{\det \hat{\chi}_a^{\;\;\mu}(x) }\equiv \frac{1}{\hat{\chi}(x)},\quad \hat{\chi}(x)\equiv \det \hat{\chi}_{a}^{\;\; \mu}(x)\neq 0  \; .
\end{eqnarray}
Obviously, the bicovariant vector field $\chi_{\mu}^{\;\; a}(x)$ transforms as 
\be
& &  \chi_{\mu}^{\;\; a}(x) \to \chi_{\mu}^{'\; a}(x) =  \chi_{\mu}^{\;\; b}(x) \Lambda_{b}^{\;\; a} (x) \; , \nonumber  \\
& &  \chi_{\mu}^{\;\; a}(x) \to  \chi_{\mu}^{'\; a}(x) = \xi^{-1}(x) \chi_{\mu}^{\;\; a}(x)
\ee
under the local spin and scaling gauge transformations, respectively, and 
\begin{eqnarray}
& &  \chi_{\mu}^{\;\; a} (x) \to  \chi_{\mu}^{'\; a}(x') = L_{\mu}^{\; \; \nu}\, \chi_{\nu}^{\;\; a} (x)\, , \quad 
 x^{\mu} \to x^{'\mu} = L^{\mu}_{\; \;\; \nu}\; x^{\nu}\; , \nonumber \\
& & \chi_{\mu}^{\;\; a}(x) \to  \chi_{\mu}^{'\;a} (x') = \chi_{\mu}^{\;\; a} (x)\, , \quad 
 x^{\mu} \to x^{'\mu} = \lambda^{-1}\; x^{\mu}\; ,
\end{eqnarray}
under the global Lorentz and scaling transformations, respectively.  $ \chi_{\mu}^{\;\; a} (x) $ can be thought of as a bicovariant vector field defined on the globally flat Minkowski spacetime and valued in the vector representation of SP(1,3) in the locally flat noncoordinate spacetime,
\be
\chi_{\mu} =  \chi_{\mu}^{\;\; a} (x)\, \frac{1}{2} \gamma_a \, .
\ee

It is seen that the bicovariant vector field $\chi_{\mu}^{\;\; a} (x)$ or dual bicovariant vector field $ \hat{\chi}_a^{\;\;\mu}(x)$ transforms in a covariant form on both the globally flat Minkowski spacetime of coordinates and the locally flat noncoordinate spacetime. Here we would like to address that the bicovariant vector field $\chi_{\mu}^{\;\; a} (x) $ is introduced to distinguish from the so-called tetrad denoted usually by $e_{\mu}^a(x)$, which is required to be a general covariant vector field under the general coordinate transformations in the Einstein theory of general relativity. The bicovariant vector field $\chi_{\mu}^{\;\; a} (x) $ will be shown to constitute a basis for the locally flat noncoordinate spacetime and a basic gravitational field in the globally flat Minkowski spacetime of coordinates.

With the above analyses, we shall be able to construct the action which is invariant under the local spin and scaling gauge transformations for the quantum fields, and also under the global Lorentz and scaling transformations for the coordinates.  Let us first examine an action for Dirac fermions as the currently known building blocks of nature are quarks and leptons. The action can simply be written as follows:
\begin{eqnarray}
\label{actionF}
S_{F}  & = & \int d^{4}x\; \chi(x)\, \frac{1}{2} \{ \bar{\Psi}(x) \hat{\chi}^{\mu}(x) i{\mathcal D}_{\mu} \Psi(x) + H.c.\}  
\end{eqnarray}
with the definitions
\begin{eqnarray}
 \hat{\chi}^{\mu}(x)i {\mathcal D}_{\mu} & = &  \frac{1}{2} \gamma^a \hat{\chi}_a^{\;\;\mu}(x) {\mathcal D}_{\mu} = \frac{1}{2} \gamma_a \eta^{ab} \hat{\chi}_{b\nu}(x)  \eta^{\mu\nu} [\, i\partial_{\mu}  + {\cal A}_{\mu}(x)  +  \Om_{\mu}(x)  \, ]  \; 
\end{eqnarray}
and 
\begin{eqnarray}
 i {\mathcal D}_{\mu} & = & i\partial_{\mu}  + {\cal A}_{\mu}(x) + \Om_{\mu}(x) = i D_{\mu} + \Om_{\mu}(x)  \; ,\quad  i D_{\mu}  = i\partial_{\mu}  + {\cal A}_{\mu}(x) \, , 
 \end{eqnarray}
where ${\mathcal D}_{\mu}$ is the covariant derivative for the usual internal gauge and spin gauge symmetries, while $D_{\mu}$ is the derivative only for the usual internal gauge symmetries. It is easy to show that the above action is invariant under the usual internal and spin gauge transformations as well as under the global Lorentz and scaling transformations in the flat Minkowski spacetime. It is interesting to note that the Hermiticity of the action automatically ensures invariance under the scaling gauge transformation, which means that the scaling gauge field has actually no interaction with the fermion fields due to Hermiticity.


\section{ Gravifield Spacetime  and Gauge Theory of Gravity } 

It is unlike gauging the usual internal symmetries of fermion fields, such as the isospin and color symmetries of the quarks,  when gauging the spin and scaling symmetries of fermion fields in the flat Minkowski spacetime of coordinates, the bicovariant vector field $\chi^{\;\;a}_{\mu}(x)$ or its inverse $\hat{\chi}_a^{\;\;\mu}(x)$ is an essential ingredient in order to distinguish the local spin and scaling gauge transformations of fermion fields from the  global Lorentz and scaling transformations of coordinates. We will show that the bicovariant vector field $\chi^{\;\;a}_{\mu}(x)$ is a basic field in characterizing the gravitational interactions.  

\subsection{Gravifield spacetime}

Geometrically,  the bicovariant vector field $\hat{\chi}_a^{\;\;\mu}(x)$ is defined on both the locally flat noncoordinate spacetime and globally flat Minkowski spacetime of coordinates, and transforms as a bicovariant vector field under both the local spin gauge transformation and global Lorentz transformation. Explicitly, $\hat{\chi}_a^{\;\;\mu}(x)$ can be expressed in terms of the basis of coordinate spacetime with valued in the Dirac $\gamma$-matrix basis $\{\gamma^a/2\}$:
\be
& & \frac{1}{2} \gamma^a\, \hat{\chi}^{\;\;\mu}_a(x)\partial_{\mu} = \frac{1}{2} \gamma^a\, \hat{\chi}_a  = \hat{\chi}^{\;\mu}\partial_{\mu} \; ,
\ee
with
\be
& & \hat{\chi}_a =  \hat{\chi}^{\;\;\mu}_a(x)\partial_{\mu} \, , \quad \hat{\chi}^{\;\mu} = \frac{1}{2} \gamma^a \hat{\chi}^{\;\;\mu}_a(x)\, ,
\ee
where the derivative operator $\partial_{\mu} \equiv \partial/\partial x^{\mu}$ defines a basis $\{\partial_{\mu}\}\equiv \{\partial/\partial x^{\mu}\} $ in the globally flat Minkowski spacetime of coordinates.  Accordingly, the vector field $ \hat{\chi}_a$ forms a basis $ \{\hat{\chi}_a \} $ for the locally flat noncoordinate spacetime. We shall call $\{\partial_{\mu}\}$ the coordinate basis and $ \{\hat{\chi}_a \} $ the corresponding noncoordinate basis. In the coordinate basis, there is a dual basis $\{dx^{\mu}\}$ that satisfies the conditions
\be
 \langle dx^{\nu},\, \partial/\partial x^{\mu}  \rangle = \frac{\partial x^{\nu}}{\partial x^{\mu}} = \eta_{\mu}^{\; \nu}
\ee

The bicovariant vector field $\chi_{\mu}^{\;\; a}(x)$ as the inverse of $\hat{\chi}_a^{\;\;\mu}(x)$ can be expressed in terms of the dual coordinate basis $\{dx^{\mu}\}$ valued in the Dirac $\gamma$-matrix basis $\{\gamma^a/2\}$:
\be
& & \frac{1}{2} \gamma_a\, \chi_{\mu}^{\;\; a} (x) dx^{\mu}  =  \frac{1}{2} \gamma_a\, \chi^{\; a} (x) =  \chi_{\mu}(x) dx^{\mu}\; , \nonumber \\
& & \chi^{\; a}   = \chi_{\mu}^{\;\; a} (x) dx^{\mu} \; , \quad  \chi_{\mu} = \chi_{\mu}^{\;\; a} (x) \frac{1}{2} \gamma_a \, , 
\ee
where the vector field $\chi^{\, a}$ defines a dual basis $\{\chi^{\, a}\}$ for the locally flat noncoordinate spacetime, since
 \be
 \langle \chi^{\, b},\,  \hat{\chi}_a \rangle = \chi_{\nu}^{\;\; b}(x)  \hat{\chi}_a^{\;\; \mu} (x)  \angle dx^{\nu} ,\, \partial_{\mu} \rangle =  \chi_{\nu}^{\;\; b}(x)  \hat{\chi}_a^{\;\; \mu} (x)  \eta_{\mu}^{\; \nu} = \eta_a^{\;\, b} \, .
\ee

For the Dirac $\gamma$-matrix basis $\{\gamma^a/2\}$, we have
 \be
 \langle \frac{1}{2}\gamma^a, \, \frac{1}{2}\gamma^b \rangle = \frac{1}{4}Tr\gamma^a\gamma^b = \eta^{ab} \, ,
\ee
which leads to the orthogonal property
 \be
 \langle \chi_{\mu}  , \, \hat{\chi}^{\; \nu} \rangle = \chi_{\mu}^{\;\; a} (x)\hat{\chi}_b^{\;\; \nu}(x) 
 \langle \frac{1}{2}\gamma_a, \, \frac{1}{2}\gamma^b \rangle = Tr \chi_{\mu}(x)\hat{\chi}^{\; \nu}(x)   = \eta_{\mu}^{\; \, \nu} \, .
\ee

The vector field $\chi^{\, a}(x) $ may be regarded as a one-form gauge potential in the flat Minkowski spacetime. It is natural to speculate that the bicovariant vector field $\chi_{\mu}^{\;\; a}(x)$, which characterizes the locally flat noncoordinate spacetime with the basis $\{\chi^{\, a}\}$, should describe the gravitational interactions as a gauge-type potential field of gravity. We shall adopt  an alternative notation when emphasizing the bicovariant vector field to be a gauge-type potential field of gravity 
\be
{\mathsf G}_{\mu}(x) \equiv \chi_{\mu}^{\;\; a} \frac{1}{2} \gamma_a \, , \qquad  {\mathsf G} = -i {\mathsf G}_{\mu} dx^{\mu} \, .
\ee 

In this sense, $\chi^{\;\;a}_{\mu}(x)$ or $\hat{\chi}^{\;\;\mu}_{a}(x)$ is identified to be the gravitational bicovariant vector field and referred as {\it gravifield} for short in the following discussions.
The noncoordinate bases $\{\chi^{a}\}$ and $\{\hat{\chi}_{a}\}$ characterized by the dual {\it gravifield} $\chi^{\;\; a}_{\mu}(x)$ and $\hat{\chi}^{\;\;\mu}_{a}(x)$ are correspondingly called {\it gravifield bases}. Accordingly, the locally flat noncoordinate spacetime spanned by the vector gravifield $\chi^a (x)$ is mentioned as the locally flat {\it gravifield spacetime}. 

In the locally flat {\it gravifield spacetime}, it can be checked that the {\it gravifield basis} ${\hat{\chi}_a }$ does not commute; it satisfies the following commutation relation:
\be
[ \hat{\chi}_a ,\; \hat{\chi}_b] = \chi_{ab}^c\,  \hat{\chi}_c, \qquad  \chi_{ab}^c \equiv - \hat{\chi}_a^{\;\; \mu} \hat{\chi}_b^{\;\; \nu} \chi_{\mu\nu}^c \, ; \qquad \chi_{\mu\nu}^c  = \partial_{\mu}\chi_{\nu}^{\;\; c} - \partial_{\nu}\chi_{\mu}^{\;\; c} 
\ee 
which indicates that the locally flat gravifield spacetime is in general associated with a non-commutative geometry, where the gauge-type field tensor $\chi_{\mu\nu}^c $ will be shown to reflect the gravitational field strength.

\subsection{Gauge theory of gravity}

In terms of the gravifield bases $\chi^{a}$ and $\hat{\chi}_{a}$,  we can define a coordinate-independent exterior differential operator in the locally flat gravifield spacetime
\be
d_{\chi} =  \chi^{a}\wedge \hat{\chi}_a   
\ee
Thus, instead of the usual exterior differential forms  in the flat Minkowski spacetime of coordinates, we can define coordinate-independent exterior differential forms in the locally flat gravifield spacetime. For instance, the spin gauge potential and field strength can be expressed as the following one-form and two-form:
\be 
& & \Om = -i \Om_a\, \chi^a\; , \qquad {\cal R} = d_{\chi} \, \Om + \Om\wedge \Om =\frac{1}{2i} {\cal R}_{ab}\, \chi^a\wedge \chi^b \; , 
\ee
where $\Om_a$ and ${\cal R}_{ab}$ are the spin gauge potential and field strength sided on the locally flat gravifield spacetime,respectively. From the above definitions, it can be checked that  $\Om_a$ and ${\cal R}_{ab}$ are correlated to the spin gauge potential $\Om_{\mu}$ and field strength ${\cal R}_{\mu\nu}$  defined on the flat Minkowski spacetime as follows:
\be
 \Om_a = \hat{\chi}_a^{\;\; \mu} \Om_{\mu} \; , \qquad {\cal R}_{ab} =  \hat{\chi}_a^{\;\; \mu}\,\hat{\chi}_b^{\;\; \nu} {\cal R}_{\mu\nu} 
\ee
The Hodge star is defined as 
\be
\ast {\cal R} = \frac{1}{4i} \epsilon^{ab}_{\;\;\;\; cd}\,  {\cal R}_{ab}\, \chi^c\wedge \chi^d 
\ee

Similarly, we can express other gauge potential and field strength as well as vector and tensor fields in terms of exterior differential forms in the locally flat gravifield spacetime. For the internal gauge field and Weyl gauge field, we have 
\be 
& & {\cal A} = -i {\cal A}_a\, \chi^a\; , \qquad {\cal F} = d_{\chi}\, {\cal A} + {\cal A} \wedge {\cal A} = \frac{1}{2i} {\cal F}_{ab}\, \chi^a\wedge \chi^b \; , \nonumber \\
& & W = -i W_a\, \chi^a\; , \qquad {\cal W} = d_{\chi}\, W = \frac{1}{2i} {\cal W}_{ab}\, \chi^a\wedge \chi^b \; .
\ee

For the gravifield, the corresponding gauge potential and field strength in the locally flat gravifield spacetime are defined as
\be
{\mathsf G} = -i {\mathsf G}_a \, \chi^a\; , \qquad {\cal G} = d_{\chi} \, {\mathsf G} + \Om\wedge {\mathsf G}  + g_w W \wedge {\mathsf G} = \frac{1}{2i} {\cal G}_{ab}\, \chi^a\wedge \chi^b \; . 
\ee
Their relations to the gauge potential ${\mathsf G}_{\mu}$ and field strength ${\cal G}_{\mu\nu}$ in the flat Minkowski spacetime are given by 
\be
 {\mathsf G}_a = \hat{\chi}_a^{\;\; \mu} {\mathsf G}_{\mu} \; , \qquad {\cal G}_{ab} =  \hat{\chi}_a^{\;\; \mu}\,\hat{\chi}_b^{\;\; \nu} {\cal G}_{\mu\nu} \, ,
\ee
with the field strength
\be
{\cal G}_{\mu\nu} & = & \nabla_{\mu}\chi_{\nu}  - \nabla_{\nu}\chi_{\mu} =  [\, \nabla_{\mu}\chi_{\nu}^{\;\; a} (x)  - \nabla_{\nu}\chi_{\mu}^{\;\; a}(x) \,] \, \frac{1}{2}\gamma_a \equiv {\cal G}_{\mu\nu}^a(x) \, \frac{1}{2}\gamma_a  \; , \nonumber \\
{\cal G}_{\mu\nu}^a(x)  & = & (\partial_{\mu} + g_w W_{\mu} ) \chi_{\nu}^{\;\; a}  + g_s\Om_{\mu\;\; b}^{\;a}  \chi_{\nu}^{\;\; b}  - (\partial_{\nu}+ g_w W_{\nu} )\chi_{\mu}^{\;\; a}   - g_s\Om_{\nu\;\; b}^{\; a}  \chi_{\mu}^{\;\; b} \, ,
\ee
where the covariant derivative is defined as 
\be
\nabla_{\mu} = \partial_{\mu}  + \Om_{\mu} +  g_w W_{\mu}  = \bigtriangledown_{\mu}  +  g_w W_{\mu} \, , 
\quad \td_{\mu} = \partial_{\mu}  + \Om_{\mu} \, .
\ee

The gauge-type field strength of the gravifield, ${\cal G}_{\mu\nu}(x) \equiv  \nabla_{\mu}\chi_{\nu}  - \nabla_{\nu}\chi_{\mu}$, is defined as the two-form through the one-form gauge potential. In general, one can define the covariant derivative in the locally flat gravifield spacetime as
\be
& & {\mathcal D}_a  =  \hat{\chi}_a -i {\cal A}_{a} - i \Om_{a} =    \hat{\chi}_a^{\;\;\mu} {\mathcal D}_{\mu} = \hat{\chi}_a^{\;\;\mu} (\, \partial_{\mu} -i {\cal A}_{\mu} -i \Om_{\mu}  \, )  \; 
\ee

With the exterior differential forms in the locally flat gravifield spacetime and the requirement of renormalizability of QFT in four-dimensional spacetime, the gauge-invariant action for the gravitational gauge theory is constructed as follows 
\begin{eqnarray}
\label{action2}
S_{\chi}  & = & \int \{\, \frac{1}{2} [i \bar{\Psi} \ast \chi \wedge {\mathcal D} \, \Psi\, + i \bar{\psi} \ast \chi \wedge \td \, \psi\, + H.c. ] 
+ y_s Tr\, (\chi \wedge \chi)\wedge \ast (\chi \wedge \chi )\, \bar{\psi} \phi \psi \nonumber \\
& - &  \frac{1}{g_A^2}  Tr\, {\cal F} \wedge \ast {\cal F}   - \frac{1}{g_s^2}  Tr\, {\cal R} \wedge \ast {\cal R}  -    \frac{1}{2} {\cal W} \wedge \ast {\cal W}    +  \frac{1}{2} \, \alpha_W\, \phi^2 Tr\, {\cal G}\wedge \ast {\cal G}  \nonumber  \\
& - &  \frac{1}{2}  d \phi \wedge \ast d \phi   - \alpha_E  Tr\, {\cal R} \wedge \ast (\chi \wedge \chi ) \,  \phi^2  + \lambda_s Tr\, (\chi \wedge \chi)\wedge \ast (\chi \wedge \chi )\, \phi^4 \,  +  {\cal L}' \} \, ,
\end{eqnarray}
The couplings $y_s$, $g_s$, $\alpha_W$, $\alpha_E$ and $\lambda_s$ are the constant parameters. ${\cal L}'$ denotes the Lagrangian density for possible other interactions. We have used the definitions and relations
\be
& & {\mathcal D} = \chi^a {\mathcal D}_a \; , \quad \ast \chi^a = \frac{1}{3!} \epsilon^a_{\;\; bcd} \,  \chi^b \wedge \chi^d \wedge \chi^c \; , \nonumber \\
& & (\chi \wedge \chi)  \equiv \chi^a \wedge \chi^b\, \frac{1}{2i} \Sigma_{ab} \; , \qquad \ast (\chi \wedge \chi)  = \frac{1}{2} \epsilon^{ab}_{\;\;\;\; cd}\, \chi^c \wedge \chi^d\, \frac{1}{2i} \Sigma_{ab} \; , \nonumber \\
& & d \phi =  (d_{\chi} - i g_w W )\phi\; , \qquad \ast d \phi =  \frac{1}{3!} \epsilon^{a}_{\;\; bcd} \, \chi^b \wedge \chi^d \wedge \chi^c \,  (\hat{\chi}_{a} - g_w W_{a})\phi \; ,
\ee
and the totally antisymmetric Levi-Civita tensor $ \epsilon^{abcd}$ ( $\epsilon^{0123} = 1$,  $\epsilon_{abcd} = - \epsilon^{abcd} $) satisfies the identities 
\be
& & \epsilon^{abcd}\epsilon_{ab}^{\;\;\;\; c'd'} = -2(\eta^{cc'}\eta^{dd'} - \eta^{cd'}\eta^{dc'} ) \; , \nonumber \\
& & \epsilon^{abcd}\epsilon_{abc}^{\;\;\;\;\;\; d'} = -6\eta^{dd'},\qquad \epsilon_{abcd}\epsilon^{abcd} = -24 \,  .
\ee

We have introduced a scalar field $\phi(x)$ to ensure both the global scaling and local scaling symmetries for the gauge-type gravifield interaction characterized by the coupling constant $\alpha_W$ and the scalar-type spin gauge interaction characterized by the coupling constant $\alpha_E$. A singlet fermion field $\psi(x)$ is introduced to couple with the scalar field, which will be shown to play a significant role for the quantum inflation of the Universe.  Here the scalar field $\phi(x)$ characterizes the conformal scaling property and transforms as
\begin{eqnarray}
& & \phi(x) \to \phi'(x) = \xi(x) \phi(x) \; , \nonumber \\
& & \phi(x) \to \phi'(x') = \lambda\, \phi(x)  \, , \quad  x^{\mu} \to x^{'\mu} = \lambda^{-1}\, x^{\nu} 
\end{eqnarray}
under the local scaling gauge transformation and the global scaling transformation, respectively.

The above general action for the gravitational gauge theory is given in the locally flat gravifield spacetime, the fermion fields and gauge fields all belong to the spinor representations and vector representations of the spin group SP(1,3), respectively.

\section{Field Equations and Dynamics in the QFT of Gravity}

To obtain the field equations of motion and study the dynamics of fields, it is useful to transform the action of gauge theory of gravity constructed in the locally flat gravifield spacetime into the action expressed in the globally flat Minkowski spacetime. This can simply be realized by converting the gravifield basis into the coordinate basis through the gravifield.  Taking the globally flat Minkowski spacetime as an inertial frame for a reference, we are able to describe the motions of quantum fields and make a meaningful definition for the momentum and energy. In particular, it enables us to provide a unified description for all basic forces based on the framework of QFT for gauge symmetries.

It is not difficult to check that the general action of gauge theory of gravity gets the following expression in the globally flat Minkowski spacetime
\begin{eqnarray}
\label{action3}
S_{\chi}  & = & \int d^{4}x\; \chi\,  \{\,  \frac{1}{2} [\, \hat{\chi}^{\mu\nu} ( \bar{\Psi} \chi_{\mu} i {\mathcal D}_{\nu}   \Psi  + \bar{\psi} \chi_{\mu} i \td_{\nu}   \psi  ) + H.c.\, ] - y_s \bar{\psi} \phi \psi   \nonumber \\
& - &  \frac{1}{4}  \hat{\chi}^{\mu\mu'} \hat{\chi}^{\nu\nu'} [\,  {\cal F}^I_{\mu\nu} {\cal F}^{I}_{\mu'\nu'} + {\cal R}_{\mu\nu}^{ab} {\cal R}_{\mu'\nu'ab} + {\cal W}_{\mu\nu} {\cal W}_{\mu'\nu'}  - \alpha_W\,  \phi^2\,  {\cal G}_{\mu\nu}^a {\cal G}_{\mu'\nu' a}  \, ]\nonumber \\
& + &  \frac{1}{2} \hat{\chi}^{\mu\nu} d_{\mu} \phi d_{\nu}\phi -  \alpha_E g_s \phi^2  \hat{\chi}^{\mu\mu'} \hat{\chi}^{\nu\nu'} \chi_{\mu}^{\;\,a } \chi_{\nu}^{\;\,b}  {\cal R}_{\mu'\nu' a b} - \lambda_s \phi^4 \, + {\cal L}'(x)   \}  \, ,
\end{eqnarray}
with the definitions
\be \label{tensor}
& &  \hat{\chi}^{\mu\nu}(x) = \hat{\chi}_{a}^{\;\;\mu}(x) \hat{\chi}_{b}^{\;\;\nu}(x) \eta^{ab}\, , \quad \chi_{\mu} \equiv \chi_{\mu}^{\;\, a} \gamma_a \; ,
\ee
where the tensor field $ \hat{\chi}^{\mu\nu}(x)$ couples to all fields.

The above action of gauge theory of gravity is now described within the framework of relativistic QFT in the globally flat Minkowski spacetime. Where the gravifield $\chi_{\mu}^{\;\; a}(x)$ appears as a gauge-type field, its dynamics is governed by the gauge-type interaction of the field strength ${\cal G}_{\mu\nu}^a(x)$. The antisymmetric field strength tensor  ${\cal G}_{\mu\nu}^a(x)$ is valued in the homogeneous vector representation of the spin gauge group SP(1,3). Unlike the usual internal gauge fields, the gravifield $\chi_{\mu}^{\;\; a}(x)$ couples inversely to all kinematic terms and also interaction terms of  quantum fields. It becomes manifest that the gravifield $\chi_{\mu}^{\;\; a}(x)$ is a basic gauge-type field and its corresponding field strength ${\cal G}_{\mu\nu}^a(x)$ characterizes the gravitational force.

Based on the above gauge-invariant action and in light of the least action principle,  we are able to obtain  equations of motion for all fields under an infinitesimal variation of the fields. To write down the explicit forms of equations of motion, we shall not consider the Lagrangian density $ {\cal L}' $. The equation of motion for the fermion fields $\Psi$ is easily obtained as follows
\be \label{EM0}
& & \chi \hat{\chi}_a^{\;\; \mu} \gamma^a i {\mathcal D} _{\mu} \Psi + \frac{1}{2} i\btd_{\mu} (\chi \hat{\chi}_a^{\;\;\mu}) \gamma^a \Psi = 0 \; .
\ee
Let us define a spin gauge-invariant vector field 
\be
{\mathsf V}_{\mu} (x) & \equiv &  \frac{1}{2} \hat{\chi}\chi_{\mu}^{\;\; c} \td_{\rho} (\chi \hat{\chi}_c^{\;\; \rho})  \, ,
\ee
so that the equation of motion for the fermion fields can simply be written as
\be \label{EM0-0}
& & \gamma^a \hat{\chi}_a^{\;\; \mu} i ( {\mathcal D} _{\mu} + {\mathsf V} _{\mu} ) \Psi  =   0 \, .
\ee

In a similar way, we arrive at the equation of motion for the singlet fermion field $\psi$ 
\be \label{EM0-0}
& & \gamma^a \hat{\chi}_a^{\;\; \mu} i ( \td _{\mu} + {\mathsf V} _{\mu} ) \psi  =   0 \, .
\ee

It is not difficult to yield the equation of motion for the gauge fields ${\cal A}_{\mu}^I$:
\be   \label{EM1}
D_{\nu} (\chi \hat{\chi}^{\mu\mu'} \hat{\chi}^{\nu\nu'}  {\cal F}^{I}_{\mu'\nu'} )  = J^{I\,\mu} \, ,
\ee
with the fermionic vector currents
\be
J^{I\,\mu} = g_A \chi  \bar{\Psi} \gamma^a \hat{\chi}_a^{\;\;\mu} T^I \Psi \, .
\ee
 The internal gauge covariant derivative is given by
 \be
 D_{\nu} (\chi \hat{\chi}^{\mu\mu'} \hat{\chi}^{\nu\nu'}  {\cal F}^{I}_{\mu'\nu'} ) = \chi \hat{\chi}^{\mu\mu'} \hat{\chi}^{\nu\nu'} D_{\nu} {\cal F}^{I}_{\mu'\nu'}  + \partial_{\nu} (\chi \hat{\chi}^{\mu\mu'} \hat{\chi}^{\nu\nu'} ) {\cal F}^{I}_{\mu'\nu'} 
 \ee
where the second term is caused by the gravitational interactions.

For the spin gauge field$\Om_{\mu}^{ab}$, we arrive at the following equation of motion:
\be   \label{EM2}
\btd_{\nu} ( \chi   \hat{\chi}^{\mu\mu'} \hat{\chi}^{\nu\nu'} {\cal R}_{\mu'\nu' }^{\;\; ab} ) & = &   J^{\mu\, ab}
\ee
with the vector-tensor currents
\be \label{EM2b}
J^{\mu\, ab} &  =  &\frac{1}{2} g_s \chi \bar{\Psi} \hat{\chi}_c^{\;\; \mu} \{ \gamma^c\;\;  \frac{1}{2} \Sigma^{ab} \} \Psi  
 + \frac{1}{2} g_s \chi \bar{\psi} \hat{\chi}_c^{\;\; \mu} \{ \gamma^c\;\;  \frac{1}{2} \Sigma^{ab} \} \psi \nonumber \\
& - & \alpha_E g_s \btd_{\nu} (\chi   \hat{\chi}^{\mu\mu'} \hat{\chi}^{\nu\nu'}   \phi^2 \chi_{\mu'\nu'}^{[ab]} ) +  \frac{1}{2} \chi \phi^2 \alpha_W \hat{\chi}^{\mu \mu'} \hat{\chi}^{\nu\nu'} 
 \chi_{\nu}^{\;\; [a} {\cal G}_{\mu'\nu'}^{b]}  \, ,
 \ee
where we have used the notations
\be
& & \chi_{\mu'\nu'}^{[ab]}   =   \chi_{\mu'}^{\;\; a} \chi_{\nu' }^{\;\; b}  - \chi_{\mu' }^{\;\; b} \chi_{\nu' }^{\;\; a}  \, ; \quad 
  \chi_{\nu}^{\;\; [a} {\cal G}_{\mu'\nu'}^{\;\; b]} = \chi_{\nu}^{\;\; a} {\cal G}_{\mu'\nu'}^{\;\; b }  - \chi_{\nu}^{\;\; b}  {\cal G}_{\mu'\nu' }^{\;\; a} \, .
\ee
The tensor currents as the sources for the dynamics of the spin gauge field$\Om_{\mu}^{ab}$ consist of two parts; one is  the fermionic tensor current, and the other is made of the gravifield $\chi_{\mu}^{\;\; a}$. The spin gauge covariant derivative is given by
\be
\btd_{\nu} ( \chi   \hat{\chi}^{\mu\mu'} \hat{\chi}^{\nu\nu'}  {\cal R}_{\mu'\nu' }^{\;\; ab} ) & = & \chi   \hat{\chi}^{\mu\mu'} \hat{\chi}^{\nu\nu'} \btd_{\nu}  {\cal R}_{\mu'\nu' }^{\;\; ab}  +\partial_{\nu} ( \chi   \hat{\chi}^{\mu\mu'} \hat{\chi}^{\nu\nu'} ) {\cal R}_{\mu'\nu' }^{\;\; ab} 
\ee
with the second term from the gravitational effects.

The equation of motion for the gravifield $\chi_{\mu}^{\;\; a}$ is found to be 
\be  \label{EM3}
& & \alpha_W (\td_{\nu} -g_w W_{\nu}) (\, \phi^2  \chi   \hat{\chi}^{\mu \mu'} \hat{\chi}^{\nu\nu'}
  {\cal G}_{\mu'\nu' a} \, ) = J_a^{\;\; \mu} \, ,
\ee
with the bicovariant vector currents
\be \label{EM3b}
 J_a^{\;\; \mu} & = & - \chi \hat{\chi}_a^{\;\;\mu} {\cal L} + \frac{1}{2} \chi \hat{\chi}_a^{\; \; \rho}  \hat{\chi}_c^{\;\; \mu}  [ \bar{\Psi} \gamma^c i {\mathcal D}_{\rho} \Psi  + \bar{\psi} \gamma^c i \td_{\rho} \psi   + H.c.\, ] \nonumber \\
& - & \chi \hat{\chi}_a^{\;\; \rho} \hat{\chi}^{\mu \mu'} \hat{\chi}^{\nu\nu'}  [\,  {\cal F}^I_{\rho\nu} {\cal F}^{I}_{\mu'\nu'} + {\cal R}_{\rho\nu}^{cd} {\cal R}_{\mu'\nu'\; cd} +  {\cal W}_{\rho\nu} {\cal W}_{\mu'\nu'}  - 
\alpha_W  \phi^2 {\cal G}_{\rho\nu}^{b} {\cal G}_{\mu'\nu' b}    \, ] \nonumber \\
& + &  \chi \hat{\chi}_a^{\;\; \nu'} \hat{\chi}^{\mu \mu'} d_{\mu'}\phi  d_{\nu'} \phi   - 2 \alpha_E g_s \chi \phi^2  \hat{\chi}_{c}^{\;\;\mu} \hat{\chi}_{a}^{\;\; \mu'}    {\cal R}_{\mu'\nu'}^{cd}\hat{\chi}_{d}^{\;\; \nu'} 
  \, .
\ee
The spin and scaling gauge covariant derivative can be written as
\be
& & (\td_{\nu} -g_w W_{\nu})  (\, \phi^2  \chi   \hat{\chi}^{\mu \mu'} \hat{\chi}^{\nu\nu'}
  {\cal G}_{\mu'\nu' a} \, ) \nonumber \\
  & & \quad = \chi   \hat{\chi}^{\mu \mu'} \hat{\chi}^{\nu\nu'} (\td_{\nu} -g_w W_{\nu}) (\, \phi^2
  {\cal G}_{\mu'\nu' a} \, ) + \partial_{\nu} (\, \chi   \hat{\chi}^{\mu \mu'} \hat{\chi}^{\nu\nu'} \, ) \phi^2
  {\cal G}_{\mu'\nu' a}  \, .
\ee
In obtaining the above equation of motion, we have used the identities $\delta \chi =  \chi \hat{\chi}_a^{\;\; \mu} \delta \chi_{\mu}^{\;\; a}$ and $\delta \hat{\chi}_b^{\;\;\nu} = - \hat{\chi}_b^{\;\; \mu} \hat{\chi}_a^{\;\; \nu} \delta \chi_{\mu}^{\;\; a}$ .  Alternatively, the equation of motion for the dual gravifield $\hat{\chi}_a^{\;\;\mu}$  can easily be read off: 
\be  \label{EM3.1}
& &  \chi_{\mu}^{\;\; c} \chi_{\rho}^{\;\; a} \alpha_W \tilde{\nabla}_{\sigma}  (\,  \phi^2  \chi   \hat{\chi}^{\rho \mu'} \hat{\chi}^{\sigma \nu'}  {\cal G}_{\mu'\nu' c}\,  )  
=  \hat{J}_{\mu}^{\;\; a} \, , \nonumber
\ee
with the gravifield currents $\hat{J}_{\mu}^{\;\; a}  = \chi_{\mu}^{\;\; c} \chi_{\rho}^{\;\; a} J_c^{\;\; \rho}$
\be
\hat{J}_{\mu}^{\;\; a} & = & - \chi \chi_{\mu}^{\;\; a} {\cal L} + \frac{1}{2} \chi  [ \bar{\Psi} \gamma^a i {\mathcal D}_{\mu} \Psi  +
\bar{\psi} \gamma^a i \td_{\mu} \psi  + H.c.\, ] \nonumber \\
& - & \chi \hat{\chi}^{a\mu'} \hat{\chi}^{\nu\nu'}  [\,  {\cal F}^I_{\mu\nu} {\cal F}^{I}_{\mu'\nu'} + {\cal R}_{\mu\nu}^{cd} {\cal R}_{\mu'\nu'\; cd} +  {\cal W}_{\mu\nu} {\cal W}_{\mu'\nu'}  - \alpha_W \phi^2\,  {\cal G}_{\mu\nu}^b {\cal G}_{\mu'\nu' b} \, ] \nonumber \\
 & + &  \chi \hat{\chi}^{a\mu'} d_{\mu} \phi d_{\mu'}\phi  - 2 \alpha_E g_s \chi \phi^2   {\cal R}_{\mu\nu}^{ab} \hat{\chi}_{b}^{\;\;\nu} 
  \, . \nonumber
\ee

For the scaling gauge field $W_{\mu}$, we obtain the following equation of motion:
\be   \label{EM4}
\partial_{\nu} (\chi   \hat{\chi}^{\mu\mu'} \hat{\chi}^{\nu\nu'}  {\cal W}_{\mu'\nu'})  =   J^{\mu} \, ,
\ee
with the bosonic vector current
\be
J^{\mu} =  - g_w\chi \hat{\chi}^{\mu\nu} \phi d_{\nu} \phi 
- g_w   \chi \phi^2  \alpha_W \hat{\chi}^{\mu \mu'} \hat{\chi}^{\nu\nu'} \chi_{\nu}^{\;\; a}  {\cal G}_{\mu'\nu' a}\, .
\ee
The derivative can be written as 
\be  
\partial_{\nu} (\chi   \hat{\chi}^{\mu\mu'} \hat{\chi}^{\nu\nu'}  {\cal W}_{\mu'\nu'})  =   \chi   \hat{\chi}^{\mu\mu'} \hat{\chi}^{\nu\nu'}  \partial_{\nu} {\cal W}_{\mu'\nu'}  + \partial_{\nu} (\chi   \hat{\chi}^{\mu\mu'} \hat{\chi}^{\nu\nu'} ) {\cal W}_{\mu'\nu'}  \, ,
\ee
where the second term arises from the gravitational effects.

For the scalar field $\phi$, the equation of motion is simply yielded as follows:
\be   \label{EM5}
(\partial_{\mu} + g_w W_{\mu} ) ( \chi \hat{\chi}^{\mu\nu}  d_{\nu}\phi ) =  J \, ,
\ee
with the scalar current
\be  \label{EM5b}
J = -\chi y_s\bar{\psi} \psi + \chi \phi [\,  \alpha_W\, \hat{\chi}^{\mu\mu'}  \hat{\chi}^{\nu\nu'}  {\cal G}_{\mu\nu}^a {\cal G}_{\mu'\nu' a}- 2\alpha_E g_s \hat{\chi}_{a}^{\;\;\mu} \hat{\chi}_{b}^{\;\;\nu}  {\cal R}_{\mu\nu}^{ab} - 4\lambda_s \phi^2\, ]\, .
\ee
The scaling gauge covariant derivative reads off 
\be 
(\partial_{\mu} + g_w W_{\mu} ) ( \chi \hat{\chi}^{\mu\nu}  d_{\nu}\phi ) =   \chi \hat{\chi}^{\mu\nu}  d_{\mu} d_{\nu}\phi + 
(\partial_{\mu} + 2g_w W_{\mu} ) ( \chi \hat{\chi}^{\mu\nu})  d_{\nu}\phi \, ,
\ee
where the second term reflects the gravitational effects.

\section{Conservation Laws and Equation of Motion for Gravifield}

The gauge theory of gravity is built within the framework of relativistic QFT in the flat Minkowski spacetime, which enables us to discuss in general the basic conservation laws of the theory. 

\subsection{Conservation law for internal gauge invariance}   

For the gauge invariance of internal gauge symmetry, the resulting well-known conservation law is the vector current conservation in the absence of gravitational interactions. In the QFT of gravity,  it can be shown from the equation of motion for the internal gauge field ${\cal A}_{\mu}^I$ in Eq.(\ref{EM1}) that the conservation law still holds  
\be
D_{\mu}J^{I\mu} & = & D_{\mu} D_{\nu} (\chi \hat{\chi}^{\mu\mu'} \hat{\chi}^{\nu\nu'}  {\cal F}^{I}_{\mu'\nu'} ) \nonumber \\
& = & \chi \hat{\chi}^{\mu\mu'} \hat{\chi}^{\nu\nu'} D_{\mu} D_{\nu} ( {\cal F}^{I}_{\mu'\nu'} ) +  \partial_{\mu} \partial_{\nu} (\chi \hat{\chi}^{\mu\mu'} \hat{\chi}^{\nu\nu'} ) {\cal F}^{I}_{\mu'\nu'} \nonumber \\
& + & \partial_{\nu}  (\chi \hat{\chi}^{\mu\mu'} \hat{\chi}^{\nu\nu'} ) D_{\mu} ( {\cal F}^{I}_{\mu'\nu'} ) + 
\partial_{\mu}  (\chi \hat{\chi}^{\mu\mu'} \hat{\chi}^{\nu\nu'} ) D_{\nu} ( {\cal F}^{I}_{\mu'\nu'} )  \nonumber \\
& = &  \frac{1}{2} \chi \hat{\chi}^{\mu\mu'} \hat{\chi}^{\nu\nu'}  f^{IJK} {\cal F}^{J}_{\mu\nu}   {\cal F}^{K}_{\mu'\nu'}  \equiv 0 \, , \nonumber 
\ee
where the symmetric and antisymmetric properties have been used and the gravitational effects are eliminated. 

From the definitions of the fermion vector currents and the covariant derivative, we have
\be
D_{\mu}J^{I\mu} & \equiv &  (\delta^{IK} \partial_{\mu} + f^{IJK} {\cal A}_{\mu} ) (g_A \chi  \bar{\Psi} \gamma^a \hat{\chi}_a^{\;\;\mu} T^K \Psi  )\nonumber \\
& = & g_A \chi \hat{\chi}_a^{\;\;\mu} D_{\mu} (\bar{\Psi} \gamma^a  T^I \Psi ) + g_A \partial_{\mu} (\chi \hat{\chi}_a^{\;\;\mu} ) \bar{\Psi} \gamma^a  T^I \Psi  = 0 \,  ,
\ee
where the second term reflects the gravitational effects.  By requiring the derivatives in the second equality to be gauge covariant for the spin gauge symmetry, the conservation law for the fermion vector currents can be rewritten as
\be
D_{\mu}J^{I\mu} & \equiv & g_A \chi \hat{\chi}_a^{\;\;\mu} {\mathcal D}_{\mu} (\bar{\Psi} \gamma^a  T^I \Psi ) + g_A \btd_{\mu} (\chi \hat{\chi}_a^{\;\;\mu} ) \bar{\Psi} \gamma^a  T^I \Psi =0\, ,
\ee
which can be proved to be held directly by applying for the equation of motion of the fermion field in Eq.(\ref{EM0}).

We then come to the conclusion that in the presence of gravity the conservation law for the fermion vector currents due to internal gauge symmetry holds when the gravitational effects are included.

\subsection{Conservation laws for spin and scaling gauge invariances}

Similarly, for the spin gauge symmetry,  we can show the following identity from the equation of motion for the spin gauge field: 
\be   \label{EM2-2}
\btd_{\mu}J^{\mu\, ab}  & = & \btd_{\mu} \btd_{\nu} (\chi   \hat{\chi}^{\mu\mu'} \hat{\chi}^{\nu\nu'}  {\cal R}_{\mu'\nu' }^{\;\; ab} ) \nonumber \\
& = &   \chi   \hat{\chi}^{\mu\mu'} \hat{\chi}^{\nu\nu'}\btd_{\mu} \btd_{\nu} {\cal R}_{\mu'\nu' }^{\;\; ab}  + \partial_{\mu} \partial_{\nu} (\chi   \hat{\chi}^{\mu\mu'} \hat{\chi}^{\nu\nu'} ) {\cal R}_{\mu'\nu' }^{\;\; ab}  \nonumber \\
& + & \partial_{\mu} (\chi   \hat{\chi}^{\mu\mu'} \hat{\chi}^{\nu\nu'} ) \btd_{\nu}  {\cal R}_{\mu'\nu' }^{\;\; ab} + \partial_{\nu} (\chi   \hat{\chi}^{\mu\mu'} \hat{\chi}^{\nu\nu'} ) \btd_{\mu} {\cal R}_{\mu'\nu' }^{\;\; ab}  \nonumber \\
& = &  \chi   \hat{\chi}^{\mu\mu'} \hat{\chi}^{\nu\nu'} (  {\cal R}_{\mu\nu c }^{\;\; \;  a}  {\cal R}_{\mu'\nu' }^{\;\; cb}  + {\cal R}_{\mu\nu c }^{\;\; \;  b}  {\cal R}_{\mu'\nu' }^{\;\; ac} ) = 0 \, , \nonumber 
\ee
where the symmetric and antisymmetric properties lead to a cancellation for all terms.

As the spin gauge field$\Om_{\mu}^{ab}$ and gravifield $\chi_{\mu}^{\;\; a}$ are introduced simultaneously to ensure the spin gauge invariance for the gauge theory of gravity, the tensor current $J^{\mu\, ab}$ and the gravifield current $J_a^{\; \mu}$ are generally correlated. It is not difficult to show that from the definition of the tensor current $J^{\mu\, ab}$,  the conservation law for the spin gauge invariance can be expressed as follows: 
\be \label{CLS}
\btd_{\mu}J^{\mu}_{\;\; ab}  = \frac{1}{2} \btd_{\mu} \Sm^{\mu}_{\;\; ab}   + \frac{1}{2} J_{[ab]} - \alpha_E g_s \phi^2 \chi  (  \hat{\chi}_a^{\;\;\mu} {\cal R}_{\mu\nu b}^{\;\;\;\;\;\, c}   -  \hat{\chi}_b^{\;\;\mu} {\cal R}_{\mu\nu a}^{\;\;\;\;\;\, c}  )\hat{\chi}_c^{\;\;\nu} =0 \, ,
\ee
with the definitions
\be 
& & \Sm^{\mu}_{\;\; ab} = g_s \chi [\,  \bar{\Psi} \hat{\chi}_c^{\;\; \mu} \{ \gamma^c\;\;  \frac{1}{2} \Sigma_{ab} \} \Psi  + \bar{\psi} \hat{\chi}_c^{\;\; \mu} \{ \gamma^c\;\;  \frac{1}{2} \Sigma_{ab} \} \psi \,] \; , \nonumber \\
& & J_{[ab]} = J_a^{\mu} \chi_{\mu b} - J_b^{\mu}\chi_{\mu a} \, .
\ee
It will be shown that the tensor $\Sm^{\mu}_{\;\; ab} $ corresponds to the spin angular momentum tensor, and $J_{[ab]}$ is related to the energy-momentum tensor. 

For the local scaling gauge symmetry, from the equation of motion for the Weyl gauge field $W_{\mu}$, it is easy to show that
\be   \label{EM4-4}
  \partial_{\mu} J^{\mu}  = \partial_{\mu} \partial_{\nu} (\chi   \hat{\chi}^{\mu\mu'} \hat{\chi}^{\nu\nu'}  {\cal W}_{\mu'\nu'}) \equiv 0 \, , \nonumber 
\ee
which leads to the conservation law for the bosonic current
\be
\partial_{\mu} J^{\mu} = \partial_{\mu} ( \chi \hat{\chi}^{\mu\nu} \phi d_{\nu} \phi 
+ \alpha_W \chi \phi^2  \hat{\chi}^{\mu \mu'} \hat{\chi}^{\nu\nu'} \chi_{\nu}^{\;\; a}  {\cal G}_{\mu'\nu' a} )= 0 \, .
\ee

We now come to discuss the bicovariant vector current, as the gravifield $\chi_{\mu}^{\;\; a}$ behaves not like the usual gauge field. From the equation of motion, the gauge and Lorentz covariant derivative to the current is given by  
\be  \label{EM3-3}
& & (\td_{\mu} -g_w W_{\mu}) J_a^{\;\; \mu} = \alpha_W (\td_{\mu} -g_w W_{\mu})  (\td_{\nu} -g_w W_{\nu})  (\,  \phi^2  \chi  \hat{\chi}^{\mu \mu'} \hat{\chi}^{\nu\nu'} {\cal G}_{\mu'\nu' a}\,  )   \, \nonumber  \\
& &  = \frac{1}{2}  \alpha_W  \phi^2  \chi  ( g_s {\cal R}_{\mu\nu a }^{\;\;\;\;\; \; b}  - g_w {\cal W}_{\mu\nu} \eta_{a}^{\;\; b}   )  \hat{\chi}^{\mu \mu'} \hat{\chi}^{\nu\nu'}  {\cal G}_{\mu'\nu' b} \, ,
\ee
which shows that such a defined bicovariant vector current is not conserved homogeneously. It may not be difficult to understand with the fact that the gravifield is actually introduced as an accompaniment of spin and scaling gauge fields to ensure the spin and scaling gauge symmetries of the action.  It will be demonstrated below that the bicovariant vector current is related to the energy-momentum tensor and an alternative conservation law for the gravifield current results from the conservation of the energy-momentum tensor.

\subsection{Energy-momentum conservation in the QFT of gravity}

So far, we have discussed the conservation laws concerning gauge symmetries in the gauge theory of gravity. As such a theory is built based on the framework of relativistic QFT in the flat Minkowski spacetime,  which implies that the differences of the spatial coordinates or time coordinates can, in principle, be measured by the standard ways proposed in special relativity. This allows us to make a meaningful definition for momentum and energy as well as angular momentum.

Let us first investigate the conservation law under the translational transformation of coordinates $x^{\mu} \to x^{'\mu} = x^{\mu} + a^{\mu}$. The variation of action is given by 
\be 
\delta S_{\chi} = \int d^4x\, \partial_{\mu} ( {\cal T}^{\;\, \mu}_{\nu}) a^{\nu} = 0 \, , \nonumber 
\ee
where the surface term has been ignored as all the fields are assumed to be vanishing at infinity. For arbitrary displacement $a^{\nu}$, it leads to the well-known energy-momentum conservation
\begin{equation} \label{EMC}
\partial_{\mu} {\cal T}^{\;\, \mu}_{\nu}= 0 
\end{equation}
with the energy-momentum tensor
\be
{\cal T}^{\;\, \mu}_{\nu}& \sim & - \eta^{\mu}_{\; \nu}\chi {\cal L} + \frac{1}{2}\chi \hat{\chi}^{\;\; \mu}_{a} [\, i\bar{\Psi} \gamma^{a}\partial_{\nu}\Psi  + i\bar{\psi} \gamma^{a}\partial_{\nu}\psi + H.c. \, ]  \nonumber \\
& - & \chi \hat{\chi}^{\mu\mu'} \hat{\chi}^{\rho\sigma}  [\,  {\cal F}^I_{\mu'\rho} \partial_{\nu} {\cal A}^{I}_{\sigma} + {\cal R}_{\mu'\rho}^{ab} \partial_{\nu}\Om_{\sigma\, ab} + {\cal W}_{\mu'\rho} \partial_{\nu}W_{\sigma}  \, ] \nonumber \\
& + & \alpha_W  \chi \hat{\chi}^{\mu \mu'}  \hat{\chi}^{\rho\sigma} \phi^2 {\cal G}_{\mu'\rho a} \partial_{\nu} \chi_{\sigma}^{\;\; a}  + \chi \hat{\chi}^{\mu\mu'} d_{\mu'} \phi \partial_{\nu}\phi  - 2 \alpha_Eg_s \chi \phi^2 \hat{\chi}_{a}^{\;\;\mu} \hat{\chi}_{b}^{\;\;\rho}  \partial_{\nu}\Om_{\rho}^{ab} \, . \nonumber 
\ee
Note that such a form of energy-momentum tensor is not explicitly gauge-invariant under gauge transformations. To obtain a manifest gauge invariant energy-momentum tensor,  it is useful to take the equations of motion for the gauge fields. By adding total derivative terms and adopting  the equations of motion for the gauge fields ${\cal A}_{\mu}$, $\Om_{\mu}$, $W_{\mu}$, and $\chi_{\mu}$, the energy-momentum tensor is found to be 
\be
{\cal T}^{\;\, \mu}_{\nu}& \sim & 2\{\, - \eta^{\mu}_{\; \nu}\chi {\cal L} + \frac{1}{2}\chi \hat{\chi}^{\;\; \mu}_{a} [\, i \bar{\Psi} \gamma^{a}{\mathcal D}_{\nu}\Psi + i \bar{\psi} \gamma^{a}\td_{\nu}\psi  + H.c. \,  ] \nonumber \\
& - & \chi \hat{\chi}^{\mu\mu'} \hat{\chi}^{\rho\sigma}  [\,   {\cal F}^I_{\mu'\rho} {\cal F}^{I}_{\nu \sigma} 
+ {\cal R}_{\mu'\rho}^{ab}  {\cal R}_{\nu \sigma\, ab} + {\cal W}_{\mu'\rho} {\cal W}_{\nu \sigma}
-  \alpha_W \phi^2 {\cal G}_{\mu'\rho}^a {\cal G}_{\nu \sigma a}  \, ]   \nonumber \\ 
& + &  \chi \hat{\chi}^{\mu\mu'} d_{\mu'} \phi d_{\nu}\phi - 2\alpha_E \chi \phi^2 \hat{\chi}_{a}^{\;\;\mu} \hat{\chi}_{b}^{\;\;\rho} 
 {\cal R}_{\nu \rho}^{ab}\, \} \nonumber \\
 & + &    \partial_{\sigma} \{ \chi \hat{\chi}^{\mu\mu'} \hat{\chi}^{\sigma\rho} ( {\cal F}^I_{\mu'\rho}  {\cal A}^{I}_{\nu}  + {\cal R}_{\mu'\rho}^{ab} \Om_{\nu\, ab}   + {\cal W}_{\mu'\rho}W_{\nu}  \, ) \} \nonumber \\
 & + &  \partial_{\sigma} \{\, \alpha_W  \phi^2 \chi   \hat{\chi}^{\mu \mu'}  \hat{\chi}^{\rho\sigma}   {\cal G}_{\mu'\rho a} \chi_{\nu}^{\;\; a}  - 2\alpha_E g_s \phi^2 \chi  \hat{\chi}^{a\mu}\hat{\chi}^{b\sigma}  \Om_{\nu\, ab} \,  \}   \nonumber
 \ee
where the total derivative terms become vanishing in the derivative form $\partial_{\mu} ( {\cal T}^{\;\, \mu}_{\nu}) $ due to the antisymmetric property. Thus, the gauge-invariant energy-momentum tensor reads
\be \label{EMT}
{\cal T}^{\;\, \mu}_{\nu}& = & - \eta^{\mu}_{\; \nu}\chi {\cal L}  + \frac{1}{2}\chi \hat{\chi}^{\;\; \mu}_{a} [\, 
i\bar{\Psi} \gamma^{a}{\mathcal D}_{\nu}\Psi + i\bar{\psi} \gamma^{a} \td_{\nu}\psi  + H.c.\,  ] \nonumber \\
& - & \chi \hat{\chi}^{\mu\mu'} \hat{\chi}^{\rho\sigma}  [\,  {\cal F}^I_{\mu'\rho} {\cal F}^{I}_{\nu \sigma} 
+  {\cal R}_{\mu'\rho}^{ab}  {\cal R}_{\nu \sigma\, ab} + {\cal W}_{\mu'\rho} {\cal W}_{\nu \sigma} 
-  \alpha_W \phi^2 {\cal G}_{\mu'\rho}^a {\cal G}_{\nu \sigma a}   \, ] \nonumber \\
& + & \chi \hat{\chi}^{\mu\mu'} d_{\mu'} \phi d_{\nu}\phi  - 2 \alpha_E g_s\chi \phi^2 \hat{\chi}_{a}^{\;\;\mu}  {\cal R}_{\nu \rho}^{ab} \hat{\chi}_{b}^{\;\;\rho} 
\ee
Note that the gauge-invariant energy-momentum tensor ${\cal T}_{\mu\nu}$ is, in general, not symmetric
\be 
{\cal T}_{\mu\nu} \neq {\cal T}_{\nu\mu} \, .
\ee
Even if freezing out the gravitational interactions, the energy-momentum tensor for the fermionic fields remains asymmetric though the energy-momentum tensor for gauge and scalar fields becomes symmetric.

\subsection{ Conservation laws under the global Lorentz and scaling transformations}

Let us now discuss the conservation law under the global Lorentz transformation of coordinates. For an infinitesimal transformation $x'^{\mu} = x^{\mu} + \delta L^{\mu}_{\; \nu} x^{\nu}$, similar to the energy-momentum conservation of translational invariance, we arrive at the following conservation law for the Lorentz transformation invariance
\be
\partial_{\mu} \Lm^{\mu}_{\;\, \rho\sigma}- {\mathsf T}_{[\rho\sigma]} =0 \, ,
\ee
 with the definitions
 \be \label{AMC}
  \Lm^{\mu}_{\;\, \rho\sigma} \equiv {\cal T}^{\;\,\mu}_{\rho}\, x_{\sigma} - {\cal T}^{\;\, \mu}_{\sigma}\, x_{\rho}  \, , \qquad {\mathsf T}_{[\rho\sigma]} \equiv{\cal T}_{\rho}^{\; \, \sigma'} \eta_{\sigma'\sigma} - {\cal T}_{\sigma}^{\; \, \rho'} \eta_{\rho'\rho} \, ,
 \ee
 where $\Lm^{\mu}_{\;\, \rho\sigma}$ corresponds to the orbital angular momentum tensor of spacetime rotation. As the energy-momentum tensor ${\mathsf T}_{\rho\sigma}$ is not symmetric, the orbital angular momentum tensor is, in general, not conserved homogeneously, i.e.,  $\partial_{\mu} \Lm^{\mu}_{\;\, \rho\sigma}\neq 0$.
 
On the other hand,  by formulating the conservation law of the spin gauge invariance in terms of the covariant form in the Minkowski spacetime,  we obtain the following expression for the conservation law of spin gauge invariance
\be \label{spinningC}
 & & \partial_{\mu} \Sm^{\mu}_{\;\; \rho\sigma}  + {\cal T}_{[\rho\sigma]} - \Sm^{\mu}_{\;\; ab} (\nabla_{\mu}\chi_{\rho}^{\;\, a}\,  \chi_{\sigma}^{\;\, b}  + \chi_{\rho}^{\;\, a}\,  \nabla_{\mu} \chi_{\sigma}^{\;\, b}  ) \nonumber \\
 & & - 2\alpha_E g_s \phi^2 \chi  (  \hat{\chi}_a^{\;\,\mu} {\cal R}_{\mu\nu b}^{\;\;\;\;\;\,c} - \hat{\chi}_b^{\;\,\mu} {\cal R}_{\mu\nu a}^{\;\;\;\;\;\,c}  )\hat{\chi}_c^{\;\,\nu} \chi_{\rho}^{\;\, a} \chi_{\sigma}^{\;\, b} =0 \, ,
\ee
with the definitions
\be 
\Sm^{\mu}_{\;\; \rho\sigma}  & = & \Sm^{\mu}_{\;\; ab} \chi_{\rho}^{\;\, a} \chi_{\sigma}^{\;\, b} = g_s \chi \hat{\chi}_c^{\;\; \mu} [\, \bar{\Psi}  \{ \gamma^c\;\;  \frac{1}{2} \Sigma_{ab} \} \Psi  + \bar{\psi}  \{ \gamma^c\;\;  \frac{1}{2} \Sigma_{ab} \} \psi\, ]  \chi_{\rho}^{\;\, a} \chi_{\sigma}^{\;\, b}  \, , \nonumber \\
{\cal T}_{[\rho\sigma]}  & = & J_{[ab]} \chi_{\rho}^{\;\, a} \chi_{\sigma}^{\;\, b} = {\cal T}_{\rho}^{\;\; \mu} \chi_{\mu\sigma}  - {\cal T}_{\sigma}^{\;\; \mu} \chi_{\mu\rho}   \, .
\ee 
In defining the antisymmetric tensor field ${\cal T}_{[\rho\sigma]} $, we have introduced the symmetric tensor field
\be \label{dtensor}
\chi_{\mu\nu} = \chi_{\mu}^{\;\; a}\chi_{\nu}^{\;\; b} \eta_{ab} \, ,
\ee
which is dual to the symmetric tensor field $\hat{\chi}^{\mu\nu}$ given in Eq.(\ref{tensor}).
 
Let us introduce a total angular momentum tensor 
\be
 {\cal J}^{\mu}_{\;\; \rho\sigma}  & \equiv & \Lm^{\mu}_{\;\; \rho\sigma}  + \Sm^{\mu}_{\;\; \rho\sigma}  \, ,
\ee
with $\Lm^{\mu}_{\;\; \rho\sigma}$  and $\Sm^{\mu}_{\;\; \rho\sigma}$ being the rotational and spinning angular momentum tensors, respectively.  By combining the conservation law of Lorentz invariance Eq.(\ref{AMC}) with the conservation law of the spin gauge invariance Eq.(\ref{spinningC}), we obtain a new form of conservation law:
\be 
 & & \partial_{\mu} {\cal J}^{\mu}_{\;\; \rho\sigma}  -({\mathsf T}_{[\rho\sigma]} -  {\cal T}_{[\rho\sigma]} ) - \Sm^{\mu}_{\;\; ab} (\nabla_{\mu}\chi_{\rho}^{\;\, a}\,  \chi_{\sigma}^{\;\, b}  + \chi_{\rho}^{\;\, a}\,  \nabla_{\mu} \chi_{\sigma}^{\;\, b}  ) 
 \nonumber \\
 & &  -   2\alpha_E g_s \phi^2 \chi  (  \hat{\chi}_a^{\;\,\mu} {\cal R}_{\mu\nu b}^{\;\;\;\;\;\,c}   -  \hat{\chi}_b^{\;\,\mu} {\cal R}_{\mu\nu a}^{\;\;\;\;\;\,c}  )\hat{\chi}_c^{\;\,\nu} \chi_{\rho}^{\;\, a} \chi_{\sigma}^{\;\, b} = 0 \, ,
\ee
which shows that in the presence of gravitational interactions the total angular momentum tensor defined above is not conserved homogeneously, i.e., $\partial_{\mu} {\cal J}^{\mu}_{\;\; \rho\sigma} \neq 0 $.
 
When turning the spin and scaling gauge symmetries into global Lorentz and scaling symmetries, i.e.,  all the quantum fields relevant to the gauge theory of gravity will be absent,
\be
 & & \Om_{\mu}^{ab} \to 0\, , \quad \hat{\chi}_a^{\;\;\mu} \to \eta_a^{\;\; \mu}\, , \quad W_{\mu} \to 0 \, , 
 \ee
 we get the following relations:
 \be
 & &  \partial_{\mu} \Sm^{\mu}_{\;\; \rho\sigma}  = -  {\mathsf T}_{[\rho\sigma]}\,  \quad ( \partial_{\mu} \Lm^{\mu}_{\;\; \rho\sigma}  =  {\mathsf T}_{[\rho\sigma]}) \, ,
 \ee
 where the conservation law of the spinning momentum tensor is governed by the asymmetric part of the energy-momentum tensor ( which is seen to have an opposite sign in comparison with the conservation law of the angular momentum tensor). As a consequence, we arrive at the conservation law for the total angular momentum tensor: 
\be
\partial_{\mu} {\cal J}^{\mu}_{\;\; \rho\sigma}  & = & \partial_{\mu} ( \Lm^{\mu}_{\;\; \rho\sigma}  + \Sm^{\mu}_{\;\; \rho\sigma}  )  =0 \, ,
\ee
which reproduces the result for the gravity-free theory in the Minkowski spacetime. 

We then come to the conclusion that in the presence of fermion field, the energy-momentum tensor is, in general, asymmetric, neither the angular momentum tensor nor the spinning momentum tensor is  conserved homogeneously due to the asymmetric part of the energy-momentum tensor, and only the total angular momentum becomes homogeneously conserved due to the cancellation in their asymmetric part of the energy-momentum tensor.

For the global scaling invariance, we obtain the following conservation law: 
\be
\left(x^{\mu} \frac{\partial}{\partial x^{\mu}} + 4) (\chi\, {\cal L} \right) + \partial_{\mu} {\cal T}^{\mu}  - {\cal T} = 0 \, ,
\ee
 with the definitions
 \be \label{SCL}
  {\cal T}^{\mu} \equiv{\cal T}^{\;\, \mu}_{\nu}\, x^{\nu}  \, , \qquad {\cal T} \equiv {\cal T}_{\nu}^{\;\, \mu} \eta_{\mu}^{\;\, \nu}
 = {\cal T}_{\mu}^{\; \mu}  \, ,
 \ee
 where ${\cal T}^{\mu}$ represents a scaling current and ${\cal T} = {\cal T}_{\mu}^{\; \mu}$ is the trace of the energy-momentum tensor. 
 As the integral $\int d^4x \lambda^4\,  \chi(\lambda x)\, {\cal L}(\lambda x) $ is independent of $\lambda$, the differentiation with respect to $\lambda$ at $\lambda =1$ leads to the following identity:
 \be
 \int d^4 x \left(x^{\mu} \frac{\partial}{\partial x^{\mu}} + 4) (\chi {\cal L} \right) = 0 \, ,\nonumber 
 \ee  
 which results in the conservation law:
 \be
 \partial_{\mu} {\cal T}^{\mu}  - {\cal T} = 0 \, .
 \ee
When the energy-momentum tensor becomes traceless, ${\cal T} = 0$, for the case that the scalar field is freezed out in the gravity-free theory, one then yields the homogeneous conservation law for the scaling invariance: 
\be
\partial_{\mu} {\cal T}^{\mu} = 0\, . 
 \ee
 
 \subsection{Equation of motion and conservation law for gravifield tensor}
 
From the definition of the bicovariant vector current shown in the equation of motion for the gravifield $\chi_{\mu}^{\;\; a}$  given in Eq.(\ref{EM2}), we observe the following relation between the energy-momentum tensor and the bicovariant vector current for the gravifield: 
\be 
{\cal T}^{\;\, \mu}_{\nu} & = & \chi_{\nu}^{\;\; a} J^{\; \mu}_a  \, ,
\ee
which enables us to obtain the equation of motion for the gravifield in connection directly with the energy-momentum tensor
\be \label{GEM1}
 \partial_{\rho} {\cal G}^{\;\mu\rho}_{ \nu}  - {\cal G}_{\nu}^{\;\; \mu} ={\cal T}^{\;\, \mu}_{\nu} \, 
\ee
which is a general gravity equation alternative to Einstein's equation of general relativity. Here we have introduced the following definitions
\be \label{GEM2}
& & {\cal G}^{\; \mu\rho}_{ \nu}  \equiv \alpha_W \phi^2  \chi  \hat{\chi}^{\mu \mu'} \hat{\chi}^{\rho\nu'}  \chi_{\nu}^{\;\; a}\, {\cal G}_{\mu'\nu' a}  = - {\cal G}^{\; \rho\mu}_{ \nu} \, , \nonumber \\
& &   {\cal G}_{\nu}^{\;\; \mu} \equiv  \alpha_W \phi^2  \chi   \hat{\chi}^{\mu \mu'} \hat{\chi}^{\rho\nu'} (\nabla_{\rho} \chi_{\nu}^{\;\; a})  \, {\cal G}_{\mu'\nu' a}  =  (\hat{\chi}_a^{\; \sigma}\nabla_{\rho} \chi_{\nu}^{\;\; a} )\,
{\cal G}^{\; \mu\rho}_{ \sigma}   \, .
 \ee 
Here ${\cal G}^{\; \mu\rho}_{ \nu}$ may be referred as the gauge-invariant {\it gravifield tensor} and $ {\cal G}_{\nu}^{\;\; \mu} $ as the gauge-invariant {\it gravifield tensor current}. 

In light of the energy-momentum conservation $\partial_{\mu} {\cal T}^{\;\, \mu}_{\rho}  =  \partial_{\mu} ( J^{\;\mu}_a \chi_{\rho}^{\;\; a} )  =0$, we come to the following conserved current
\be
\partial_{\mu}{\cal G}_{\nu}^{\;\; \mu} =  \partial_{\mu}( \hat{\chi}_a^{\; \sigma}\nabla_{\rho} \chi_{\nu}^{\;\; a} 
{\cal G}^{\; \mu\rho}_{ \sigma} ) =\partial_{\mu}  (\,  \phi^2  \chi  \hat{\chi}^{\mu \mu'} \hat{\chi}^{\rho\nu'} \nabla_{\rho} \chi_{\nu}^{\;\; a}\, {\cal G}_{\mu'\nu' a} \, )=  0 \, ,
 \ee 
which is considered to be an alternative conservation law for the {\it gravifield tensor current}.


\section{Gravitational Gauge Symmetry Breaking and Dynamics of background Fields }

In establishing the gauge theory of gravity, we have postulated the spin and scaling gauge symmetries and introduced correspondingly the spin gauge field $\Om_{\mu}^{ab}(x)$ associated with the essential gravifield $\chi_{\mu}^{\;\; a}(x)$ and the scaling gauge field $W_{\mu}(x)$ as well as the scalar field $\phi(x)$. These fields are, in general, massless without considering gauge symmetry breaking.  Based on the fact that these particles are not yet observed experimentally, it happens that either they are very heavy or their interactions are very weak. To generate massive spin and scaling gauge fields and study how weak the gravitational interactions are, we shall take into account the gravitational gauge symmetry breaking and the evolution of the Universe. 

\subsection{Gravitational gauge symmetry breaking }

As the scaling gauge symmetry is directly related to the mass scale, its gauge symmetry breaking can be analyzed by making a special scaling gauge transformation to fix the local scaling gauge symmetry, for instance, by choosing a gauge-fixing condition, so that the scalar field $\phi(x)$ is rescaled to be a constant mass scale. Let us consider a special scaling gauge transformation $\xi_a(x) =1/a_{\chi}(x)$, which leads to
\be
\label{GFC}
\phi(x) \to \phi'(x) = \xi_a(x) \phi(x) =  \phi(x)/a_{\chi}(x) = M_S \, , 
\ee
with $M_S$ being regarded as the basic {\it scaling energy scale} for fixing the scaling gauge transformation. 

On the other hand, one can always choose an alternative gauge condition to fix the scaling gauge symmetry, so that the determinant of the gravifield $\chi = \det \chi_{\mu}^{\;\; a}$ is rescaled into unity:
\be
\chi(x) \to \chi'(x) = \xi_{\chi}^{-4}(x) \chi(x) =  1 \,\quad a_{\chi}(x)\to a(x) = \xi_{\chi}(x) a_{\chi}(x) 
= \chi^{1/4}(x)a_{\chi}(x) \, . 
\ee
In such a fixing gauge condition $\chi = \det \chi_{\mu}^{\;\; a}=1$, the scalar field can be written as the following  general form-
\be
\phi(x) \equiv M_S\, a(x) \, ,
\ee 
which will be shown to be a useful choice for discussing the gravitational gauge theory within the framework of QFT. 

Let us now discuss the gravitational gauge symmetry breaking under the scaling gauge-fixing condition  $\chi(x) =  1$. We shall make a reliable postulate that the gravitational gauge symmetry is broken down in such a way that the theory still possesses a global Lorentz symmetry.  With this postulate, we are led to the following simple background structure: 
\be 
& & \langle \chi_{\mu}^{\;\; a}(x) \rangle = \chib_{\mu}^{\;\; a}(x) = \eta_{\mu}^{\;\; a}  \, , \quad \langle \phi(x) \rangle = \bar{\varphi}(x) \equiv \ab(x)\, M_S\, , \quad \langle W_{\mu} \rangle = \bar{w}_{\mu}(x) \, , \nonumber \\
& &  \langle \Om_{\mu}^{ab}(x) \rangle =   \eta_{\mu\rho}^{[ab]} \bar{\omega}^{\rho}(x)\, , \quad \eta_{\mu\rho}^{[ab]} \equiv  \eta_{\mu}^{\;\; a} \eta_{\rho}^{\;\;b} - \eta_{\mu}^{\;\; b} \eta_{\rho}^{\;\;a}
 \, , \nonumber \\
& & \langle \Psi(x) \rangle = 0\, , \quad  \langle \psi(x) \rangle = 0\, ,\quad \langle {\cal A}_{\mu}^I(x) \rangle = 0 \, ,
\ee
where $\pb(x)$($\ab(x)$), $\ob_{\mu}(x)$, and $\wb_{\mu}(x)$ are the gravitational background fields.

\subsection{Equations of motion and dynamics for the background fields}

To analyze the properties of the background fields, we shall find solutions of the background fields $\pb(x)$, $\ob(x)$, and $\wb(x)$ by solving their field equations of motion. 

Let us first check the field strength for the background fields. The field strength for the background spin gauge field is found to be
\be
\bar{R}_{\mu\nu}^{ab} = d_{\mu}\ob^{\rho} \eta_{\nu\rho}^{[ab]} - d_{\nu}\ob^{\rho} \eta_{\mu\rho}^{[ab]} - g_s \ob_{\rho}^2 \eta_{\mu\nu}^{[ab]}    \, .
\ee
It is then not difficult to yield the following results: 
\be
& & \bar{R} = \eta_a^{\mu} \eta_b^{\nu} \bar{R}_{\mu\nu}^{ab}  = - 6 (\partial_{\mu}\ob^{\mu} 
+ g_s \ob_{\mu}\ob^{\mu} ) \, , \nonumber \\
& & \bar{R}_{\mu\nu}^{ab} \bar{R}^{\mu\nu}_{ab} = 4 (d_{\mu}\ob^{\mu} )^2 + 8 d_{\mu}\ob_{\nu}  d^{\mu}\ob^{\nu} + 24 g_s ( d_{\mu}\ob^{\mu} + g_s \ob_{\mu}\ob^{\mu} )\ob_{\nu}\ob^{\nu} \, .
\ee
The field strength for the background gravifield is simply given by
\be
\bar{G}_{\mu\nu}^a = \nabla_{\mu} \eta_{\nu}^a - \nabla_{\nu}\eta_{\mu}^a = \eta_{\mu}^a \Ob_{\nu} - \eta_{\nu}^a\Ob_{\mu} \, , \quad \Ob_{\mu} \equiv g_s \ob_{\mu} - g_w \wb_{\mu}\, , 
\ee
and the field strength for the background scaling gauge field has the usual form 
\be
\bar{W}_{\mu\nu} = \partial_{\mu} \wb_{\nu} - \partial_{\nu} \wb_{\mu} \, .
\ee
The corresponding Lagrangian for the background fields reads 
\be
\bar{\cal L} & = &  - (d_{\rho}\ob^{\rho} )^2 -2 d_{\rho}\ob_{\sigma} d^{\rho}\ob^{\sigma} -6g_s \partial_{\rho}\ob^{\rho} \ob_{\sigma}\ob^{\sigma} + \frac{3}{2} \alpha_W  \Ob_{\rho}\Ob^{\rho} \pb^2 \nonumber \\
& & -\frac{1}{4} \bar{W}_{\rho\sigma}\bar{W}^{\rho\sigma} + \frac{1}{2} d_{\rho}\pb d^{\rho}\pb + 6\alpha_E g_s (\partial_{\rho}\ob^{\rho} + g_s \ob_{\rho}\ob^{\rho} )\pb^2 - \lambda_s \pb^4 \, ,
 \ee
where we have used the definitions 
\be \label{CDV}
& & d_{\mu}\ob^{\rho}  \equiv (\partial_{\mu} - g_s \ob_{\mu}) \ob^{\rho} \, , \nonumber \\
& & d_{\mu}\pb \equiv (\partial_{\mu} - g_w \wb_{\mu}) \pb \, .
\ee

From Eqs.(\ref{EM2})-(\ref{EM5b}), the equations of motion can be written as follows:
\be  \label{BEM1}
& & [\, - \partial_{\sigma}\partial^{\sigma} \ob^{\rho} + g_s \ob^{\sigma}\partial_{\sigma} \ob^{\rho} + 2 g_s \ob^{\rho}\partial_{\sigma}\ob^{\sigma} - 3 g_s \ob^{\sigma}\partial^{\rho}\ob_{\sigma} + 2 g_s^2 \ob_{\sigma}\ob^{\sigma}\ob^{\rho} \nonumber \\
& &  + \frac{1}{2} \alpha_W  g_s \Ob^{\rho} \pb^2 + \alpha_E g_s \partial^{\rho} \pb^2 - 2\alpha_E g_s^2 \ob^{\rho} \pb^2 \, ] \eta_{\mu\rho}^{[ab]} \nonumber \\
& & + [ \,  \partial_{\mu} \partial^{\sigma}\ob^{\rho} - 2g_s \ob_{\mu} \partial^{\sigma}\ob^{\rho} - g_s \ob^{\rho} \partial^{\sigma} \ob_{\mu} - g_s \ob^{\sigma} \partial_{\mu} \ob^{\rho} \, ] \eta_{\sigma\rho}^{[ab]} = 0 \, , 
\ee
for the background spin gauge field $\langle \Om_{\mu}^{ab}(x)\rangle$, and 
\be  \label{BEM2}
& & \eta_{\mu}^a \alpha_W d^{\rho}(\pb^2\Ob_{\rho}) - \eta_{\rho}^a \alpha_W  [\, d^{\rho} (\pb^2\Ob_{\mu}) + g_s (\ob_{\mu} \Ob^{\rho} + 2 \Ob_{\mu}\ob^{\rho} ) \pb^2 \, ] \nonumber \\
& & = - \eta_{\mu}^a 2[\, d_{\rho}\ob_{\sigma} d^{\rho}\ob^{\sigma} + 2 g_s d_{\rho}\ob^{\rho} \ob_{\sigma}\ob^{\sigma} 
+ 3 g_s^2  (\ob_{\sigma}\ob^{\sigma})^2 - \alpha_W \Ob_{\sigma}\Ob^{\sigma}\pb^2 \, ] \nonumber \\
& & \;\;\; - \eta_{\mu}^a 2 \alpha_Eg_s (d_{\rho}\ob^{\rho} + 3g_s \ob_{\rho}\ob^{\rho})  \pb^2 
-\eta_{\rho}^a [ \bar{W}_{\rho\sigma}\bar{W}^{\mu\sigma} -  d^{\rho}\pb d_{\mu}\pb - 2 \alpha_W  \Ob^{\rho}\Ob_{\mu} \pb^2 ]\nonumber \\
& &\;\;\; -\eta_{\rho}^a 2[\, d_{\rho}\ob_{\sigma} d_{\mu}\ob^{\sigma}  +  (d_{\rho}\ob_{\mu} + d_{\mu}\ob_{\rho} ) d_{\sigma}\ob^{\sigma} - d_{\sigma}\ob_{\mu} d^{\sigma}\ob^{\rho} + 2 g_s  (d_{\rho}\ob_{\mu} + d_{\mu}\ob_{\rho} ) \ob_{\sigma}\ob^{\sigma} \, ] \nonumber \\
& & \;\;\; + \eta_{\rho}^a  4 \alpha_Eg_s \pb^2 d^{\rho}\ob_{\mu} -\eta_{\mu}^a \bar{\cal L} \, ,
\ee
for the background gravifield $\langle \chi_{\mu}^{\;\;a}(x)\rangle$. Here we have adopted the definition 

\[ d^{\rho}(\pb^2\Ob_{\rho}) \equiv (\partial^{\rho} - g_w \wb^{\rho}) (\pb^2\Ob_{\rho}) \, . \]

The equations of motion for the background scaling gauge field and scalar field are given correspondingly by 
\be   \label{BEM3}
\partial_{\nu} \bar{W}^{\mu\nu} = - g_w \pb d^{\mu} \pb - 3 \alpha_W  \Ob^{\mu} \pb^2 \,  
\ee
and 
\be  \label{BEM4}
(\partial_{\mu} + g_w \wb_{\mu}) d^{\mu} \pb = 6 \alpha_W \Ob_{\mu}\Ob^{\mu} \pb^2 
+ 12 \alpha_Eg_s (\partial_{\mu}\ob^{\mu} + g_s \ob_{\mu}\ob^{\mu})  \pb - 4 \lambda_s \pb^3  \, .
\ee

It appears difficult to solve exactly the equations of motion. Especially, the equations of motion for the background  gravifield and spin gauge field look more complicated. To simplify the equations, let us make a rational {\it ansatz} that the conformally covariant derivative defined in Eq.(\ref{CDV}) for the background scalar field vanishes
\be  \label{BSL}
 d_{\mu} \pb = 0  \, , \quad \mbox{i.e. } \quad g_w\wb_{\mu}(x) = \partial_{\mu} \ln \pb(x) \, , \quad \bar{W}_{\mu\nu} = 0\, ,
\ee                                                                                                                                                                                                                                                                                                                                                                                                                                                                                                                                                                                                                                                                                                                                                            
which indicates that the background scaling gauge field $\wb(x)$ is a pure gauge field and solely determined by the background scalar field $\pb(x)$. In other words, such an ansatz is equivalent to the postulate that the background scalar field has a vanishing conformally covariant kinetic energy. From the equation of motion Eq.(\ref{BEM3}) for the background scaling gauge field, such an ansatz leads to the following relation
\be  \label{BSL0}
& & \bar{G}_{\mu\nu} = 0, \quad {\mbox{i.e.}} \quad  g_s \ob_{\mu}(x) = g_w \wb_{\mu}(x) = \partial_{\mu} \ln \pb(x)\, ,
\ee
which shows that the background spin gauge field is also governed by the background scalar field.  

Applying the above ansatz and relation, i.e., $ g_s \ob_{\mu}(x) = g_w \wb_{\mu}(x) = \partial_{\mu} \ln \pb(x)$, to the equations of motion Eqs.(\ref{BEM1}) and (\ref{BEM2}) as well as  Eq.(\ref{BEM4}) for the spin gauge field and gravifield as well as background scalar field, we arrive at the following simplified equations  of motion, respectively:
\be \label{BSL2}
3 (\partial_{\mu}\pb(x) ) \partial_{\nu}^2 \pb(x) = \pb(x) \partial_{\mu} (\partial_{\nu}^2 \pb(x) )\, ,
\ee
and
\be 
& & \left( \frac{1}{\pb^3} \partial_{\rho}^2\pb - \alpha_E \right)  [\, 2 \partial_{\mu}\pb \partial^{\mu}\pb  - \pb \partial_{\mu}\partial^{\mu} \pb \, ] + 3\alpha_E \pb \partial_{\mu}^2 \pb - \lambda_s \pb^4 = 0  \, , \label{BSL3}
 \\ 
& &  \left( \frac{1}{\pb^3} \partial_{\rho}^2\pb - \alpha_E\right) [\, 2 \partial_{\mu}\pb \partial_{\nu}\pb - \pb \partial_{\mu}\partial_{\nu} \pb \, ]  = 0  \, , \label{BSL4}
\ee
as well as 
\be \label{BSL1}
& & 3\alpha_E \partial_{\nu}^2 \pb(x) = \lambda_s \pb^3(x)\, .
\ee

The above four equations are actually not independent. Equation (\ref{BSL2}) can be obtained from Eq.(\ref{BSL1}), while Eq.(\ref{BSL3}) can be derived from Eqs.(\ref{BSL4}) and (\ref{BSL1}). Equation(\ref{BSL4}) is equivalent to 
\be
 2\partial_{\mu}\pb \partial_{\nu}\pb = \pb \partial_{\mu}\partial_{\nu} \pb \, , \quad \mbox{i.e.} \quad  d_{\mu}\ob_{\nu} = 0 \, ,\quad \lambda_s \neq 3 \alpha_E^2  \, , \label{BSL5}
\ee
which shows that the covariant derivative for the background field $\ob_{\mu}(x)$ also vanishes. Thus, taking Eqs.(\ref{BSL5}) and (\ref{BSL1}) as independent equations of motion, we arrive at the following exact solution for the background scalar field:
\be  \label{BSL6}
\pb(x) = \frac{m_{\kappa} }{\alpha_S(1 - x^{\mu}\kappa_{\mu})}\, , \quad m_{\kappa} = \sqrt{ \kappa_{\mu}\kappa^{\mu} } \, .
\ee
Here $\kappa_{\mu}$ is regarded as a constant {\it cosmic vector} with the {\it cosmological mass} scale $m_{\kappa} $. $\alpha_S$ is a constant parameter. For nonzero constants $\alpha_E$ and 
$\lambda_s$,  we obtain the following relation:
\be  \label{BSL7}
\lambda_s =  6\alpha_E \alpha_S^2\, .
\ee 

It is noticed that the equations of motion possess a mirror symmetry of $Z_2$: $\pb \to - \pb$, and also the reflection symmetry of spacetime: $x^{\mu}\to - x^{\mu}$, $\pb(x)\to \pb(-x)$. The solution is actually invariant under the transformation: $x^{\mu}\to - x^{\mu}$, $\kappa_{\mu}\to - \kappa_{\mu}$. Thus, there are, in general, four solutions for the background fields:
\be
\pb(x) \to \pb_{\kappa\pm}^{(\pm)} (x) = \pm \pb_{\kappa\pm} (x) = \pm \frac{ \mk}{\alpha_S(1\mp x^{\mu}\kappa_{\mu})} \, ,
\ee
which all satisfy the equations of motion.

\section{Geometry of Gravifield spacetime and Evolution of Early Universe With Conformal Inflation and Deflation }

To understand the evolution of the Universe, we shall study the geometry of gravifield spacetime. Obviously, the background structure after gravitational gauge symmetry breaking forms a {\it background gravifield spacetime}. With the solution obtained above for the background scalar field, it enables us to explore the properties of the {\it background gravifield spacetime} and the evolution of the early Universe. 

\subsection{ Line element of gravifield spacetime and scalinon field}

Before discussing the background gravifield spacetime, let us first define a spin and scaling gauge-invariant line element in the locally flat gravifield spacetime with the gravifield basis $\chi^a$:
\be
l^2_{\chi} = a^2_{\chi} \, \eta_{ab}\, \chi^a \chi^b  \, ,
\ee
where the multiplying factor $a_{\chi}$ is given by the scalar field, $a_{\chi}\equiv \phi^2/M_S^2 $, which ensures an invariant line element under both the local and global scaling transformations.  

In light of the coordinate basis in the flat Minkowski spacetime, the above invariant line element can be rewritten as
\be
l_{\chi}^2 = a_{\chi}^2(x)\, \eta_{ab} \chi_{\mu}^{\;\; a}(x) \chi_{\nu}^{\;\; b}(x)\, dx^{\mu} dx^{\nu}  \equiv a_{\chi}^2(x) \chi_{\mu\nu}(x) dx^{\mu} dx^{\nu}  \, ,
\ee
which sets  a {\it conformal basis } with $a_{\chi}^2(x)$ as a {\it conformal scale field}, where we have introduced a tensor field in the flat Minkowski spacetime
\be
\chi_{\mu\nu}(x) =\chi_{\mu}^{\;\; a}(x) \chi_{\nu}^{\;\; b}(x) \eta_{ab} \, ,
\ee
which defines a Lorentz covariant metric tensor field  and is referred as a {\it gravimetric field} for short and convenience. Such a gravimetric field characterizes the geometric property of the gravifield spacetime. 

One can always make a scaling gauge transformation to choose a special scaling gauge condition, for instance, by taking the scaling gauge transformation,  $a_{\chi} (x) \to a_{\chi}^E (x) = \xi_a(x) a_{\chi}(x) =  1$, $\chi_{\mu}^{\;\; a}(x) \to \chi_{\mu}^{E\, a}(x) = \xi_a^{-1}(x) \chi_{\mu}^{\;\; a}(x) = a_{\chi}(x) \chi_{\mu}^{\;\; a}(x) $,  so that  the line element can be expressed as follows:
\be
l_{\chi}^2 = a_{\chi}^2(x) \chi_{\mu\nu}(x) dx^{\mu} dx^{\nu}  \equiv  \chi_{\mu\nu}^E (x) dx^{\mu} dx^{\nu}  \, ,\quad 
\chi_{\mu\nu}^E (x) =\chi_{\mu}^{E\, a}(x) \chi_{\nu}^{E\, b}(x) \eta_{ab} \, .
\ee
Such a choice of the gauge-fixing condition yields an {\it  Einstein-type basis } in the flat Minkowski spacetime of coordinates.  

On the other hand, by making an alternative special scaling gauge transformation, i.e., $ a_{\chi}(x)\to a_U(x) = \chi^{1/4}(x)a_{\chi}(x)$, $\chi_{\mu}^{\;\; a}(x) \to \chi_{\mu}^{U\, a}(x) = \chi^{-1/4}(x) \chi_{\mu}^{\;\; a}(x)$, we arrive at a specific line element
\be
& & l_{\chi}^2  =   a_{\chi}^2(x) \chi_{\mu\nu}(x) dx^{\mu} dx^{\nu}  \equiv a_U^2(x) \chi_{\mu\nu}^U (x) dx^{\mu} dx^{\nu}  \, ,\nonumber \\ 
& & \chi_{\mu\nu}^U (x) =\chi_{\mu}^{U\, a}(x) \chi_{\nu}^{U\, b}(x) \eta_{ab} \, , \quad \chi^U = \det \chi_{\mu}^{U\, a} = 1
\ee
which sets another basis for the gravifield spacetime. We may call such a basis a {\it unitary basis} .

For a convenience of expression, we will omit in the following discussions the label $ ``U"$ for all the quantities in the unitary basis and express the line element as
\be
l_{\chi}^2 = a^2(x) \chi_{\mu\nu}(x) dx^{\mu} dx^{\nu} \, ; \qquad \chi = \det \chi_{\mu}^{\;\; a} = 1  \, .
\ee

It will be shown that the unitary basis is a convenient and physically meaningful basis for considering a quantum effect within the framework of QFT. The resulting conformal scale field $a(x)$ or the corresponding conformal scalar field $\phi(x) \equiv M_S a(x)$ reflects directly the physics degree of freedom in the unitary basis;  we may call such a scalar particle $\phi(x)$ a {\it scalinon} particle that characterizes the conformal scaling evolution of the early Universe.

\subsection{Background gravifield spacetime and cosmological horizon}

The background structure after gravitational gauge symmetry breaking has been found to have the following form in the unitary basis ($\chi = \det \chi_{\mu}^{\;\; a} = 1$):
\be 
& & \langle \chi_{\mu}^{\;\; a}(x) \rangle = \chib_{\mu}^{\;\; a}(x) = \eta_{\mu}^{\;\; a}  \, , \quad \langle \phi(x) \rangle = \bar{\varphi}(x) =\ab(x) M_S \, , \nonumber \\
& &  \langle \Om_{\mu}^{ab}(x) \rangle =  g_s^{-1} \eta_{\mu\rho}^{[ab]} \partial^{\rho} \ln \pb(x)\, , \quad  \langle W_{\mu} \rangle = g_w^{-1} \partial_{\mu}\ln \pb(x)\, .
\ee

Such a background structure forms a {\it background gravifield spacetime}  with the line element 
\be
\langle l_{\chi}^2 \rangle = M_S^{-2} \langle \phi^2(x) \eta_{ab}\, \chi^a(x) \chi^b(x) \rangle = \ab^2(x)\, \eta_{\mu\nu} \, dx^{\mu} dx^{\nu}  \, , 
\ee
which coincides with a conformally flat Minkowski spacetime governed by the background conformal scale field $\ab(x)$. Such a background gravifield spacetime is distinguished from the globally flat Minkowski spacetime that is introduced as the inertial reference frame of coordinates. 

Let us now demonstrate the property of the background gravifield spacetime from the solutions of background fields. $\ab(x)$ is  determined by the solutions of the background scalinon field $\pb(x)$ in the unitary basis divided by the basic scaling energy scale $M_S$:
\be
\ab(x) \to \pm \ab_{\kappa\pm}(x) = \pm \pb_{\kappa\pm}(x)/M_S = \pm \frac{\mk }{\alpha_S M_S} \frac{1}{1 \mp x^{\mu}\kappa_{\mu}}\, .
\ee
The line element is invariant under the global Lorentz and scaling transformations 
\be
& & x^{\mu} \to x^{'\mu} = L^{\mu}_{\;\; \nu} x^{\nu}\, , \quad \kappa_{\mu} \to \kappa'_{\mu}= L_{\mu}^{\;\; \nu} \kappa_{\nu}
  \nonumber \\
& & x^{\mu} \to x^{'\mu} = \lambda^{-1} x^{\mu}\, , \quad \kappa_{\mu}\to \kappa'_{\mu} = \lambda \kappa_{\mu}\,, \quad \ab(x) \to \ab'(x') = \lambda \ab(x) \, . \nonumber 
\ee
As the scalar product  $x^{\mu}\kappa_{\mu}$ is a Lorentz- and scaling-invariant quantity,  we are able to define a {\it conformal proper time} $\eta$ as follows:
\be
x^{\mu}\kappa_{\mu} \equiv \kh \, c \eta  \, ,
\ee 
where $\kh$ is regarded as a {\it conformal proper energy scale} and $c$ is the speed of light in vacuum. In general, $\eta$ and $\kh$ are allowed to obey a different conformal scaling transformation 
\be
\eta\to \eta' = \lambda^{-\alpha}\, \eta\, , \quad \kh \to \kh' = \lambda^{\alpha} \, \kh \, ,
\ee
with $\alpha$ a constant parameter.  Thus, the usual Lorentz time component $x^0=ct $ can be expressed in terms of the {\it conformal proper time} $\eta$
\be 
& & x^0 =ct = \frac{\kh}{\kappa_0}\, c \eta - \frac{\kappa_i}{\kappa_0} x^i \equiv  \frac{1}{u_0} ( c \eta - u_i x^i ) \, , \nonumber \\
& & u_0 \equiv \frac{\kappa_0}{\kh }\, , \quad  u_i \equiv  \frac{\kappa_i }{\kh}\, ,
\ee  
and the infinitesimal displacement $dx^0$ of the Lorentz time component is replaced by
\be 
dx^0 = c\, dt = \frac{1}{u_0} ( c d\eta - u_i dx^i ) \, .
\ee 
The conformal scale field is rewritten to be 
\be
\ab(x) \to \ab(\eta)\to \pm \ab_{\kappa\pm} (\eta) = \pm \frac{\mk}{\alpha_S M_S } \frac{1}{1 \mp \kh\,c \eta}\, ,
\ee 
and the line element can be expressed as follows:
\be
\langle l_{\chi}^2 \rangle 
& = & \ab^2(\eta) \chib_{\mu\nu} d\hat{x}^{\mu}d\hat{x}^{\nu}\, , \quad \hat{x}^{\mu} = (c\eta, x^i) \, , 
\ee
where $\hat{x}^{\mu}$ denotes the coordinates with the {\it conformal proper time} $\eta$ and  $\chib_{\mu\nu}$ is the corresponding background metric for the background gravifield spacetime: 
\be
\chib_{\mu\nu} = \frac{1}{u_0^2}
\left(
\begin{array}{cc}
 1 &  -u_j     \\
 -u_i  &    u_0^2  \chib_{ij}
\end{array}
\right)   \, , \quad 	 \chib_{ij} = \eta_{ij} + \frac{u_iu_j}{u_0^2 } \, , 
\ee
which is, in general, nondiagonal. 

One can, in principle, choose a special reference frame and conformal scaling factor so that $u_i =0$ ($\kappa_i =0$) and $\kh = \mk=\kappa_0$ and $u_0 =1$,  which is known as the comoving reference frame that can be made by a Lorentz and conformal scaling transformation. Namely, only by fixing the Lorentz transformation and conformal scaling condition to be the comoving reference frame can one yield an isotropic and homogeneous background gravifield spacetime: 
\be
\langle l_{\chi}^2 \rangle & = &   \ab^2(\eta)  \eta_{\mu\nu} d\hat{x}^{\mu}d\hat{x}^{\nu} =  \ab^2(t)  \eta_{\mu\nu} dx^{\mu}dx^{\nu}  \, ,
\ee
where $\eta$ coincides with the {\it comoving Lorentz time}, i.e., $\eta = t$. 

We now turn to the background conformal scale field which has a singularity when the {\it conformal proper time} or the comoving Lorentz time approaches to the epoch given by the inverse of {\it the conformal proper energy scale} 
\be
\eta \to \ek \equiv \pm \frac{1}{c\kh}\, , \quad \ab(\eta)\to \ab_{\pm}(\ek) = \infty \, ,
\ee
which leads the conformal size of  the background gravifield spacetime  to be infinitely large.  The light-traveled distance from $\eta=0$ to $\eta=|\ek|$ is given by 
\be
\Lk = c|\ek|  = 1/\kh \, ,
\ee
which defines the {\it cosmological horizon} in the background gravifield spacetime.

In conclusion, the {\it background gravifield spacetime} is a conformally flat Minkowski spacetime, which is described by the {\it cosmic vector} $\kappa_{\mu}$ with the {\it cosmological mass} scale $\mk = \sqrt{\kappa_{\mu}\kappa^{\mu}}$. Such a background gravifield spacetime is shown to be characterized by the {\it conformal proper time } $\eta = x_{\mu}\kappa^{\mu}/c\kh $ with the {\it cosmological horizon} $\Lk=1/\kh$ at which the {\it conformal scale factor} $\ab(\eta)$ becomes {\it singular}. 

\subsection{Evolution of Early  Universe with conformal inflation and deflation}

To describe the evolution of the Universe in the background gravifield spacetime, it is useful to introduce a {\it cosmic proper time } $\tau$ via the following definition:
\be
d\tau \equiv \ab(\eta) d\eta  \, , \quad  \ab(\eta) \to \pm \ab_{\kappa\pm} (\eta) = \pm \frac{\mk}{\alpha_S M_S } \frac{1}{1 \mp \kh \, c\eta} = \pm \frac{\mk}{\alpha_S M_S } \frac{1}{1 \mp c\eta/\Lk }\, .
\ee 
Performing the integration, we arrive at the relation between the {\it conformal proper time} $\eta$ and the {\it cosmic proper time} $\tau$
\be
 & & \pm \frac{1}{1 - \kh \, c\eta} = \pm e^{\Mk c \tau}\, , \quad -\infty \leq c\eta \leq \Lk \, , \quad  -\infty \leq \tau \leq \infty  \, , \nonumber \\
 & & \pm \frac{1}{1 - \kh \, c\eta} = \mp e^{-\Mk c \tau}\, , \quad \Lk  \leq c\eta \leq \infty \, , \quad  -\infty \leq \tau \leq \infty  \, , \nonumber \\
 & & \pm \frac{1}{1 + \kh \, c\eta} = \mp e^{\Mk c \tau}\, , \quad -\infty \leq c\eta \leq -\Lk \, , \quad  -\infty \leq \tau \leq \infty  \, , \nonumber \\
 & & \pm \frac{1}{1 + \kh \, c\eta} = \pm e^{-\Mk c \tau}\, , \quad -\Lk \leq c\eta \leq \infty \, , \quad  -\infty \leq \tau \leq \infty  \, .
 \ee
 The conformal scale factor in light of the {\it cosmic proper time} is given by
\be
& &  \ab(\tau) \to \pm \ab_{\kappa\pm}(\tau) = \pm \ab_0 e^{\pm \Mk c \tau} \, (\mbox{or} \; \mp \ab_0 e^{\mp\Mk c \tau} ) \, , \quad  \ab_0 = \kh/\Mk = \lk/\Lk \, ,
\ee
with the definition
 \be 
 & & \Mk \equiv \alpha_S M_S \, \frac{\kh}{\mk} \, ,  \qquad \lk = \frac{1}{\Mk} \, .
 \ee
Here $\Mk$ is regarded as the {\it primary cosmic energy scale} and $\lk$ as the {\it primary cosmic horizon}. In obtaining the above relation, we have used the conventional integration condition for the conformal scale factor $\ab(\eta=0) = \ab(\tau=0)$. 

It is interesting to notice that it needs to take an infinitely large {\it cosmic proper time} $\tau \to \infty$ for the light-traveled distance closing to the {\it cosmological horizon} $c\eta \to \Lk$.  The line element in terms of the {\it cosmic proper time} reads
\be
\langle l_{\chi}^2 \rangle & = & \gb_{\mu\nu}(\tau) d\hat{x}^{\mu}d\hat{x}^{\nu} \,  ; \qquad \hat{x}^{\mu} = (c\tau, x^i) \, ,
\ee
with the metric tensor 
\be
& & \bar{g}_{\mu\nu} (\tau) = \frac{1}{u_0^2}
\left(
\begin{array}{cc}
 1 &  - \ab(\tau)\, u_j     \\
 - \ab(\tau)\, u_i  &     \ab^2(\tau)\, u_0^2\, \bar{g}_{ij}
\end{array}
\right)   \, , \quad 	 \bar{g}_{ij} = \eta_{ij} + \frac{u_iu_j}{u_0^2 } \, .
\ee

The metric of the background gravifield spacetime is, in general, not isotropic in terms of the cosmic proper time $\tau$. Only in the comoving reference frame with $u_i =0$ and $u_0=1$, i.e., the {\it  cosmic proper time } $\tau$ is correlated to the {\it comoving Lorentz time} $t$, does the background gravifield spacetime become isotropic and homogeneous with the line element
\be
\langle l_{\chi}^2 \rangle & = &  c^2 d\tau^2 +  \ab^2(\tau) \eta_{ij}\, dx^{i} dx^{j}  \, ,
\ee
which produces the well-known form of Friedman-Lema"tre-Robertson-Walker (FLRW) metric for characterizing the inflationary $[\ab(\tau) = \ab_0 e^{\Mk c \tau} ]$  or deflationary $[\ab(\tau) = \ab_0 e^{-\Mk c \tau}] $ expansion of the Universe as the time arrow from past to future. In other words,  only in light of the {\it cosmic proper time} $\tau$, does the background gravifield spacetime appear to be a conformally inflationary or deflationary Universe.

\section{Quantization of Gravitational Interactions in Unitary Basis and Quantum Inflation of Early Universe}

Let us now discuss the quantization of gravitational gauge theory based on the background structure of gravifield spacetime after the gravitational gauge symmetry breaking.  As a direct consequence, we demonstrate how the quantum effect causes the inflation of the early Universe and leads the inflationary Universe to end at the epoch when the scaling symmetry is broken down spontaneously in a quantum induced effective background scalar potential. 

\subsection{Quantization of gravitational interactions in unitary basis}

 Based on the background gravifield spacetime characterized by the conformally flat Minkowski spacetime, the quantized fields are expressed as the following forms in the unitary basis
\be
& & \chi_{\mu}^{\;\; a} = \chib_{\mu}^{\;\; a} + h_{\mu}^{\;\; a}(x)/M_W \, , \quad \chib_{\mu}^{\;\; a} = \eta_{\mu}^{\; a} \, , \nonumber \\
& &   \phi(x) =  \bar{\varphi}(x) + \varphi(x)\, , \quad \pb(x) = M_S\, \ab(x) \, , \nonumber \\
& &  \Om_{\mu}^{ab}(x) =  \tilde{\chi}_{\mu\rho}^{[ab]} \ob^{\rho}(x) +  \Omega_{\mu}^{ab}(x) = \eta_{\mu\rho}^{[ab]} \ob^{\rho}(x)  + \tilde{H}_{\mu\rho}^{[ab]} \ob^{\rho}(x)/M_W +  \Omega_{\mu}^{ab}(x) \, , \nonumber \\
& & W_{\mu}(x) = \bar{w}_{\mu}(x)  + w_{\mu}(x)\, , \quad g_w \wb_{\mu}(x) = g_s \ob_{\mu}(x) = \partial_{\mu} \ln \pb(x) \, , 
\ee
with the definitions
\be
 & & \tilde{\chi}_{\mu\rho}^{[ab]}(x) = \chi_{\mu}^{\;\; a} \hat{\chi}^b_{\;\;\rho}- \chi_{\mu}^{\;\; b}\hat{\chi}^a_{\;\;\rho}   = \langle \tilde{\chi}_{\mu\rho}^{[ab]}(x) \rangle + \tilde{H}_{\mu\rho}^{[ab]}(x) \, ,  \quad   \langle \tilde{\chi}_{\mu\rho}^{[ab]}(x) \rangle  = \eta_{\mu\rho}^{[ab]} \, , \nonumber \\
 & & \tilde{H}_{\mu\rho}^{[ab]}(x) = h_{\mu}^{\;\; a} \eta^b_{\;\;\rho} - h_{\mu}^{\;\; b} \eta^a_{\;\;\rho}
 - \eta_{\mu}^{\;\; a}\hat{h}^b_{\;\;\rho}  + \eta_{\mu}^{\;\; b}\hat{h}^a_{\;\;\rho}  - (\, h_{\mu}^{\;\; a} \hat{h}^b_{\;\;\rho} - h_{\mu}^{\;\; b} \hat{h}^a_{\;\;\rho} \, )/M_W \, , \nonumber \\
 & & \hat{\chi}_a^{\;\; \mu}(x) = \eta_a^{\; \mu} - \hat{h}_a^{\; \mu}(x)/M_W = \eta_a^{\; \mu} - h_a^{\; \mu}(x)/M_W  +  \Nh_a^{\;\; \mu}/M_W^2   \, , \nonumber \\
& & \hat{h}_{a}^{\;\; \mu} \equiv h_a^{\;\; \mu} -  \Nh_a^{\;\; \mu}/M_W\, , \quad  \Nh_a^{\;\; \mu} = \sum_{n=1}^{\infty}  \frac{(-1)^{n-1}}{M_W^{n-1}} (h^{n+1})_{a}^{\;\; \mu}  \, ,
 \ee
where we have introduced a {\it weighting energy scale} $M_W$ to make the quantized gravifield $h_{\mu}^{\;\; a}(x)$ dimensionful. $M_W$ may be fixed via the normalization of the kinetic term for the gravifield and given by the basic scaling energy scale 
\be
M_W^2 = \alpha_W M_S^2 \, .
\ee
The gravifield must satisfy an additional condition in the unitary basis ($\chi = \det \chi_{\mu}^{\;\; a}=1$) 
\be 
& &  \ln \det \chi_{\mu}^{\;\; a} =   \Tr\ln ( \, \eta_{\mu}^{\; a} + h_{\mu}^{\;\; a}/M_W\, )  =0 \, , \;\; \mbox{or} \;\; \sum_{n=1}^{\infty} \frac{ (-1)^{n-1}}{ n M_W^{n-1}} (h^n)_{\mu}^{\;\; a} \eta_a^{\; \mu} =0 \, .
\ee

Before proceeding, we would like to point out that a similar unitary basis was actually adopted in Einstein's original paper to make a significant simplification for the equations of motion. It was emphasized by Einstein that: ``if  $-\det g_{\mu\nu} $ (in curved coordinate spacetime) is always finite and positive, it is natural to settle the choice of coordinates  {\it a posteriori } in such a way that this quantity is always equal to unity." ``Thus, with this choice of coordinates, only substitutions for which the determinant is unity are permissible. " ``But it would be erroneous to believe that this step indicates a partial abandonment of the general postulate of relativity. We do not ask : what are the laws of nature which are covariant in face of all substitutions for which the determinant is unity? But our question is: what are the general covariant laws of nature?  "  Obviously, the QFT of gravity described in our present consideration based on the spin and scaling gauge symmetries in the flat Minkowski spacetime provides an answer to the question that Einstein did not ask. The answer is manifest that it is the scaling gauge invariance that allows us to settle the choice of gauge-fixing condition, so that the determinant of the metric tensor field can always be made to be unity, i.e., $\chi = \det \chi_{\mu}^{\;\; a}=1$ or $-\det \chi_{\mu\nu} =1 $. Meanwhile, such a gravitational gauge theory is invariant, or laws of nature are covariant, in the face of all "substitutions" resulting from both the local spin gauge transformations of SP(1,3) and the global Lorentz transformations of SO(1,3) for which the determinant is unity.

In terms of the above quantized fields $h_{\mu}^{\;\; a}(x)$, $\varphi(x)$, $w_{\mu}(x)$, and $\Omega_{\mu}^{ab}(x)$ in the unitary basis,  the action for the quantum gravity gauge theory gets the following form: 
\be
\label{action4}
S_{\chi}  & = & \int d^{4}x\;  \frac{1}{2}  [\,  \hat{\chi}^{\mu\nu} (\bar{\Psi} \chi_{\mu} i {\mathsf D }_{\nu}   \Psi  + \bar{\psi} \chi_{\mu} i \td_{\nu}   \psi )  + H.c. \, ]  - y_s \bar{\psi} (\pb + \varphi) \psi \nonumber \\
& - & \frac{1}{4}  \hat{\chi}^{\mu\mu'} \hat{\chi}^{\nu\nu'} [\, {\cal F}^I_{\mu\nu} {\cal F}^{I}_{\mu'\nu'}  + {\mathsf R}_{\mu\nu}^{ab}  {\mathsf R}_{\mu'\nu' ab}  + W_{\mu\nu} W_{\mu'\nu'}   \, ]   \nonumber \\
& + &  \alpha_W (\, \pb + \varphi \,)^2 \frac{1}{4}  \hat{\chi}^{\mu\mu'} \hat{\chi}^{\nu\nu'}  {\mathsf G}_{\mu\nu}^a {\mathsf G}_{\mu'\nu' a}  + \frac{1}{2} \hat{\chi}^{\mu\nu} d_{\mu} \varphi d_{\nu}\varphi  -  \lambda_s ( \pb + \varphi  )^4    \nonumber \\
& - & [\, \alpha_E ( \pb + \varphi )^2 - \hat{\chi}^{\rho\sigma} \ob_{\rho}\ob_{\sigma} \, ]  g_s \hat{\chi}_a^{\;\;\mu} \hat{\chi}_b^{\;\;\nu}  {\mathsf R}_{\mu\nu}^{ab} \nonumber \\
& + & 6 \alpha_E g_s ( \pb + \varphi  )^2  [\,  \hat{\chi}_a^{\;\mu}  \td_{\mu} (\hat{\chi}^{a\nu} \ob_{\nu} ) + g_s \hat{\chi}^{\mu\nu} \ob_{\mu}\ob_{\nu}  \, ]  \nonumber \\
&- & 2 \alpha_E g_s ( \pb + \varphi )^2 \hat{\chi}^{\nu\rho} \ob_{\rho} \hat{\chi}_a^{\;\;\mu}  G_{\mu\nu}^a  - 3 g_s^2  (\, \hat{\chi}^{\mu\nu} \ob_{\mu}\ob_{\nu} )^2  \nonumber \\
& + &  \frac{1}{2} \hat{\chi}^{\mu\nu}[\, g_w^2 ( w_{\mu}\pb  + \wb_{\mu} \varphi) ( w_{\nu}\pb  + \wb_{\nu} \varphi) - 2g_w (\pb w_{\mu} + \wb_{\mu} \varphi) d_{\nu} \varphi \, ] \nonumber \\
& - & \hat{\chi}_b^{\;\rho}\, \ob_{\rho} \,  \hat{\chi}^{\mu\mu'} \hat{\chi}^{\nu\nu'} G_{\mu'\nu' a}  {\mathsf R}_{\mu\nu}^{ab}  -2\hat{\chi}_a^{\;\;\mu} \hat{\chi}^{\nu\nu'}   [ \, \td_{\nu'}(\hat{\chi}_b^{\;\; \rho}\ob_{\rho})  - 
g_s \ob_{\nu'}  \hat{\chi}_b^{\;\; \rho}\ob_{\rho}  \, ]  {\mathsf R}_{\mu\nu}^{ab}    \nonumber \\
& - & \frac{1}{4}  \hat{\chi}^{\mu\mu'} \hat{\chi}^{\nu\nu'}[\,  \hat{\chi}^{\rho\sigma}\ob_{\rho}\ob_{\sigma}\eta_{ab} -  \hat{\chi}_a^{\;\;\rho} \hat{\chi}_b^{\;\;\sigma} \ob_{\rho}\ob_{\sigma}\, ] G_{\mu\nu }^a G_{\mu'\nu'}^b \nonumber \\ 
& - & 2 \hat{\chi}^{\mu\nu} \td_{\mu} ( \hat{\chi}_a^{\;\;\rho} \ob_{\rho}) \td_{\nu} ( \hat{\chi}^{a \sigma} \ob_{\sigma}) - 
[\hat{\chi}_a^{\;\mu}  \td_{\mu} ( \hat{\chi}^{a \rho} \ob_{\rho}) ]^2  \nonumber \\
& - & \hat{\chi}^{\mu\mu'} \hat{\chi}^{\nu\nu'}  G_{\mu\nu }^a  \ob_{\mu'} \td_{\nu'}(\hat{\chi}_a^{\;\; \rho}\ob_{\rho}) 
 + \frac{1}{2} \hat{\chi}_a^{\;\;\mu}  \hat{\chi}^{\nu\nu'}  G_{\mu\nu }^a  \partial_{\nu'}(\hat{\chi}^{\rho\sigma}\ob_{\rho} \ob_{\sigma})  \nonumber \\
 & + & 2 g_s \hat{\chi}^{\mu\nu}  \partial_{\mu}(\hat{\chi}^{\rho\sigma}\ob_{\rho} \ob_{\sigma}) \ob_{\nu} 
 - 4g_s \hat{\chi}_a^{\;\mu}  \td_{\mu}(\hat{\chi}^{a \nu}\ob_{\nu})  \hat{\chi}^{\rho\sigma}\ob_{\rho} \ob_{\sigma}    +  {\cal L}'(x)   \, , 
\ee
with 
\be
& & i{\mathsf D}_{\mu} = i\partial_{\mu} + \Omega_{\mu} + {\cal A}_{\mu} \, , \quad \chi_{\mu} \equiv \chi_{\mu}^{\;\; a}\gamma_a  \nonumber \\
& & {\mathsf R}_{\mu\nu}^{ab}  = \partial_{\mu}\Omega_{\nu}^{ab} -   \partial_{\nu}\Omega_{\mu}^{ab} + g_s( \Omega_{\mu c}^{a} \Omega_{\nu}^{cb}  - \Omega_{\nu c}^{a} \Omega_{\mu}^{cb} )  \nonumber \\
& & W_{\mu\nu} = \partial_{\mu} w_{\nu} - \partial_{\nu} w_{\mu}  \, , \quad d_{\mu}\varphi = (\partial_{\mu} - g_w w_{\mu} ) \varphi \nonumber \\
& & G_{\mu\nu}^a = \td_{\mu} \chi_{\nu}^{\;\; a}  - \td_{\nu} \chi_{\mu}^{\;\; a}   = \chi_{\mu\nu}^a +    g_s (\Omega_{\mu b}^a \chi_{\nu}^{\;\; b}  -  \Omega_{\nu b}^a \chi_{\mu}^{\;\; b} )\, , \quad \chi_{\mu\nu}^a = \partial_{\mu} \chi_{\nu}^{\;\; a}  - \partial_{\nu} \chi_{\mu}^{\;\; a}   \nonumber \\
& & {\mathsf G}_{\mu\nu}^a = \nabla_{\mu} \chi_{\nu}^{\;\; a}  - \nabla_{\nu} \chi_{\mu}^{\;\; a}   =  G_{\mu\nu}^a +   g_w (w_{\mu} \chi_{\nu}^{\;\; a} -  w_{\nu} \chi_{\mu}^{\;\; a} )   \, ,\nonumber \\
& & \hat{\chi}^{\mu\mu'} \hat{\chi}^{\nu\nu'}  {\mathsf G}_{\mu\nu}^a {\mathsf G}_{\mu'\nu' a} = \hat{\chi}^{\mu\mu'} \hat{\chi}^{\nu\nu'}  G_{\mu\nu}^a G_{\mu'\nu' a} + 4g_w \hat{\chi}^{\mu\rho}  \hat{\chi}_a^{\;\;\nu} G_{\mu\nu}^a w_{\rho} 
+ 6 g_w^2 \hat{\chi}^{\mu\nu} w_{\mu}w_{\nu} \, .
\ee
It will be useful to introduce the following notations for the tensors of the gravifield: 
\be 
& & \chi_{\mu\nu} =  \eta_{\mu\nu} + {\cal H}_{\mu\nu}/M_W \, , \quad {\cal H}_{\mu\nu}  =  H_{\mu\nu} + N_{\mu\nu}/M_W\, , \nonumber \\
& &  H_{\mu\nu} =  h_{\mu\nu} + h_{\nu\mu}\, , \quad N_{\mu\nu} = h_{\mu}^{\;\; a}h_{\nu}^{\; \; b}\eta_{ab} \, , \quad h_{\mu\nu} = h_{\mu}^{\;\; a} \eta_{\nu a}   \nonumber \\
& &  \chih^{\mu\nu} =  \eta^{\mu\nu} - {\cal \Hh}^{\mu\nu} /M_W \, , \quad {\cal  \Hh}^{\mu\nu} \equiv H^{\mu\nu} - \hat{N}^{\mu\nu}/M_W =  \sum_{n=1}^{\infty}\frac{ (-1)^{n-1} }{M_W^{n-1}} ({\cal H}^n)^{\mu\nu}  \nonumber \\
 & & {\cal \hat{H}}^{\mu\nu}_{ab} =  \hh_a^{\;\; \mu} \eta_b^{\nu} + \eta_a^{\mu} \hh_b^{\;\; \nu} - \hh_a^{\;\; \mu} \hh_b^{\;\;\nu} /M_W \, , \quad \hat{N}^{\mu\nu}_{ab} =  \Nh_a^{\;\; \mu} \eta_b^{\nu} + \eta_a^{\mu} \Nh_b^{\;\; \nu} + \hh_a^{\;\; \mu} \hh_b^{\;\;\nu} \, 
\ee
and express the field strength into the following form
\be
& & G_{\mu\nu}^a  = (\td_{\mu} h_{\nu}^{\;\; a}  - \td_{\nu} h_{\mu}^{\;\; a})/M_W +  g_s (\Omega_{\mu b}^a \eta_{\nu}^{\;\; b}  -  \Omega_{\nu b}^a \eta_{\mu}^{\;\; b} ) \, , \nonumber \\
& & {\mathsf G}_{\mu\nu}^a  = (\nabla_{\mu} h_{\nu}^{\;\; a}  - \nabla_{\nu} h_{\mu}^{\;\; a})/M_W  +   g_w (w_{\mu} \eta_{\nu}^{\;\; a} -  w_{\nu} \eta_{\mu}^{\;\; a} ) + g_s (\Omega_{\mu b}^a \eta_{\nu}^{\;\; b}  -  \Omega_{\nu b}^a \eta_{\mu}^{\;\; b} )  \, .
\ee

It is seen from the above quantized action Eq.(\ref{action4}) that there exist rich gravitational interactions with respect to the background fields. Note that, without specifying the background fields $\pb(x)$, $\ob_{\mu}(x) $, and $ \wb_{\mu}(x)$, the resulting action formally remains gauge invariant.  Once applying the background field solutions obtained from the equations of motion to the above action, all the linear terms of quantized fields become vanishing and the gravitational gauge symmetries are broken down to the corresponding global symmetries.

\subsection {Physical degrees of freedom with massless graviton and massive spinon}

To discuss the quantum effects, it is necessary to know the physical quantum degrees of freedom for all the quantum fields. For the Dirac fermion fields and internal gauge fields, the counting rule on the physical quantum degrees of freedom is well known. Concerning the quantum fields $\chi_{\mu}^{\;\; a}(x)$, $\Omega_{\mu}^{ab}(x)$, $W_{\mu}(x)$, and $\phi(x)$ appearing in the QFT of gauge gravity interactions, the same counting rule should be applicable to figure out the independent physical quantum degrees of freedom. In general, it should be convenient to work in Euclidean spacetime by making a Wick rotation.  

Classically, the gravifield $\chi_{\mu}^{\;\; a}(x)$ as a bicovariant vector field has 16 field components, $\Omega_{\mu}^{ab}(x)$ possesses 24 field components due to the antisymmetric feature $\Omega_{\mu}^{ab}(x)= - \Omega_{\mu}^{ba}(x)$, $W_{\mu}(x)$ is a gauge vector field with four field components, and $\phi(x)$ is a single real field.  There are totally 45 components. Before gauge symmetry breaking, they are all massless. For  the massless gauge fields, only the transverse components are physical quantum degrees of freedom. Thus, there are 12 independent physical quantum degrees of freedom for the massless spin gauge field$\Omega_{\mu}^{ab}(x)$, two independent physical quantum degrees of freedom for the massless scaling gauge field $W_{\mu}(x)$ and eight independent physical quantum degrees of freedom for the massless gauge-type gravifield $\chi_{\mu}^{\;\; a}(x)$. When fixing the scaling gauge condition to the unitary basis, i.e., $\chi = \det \chi_{\mu}^{\;\; a}(x) = 1$,  the eight physical quantum degrees of freedom for the gravifield $\chi_{\mu}^{\;\; a}(x)$ will be reduced to seven.  Thus, there are only 22 independent physical quantum degrees of freedom including the single scalar field.

After considering the gravitational gauge symmetry breaking with $\langle \chi_{\mu}^{\;\; a}(x) \rangle = \chib_{\mu}^{\;\; a} = \eta_{\mu}^{\;\; a} $, $\langle \phi(x) \rangle = \pb(x) = \ab M_S $, the leading gravifield dynamical interaction term generates masses for the spin and scaling gauge fields:   
\be 
 & & \frac{1}{4}  \ab^2 M_W^2 \eta^{\mu\mu'} \eta^{\nu\nu'}  {\mathsf G}_{\mu\nu}^a {\mathsf G}_{\mu'\nu' a}  = \frac{1}{4}  \ab^2  \eta^{\mu\mu'} \eta^{\nu\nu'}  (\nabla_{\mu} h_{\nu}^{\;\; a}  - \nabla_{\nu} h_{\mu}^{\;\; a}) (\nabla_{\mu'} h_{\nu' a} - \nabla_{\nu'} h_{\mu' a} ) \, \nonumber \\
& & \quad  + \ab^2  M_W  \eta^{\mu\mu'} \eta^{\nu\nu'}   [ \, g_w w_{\mu} \eta_{\nu}^{\;\; a} + 
g_s \Omega_{\mu b}^a \eta_{\nu}^{\;\; b}   \, ]    (\nabla_{\mu'} h_{\nu' a} - \nabla_{\nu'} h_{\mu' a} )  - \frac{1}{2} \ab^2 g_sg_w M_W^2  \Omega_{\mu}^{ab} w_{\nu} \eta^{\mu}_a\eta^{\nu}_b  \nonumber \\
& & \quad + \frac{3}{2}\ab^2 g_w^2 M_W^2 w_{\mu} w_{\nu} \eta^{\mu\nu}  + \frac{1}{4} \ab^2 g_s^2 M_W^2 
  \Omega_{[\mu a b] }  \Omega_{[\mu' a' b']} \eta^{\mu\mu'} \eta^{aa'}   \eta^{bb'}  \, ,
\ee
with $\Omega_{[\mu a b]} \equiv  \Omega_{\mu a b } - \Omega_{b a \mu }$. It indicates that the fully antisymmetric part of the spin gauge field $\Omega_{[\mu a b]}$ obtains masses. Namely, four components of the spin gauge field become massive. The spin gauge field gets additional four physical quantum degrees of freedom and possesses totally 16 physical quantum degrees of freedom. The massive scaling gauge field has three physical quantum degrees of freedom. As the gauge-type gravifield plays the role as a Higgs-type boson for causing the spin and scaling gauge symmetry breaking and generates masses for spin and scaling gauge fields, its five physical quantum degrees of freedom will be eaten by the spin and scaling gauge fields. Eventually,  the gauge-type gravifield possesses only two independent physical quantum degrees of freedom. To be specific, it is natural to choose the following two transverse components of the quantized gravifield $\chi_{\mu}^{\;\; a}(x) = \eta_{\mu}^a + h_{\mu}^{\;\; a}(x) /M_W$: 
\be \label{Graviton1}
h_{12}(x) = h_{21}(x), \quad h_{11}(x) = - h_{22}(x) 
\ee
to be independent physical quantum degrees of freedom. 

Let us now carry out an alternative analysis. It is well known that, for a gauge symmetry, one can always make a particular gauge transformation to fix the gauge. For the spin gauge symmetry, it concerns six free gauge group parameters $ \alpha_{ab}(x) = -\alpha_{ba}(x)$ ($ a,b = 0,1,2,3 $) in the spin gauge transformation $S(x) = e^{i\alpha_{ab}(x) \Sigma^{ab}/2}\in $ SP(1,3), which allows us to rotate away six components of the gauge-type gravifield $\chi_{\mu}^{\;\; a}(x)$ and reduce its 16 components into ten components. By appropriately choosing a gauge-fixing condition, such ten components of the gauge-type gravifield $\chi_{\mu}^{\;\; a}(x)$ may be made to be ten symmetric components for the gauge-type gravifield $\chi_{\mu}^{\;\; a}(x)$, i.e., $\chi_{\mu a} (x) = \chi_{a \mu}(x)$. Such a fixing gauge is referred as a {\it unitary gauge}. When further making a scaling gauge transformation to fix the scaling gauge in such a way that the determinant of gravifield $\chi_{\mu}^{\;\; a}(x)$ becomes unity, i.e., $\chi = \det \chi_{\mu}^{\;\; a}(x) = 1$, which is shown to define a unitary basis for the gravifield spacetime. Thus the symmetric gravifield $\chi_{\mu}^{\;\; a}(x)$ is constrained to be nine components. As a consequence, in the unitary basis with a unitary gauge-fixing condition, the independent physical quantum degrees of freedom for the gauge-type gravifield $\chi_{\mu a}(x)= \eta_{\mu a} + h_{\mu a}(x) /M_W $ are counted by the symmetric transverse components with a traceless condition
\be \label{Graviton2}
h_{ij} = h_{ji}\, , \quad \sum_{i} h_{ii} = 0\, ; \quad i, j = 1, 2 \, ,
\ee
which comes to the same result as Eq.(\ref{Graviton1}). Such a massless quantized transverse gravifield $h_{ij}$ should be  the {\it graviton}.

Before ending this subsection, let us illustrate the gravitational gauge interactions with fermions. The gauge-invariant Lagrangian for fermions is given by  
\be
 {\cal L}_F & = & \frac{1}{2}  [\, \eta^{\mu\nu} (\bar{\Psi} \gamma_{\mu} i D_{\nu} \Psi  + \bar{\psi} \gamma_{\mu} i \partial_{\nu}   \psi )   + H.c.\, ] 
 + \frac{1}{4} \Omega_{[\mu a b]} [\,  \bar{\Psi} \{ \gamma^{\mu}\, \;  \frac{1}{2}\Sigma^{a b} \}  \Psi  
 +  \bar{\psi} \{ \gamma^{\mu}\, \;  \frac{1}{2}\Sigma^{ab} \}   \psi \, ] \nonumber \\
 & - &  \frac{1}{2M_W} [\, ( h_a^{\; \; \mu} - \Nh_a^{\; \; \mu}/M_W )  (\bar{\Psi} \gamma^a i {\mathsf D }_{\mu}   \Psi  + \bar{\psi} \gamma^a i \td_{\mu}   \psi )   + H.c. \, ] \, ,
 \ee
with the relation
\be
\{ \gamma^{\mu}\, \;  \Sigma^{ab} \}  = \epsilon^{\mu a b\nu} \gamma_{\nu} \gamma_5 \, .
\ee

It can be seen that the leading spin gauge interaction with fermions is governed by the totally antisymmetric components of the spin gauge field $\Omega_{[\mu a b]}$, which leads to an axial-vector current interaction.  We may call such a totally antisymmetric spin gauge field $\Omega_{[\mu a b]} $  a {\it spinon}. As demonstrated above, the {\it spinon} is massive and its interaction is suppressed by a heavy mass. The interactions of the massless graviton $h_a^{\; \; \mu}$ are suppressed by the weighting mass scale $M_W$.

In conclusion, the gauge-type gravifield $\chi_{\mu}^{\;\; a}(x)$ has two independent physical quantum degrees of freedom that act as the massless graviton with two transverse polarizations $h_{ij}$. The totally antisymmetric spin gauge field$\Omega_{[\mu a b]}$ acts as the massive spinon, which leads to a torsional interaction.

\subsection{Gauge-fixing contributions to quantization of gravity theory}

We shall show explicitly how the gauge-fixing requirement contributes to the quantization of gravity theory. In the path integral approach,  as both the path integral measure of the gauge fields and the action are gauge invariant, the functionally integrating over the gauge fields will over count the degrees of freedom. To overcome this problem, a gauge-fixing condition is needed. In general, we have the freedom to choose the gauge-fixing condition in the path integral method as the gauge-fixing is realized by inserting the delta functions into the path integral. Such a formalism was initially developed by Faddeev and Popov\cite{FP}. The main step of gauge-fixing is to find an explicit expression for the Faddeev-Popov determinant. 
Suppose that a gauge-fixing condition is set to be
\be
{\cal F}(A_{\mu}) = 0 \, ; 
\ee
then the path integral is expressed as 
\be
\int {\cal D} A \,  \prod_{x} \delta ( {\cal F}(A) )\, \Delta_{F}\, \, e^{i \int d^4x\, {\cal L} }\, ,
\ee
where $\Delta_{F}$ is the Faddeev-Popov determinant given by
\be
& & \Delta_{F} = \det {\cal M} \, ; \qquad {\cal M}_{jk}  =  \frac{\delta {\cal F}_j(A^g)(x) }{\delta \alpha_k(y)}|_{g=e}
\ee
with $\alpha_k$ the group parameters. Both the $\delta$ function and the determinant can be expressed by the exponential form as follows:
\be
& & \int {\cal D} A \,  \prod_{x} \delta ( {\cal F}(A) )\, \Delta_{F}\, \, e^{i \int d^4x\, {\cal L}_G } = \int {\cal D} A \int \prod_k d\vt_k  d\bvt_k \, e^{i \int d^4x\, ( {\cal L}_G + {\cal L}_{Gf} + {\cal L}_{Gs}) }\, , \nonumber \\
& & {\cal L}_{Gf} = - \frac{1}{2} \lambda_F\,   {\cal F}_j(A)  {\cal F}^j(A) \, , \quad  {\cal L}_{Gs} =  \int d^4 y \, \bvt^{j}(x) {\cal M}_{jk}(x,y) \vt^{k}(y) \, , 
\ee 
where ${\cal L}_G$, ${\cal L}_{Gf}$, and ${\cal L}_{Gs}$ represent the Lagrangian for the gauge invariant theory of gravity, the gauge-fixing term, and the Faddeev-Popov ghost term, respectively.  $\lambda_F$ is an arbitrary parameter, and $\vt^{k}$ are the Grassman variables. 

For the gauge theory of gravity described by the action given in Eq.(\ref{action3}), the relevant gauge and tensor fields concern the spin gauge field$\Om_{\mu}^{ab}$, the scaling gauge field $W_{\mu}$, the gauge-type gravifield $\chi_{\mu}^{\;\, a}$, and the tensor field $\hat{\chi}^{\mu\nu}$. The action is, in general, invariant under the spin and scaling gauge transformations. To fix the gauge, let us take explicitly the following Lorentz-type gauge-fixing conditions: 
\be
\sqrt{\chi}\, \hat{\chi}^{\mu\nu} \partial_{\mu}\Om_{\nu}^{ab} = 0\,  , \quad \sqrt{\chi}\, \hat{\chi}^{\mu\nu} \sqrt{\alpha_w}\, \phi \partial_{\mu}\chi_{\nu}^{\;\, a} = 0\, , \quad  \sqrt{\chi}\, \hat{\chi}^{\mu\nu} \partial_{\mu}W_{\nu} =  0 \, .
\ee
Note that the tensor gravifield $\hat{\chi}^{\mu\nu}= \hat{\chi}_a^{\;\, \mu}  \hat{\chi}_b^{\;\, \nu} \eta^{ab}$ that couples to all interaction terms remains invariant under the spin gauge transformation as the spin gauge appears to be a hidden gauge for the tensor gravifield $\hat{\chi}^{\mu\nu}$. For that, we shall impose an additional gauge-fixing condition:
\be
\sqrt{\chi}\, \sqrt{\alpha_w}\, \phi \partial_{\mu} \hat{\chi}_a^{\;\, \mu} = 0\, .
\ee
Such a condition is equivalent, from the identity $\partial_{\mu} \hat{\chi}_a^{\;\, \mu} = (\partial_{\mu} \hat{\chi}^{\mu\nu} ) \chi_{\nu a } + \hat{\chi}^{\mu\nu} \partial_{\mu}\chi_{\nu a}$,  to the following gauge-fixing condition 
\be
 \sqrt{\chi}\, \sqrt{\alpha_w}\, \phi \partial_{\mu} \hat{\chi}^{\mu\nu} =0 \, ,
\ee 
which is usually taken as the gauge-fixing condition for a general coordinate invariance. 

We now turn to calculate the Faddeev-Popov determinant. Let us first consider the infinitesimal spin and scaling gauge transformations $S(x) = e^{i\alpha_{ab}(x) \Sigma^{ab}/2} \simeq 1 + i\alpha_{ab}(x) \Sigma^{ab}/2 \in $ SP(1,3) and $\xi(x) = e^{\alpha(x)}\simeq 1 + \alpha(x)$, respectively. The corresponding infinitesimal changes of gauge fields and gravifield are found to be 
\be
& & \delta \Om_{\mu}^{ab} \simeq  - g_s^{-1} \partial_{\mu} \alpha^{ab} + \frac{1}{2} \alpha^{a}_{\; a'} \Om_{\mu}^{a' b}  
- \frac{1}{2} \alpha^{b}_{\; a'} \Om_{\mu}^{a' a} \, , \nonumber \\
& & \delta \chi_{\mu}^{\;\, a} \simeq  \chi_{\mu}^{\;\, b} \alpha_{b}^{\;\, a} \, , \quad \delta \chi_{\mu}^{\;\, a} \simeq  - \alpha \chi_{\mu}^{\;\, a} \, , \nonumber \\
& & \delta \hat{\chi}_a^{\;\, \mu} \simeq  \alpha_{a}^{\;\, b} \hat{\chi}_b^{\;\, \mu}\, , \quad  \delta \hat{\chi}_a^{\;\, \mu} \simeq \alpha \hat{\chi}_a^{\;\, \mu} \, , \nonumber \\
& & \delta W_{\mu} =  - \partial_{\mu} \alpha \, , \quad \delta \hat{\chi}^{\mu\nu} = 2\alpha \hat{\chi}^{\mu\nu} \, , 
\ee 
and the Faddeev-Popov matrix elements read off 
\be
& & {\cal M}_{(ab)(cd) } (x, y) = \sqrt{\chi}\,  \hat{\chi}^{\mu\nu}  \partial_{\mu} [\, - g_s^{-1} \partial_{\nu} \eta_{ac}\eta_{bd}  
+ \frac{1}{2} \eta_{ac} \Om_{\nu d b}  - \frac{1}{2} \eta_{bc} \Om_{\nu d a}  \, ] \,\delta^4(x-y) \, , \nonumber \\
& &  {\cal M}_{(a)(cd) } (x, y) = \sqrt{\chi}\,  \hat{\chi}^{\mu\nu} \sqrt{\alpha_w}\, \phi \eta_{ad} ( \partial_{\mu} \chi_{\nu c} + \chi_{\nu c} \partial_{\mu} ) \, \delta^4(x-y)  \, , \nonumber \\
& &  {\cal M}_{(ab)(c) } (x, y) = \sqrt{\chi}\,  \eta_{ac} \sqrt{\alpha_w}\, \phi ( \partial_{\mu} \hat{\chi}_{b}^{\;\,\mu} +  \hat{\chi}_{b}^{\;\,\mu} \partial_{\mu} )\, \delta^4(x-y) \, , \nonumber \\
& &  {\cal M}_{(a)(1) } (x, y)  = - \sqrt{\chi}\,  \hat{\chi}^{\mu\nu} \sqrt{\alpha_w}\, \phi  \chi_{\nu a} \partial_{\mu}\, \delta^4(x-y)  \, , \nonumber \\
& &  {\cal M}_{(1)(a) } (x, y) = \sqrt{\chi}\,   \sqrt{\alpha_w}\, \phi  \hat{\chi}_{a}^{\;\,\mu}\partial_{\mu} \, \delta^4(x-y) \, , \nonumber \\
& &  {\cal M}_{(1)(1) } (x, y) = - \sqrt{\chi}\,  \hat{\chi}^{\mu\nu} g_w^{-1} \partial_{\mu}\partial_{\nu}\, \delta^4(x-y) \, .
\ee

From the above analyses, we arrive at the Lagrangian for the gauge-fixing and Faddeev-Popov ghost terms as follows:
\be
& & {\cal L}_{Gf} = - \chi\, \frac{1}{2}\lambda_{\omega} \, \hat{\chi}^{\mu\nu} \hat{\chi}^{\mu'\nu'} [\,\partial_{\mu}\Om_{\nu}^{ab}  \partial_{\mu'}\Om_{\nu'ab} 
+ \partial_{\mu}W_{\nu}  \partial_{\mu'}W_{\nu'} \, ] \, , \nonumber \\
& & \qquad \quad + \chi\, \frac{1}{4}\lambda_{\chi}\alpha_w \phi^2 [\,  \hat{\chi}^{\mu\nu} \hat{\chi}^{\mu'\nu'}  \partial_{\mu}\chi_{\nu}^{\;\, a} \partial_{\mu'}\chi_{\nu' a}   +  \partial_{\mu}\hat{\chi}^{\;\, \mu}_{a} \partial_{\nu}\hat{\chi}^{a \nu } \, ]  \, , \nonumber \\
& & {\cal L}_{Gs} = - \sqrt{\chi}\, \hat{\chi}^{\mu\nu}  \bvt^{ab}  \partial_{\mu} (\, \eta_{ac}\eta_{bd}  \partial_{\nu}  
+  g_s \eta_{ac} \Om_{\nu bd}  \, ) \vt^{cd} \, \nonumber \\
& & \qquad \quad -  \sqrt{\chi}\, \sqrt{\alpha_w}\, \phi \hat{\chi}^{\mu\nu} \partial_{\mu} \chi_{\nu c} \eta_{ab}  ( \bvt^{b}  \vt^{ac} -\bvt^{ac} \vt^b ) 
+  \sqrt{\chi}\,  \sqrt{\alpha_w}\, \phi \partial_{\mu} \hat{\chi}^{\mu\nu}  \eta_{ab} \bvt^{ac} \chi_{\nu c}  \vt^b \,  \nonumber \\
& & \qquad \quad  -  \sqrt{\chi}\,  \sqrt{\alpha_w}\, \phi  \eta_{ab}\hat{\chi}_{c}^{\;\,\mu} ( \bvt^{b} \partial_{\mu} \vt^{ac} - \bvt^{ac} \partial_{\mu} \vt^{b} )  \, \nonumber \\ 
& & \qquad \quad  - \sqrt{\chi}\, \sqrt{\alpha_w}\, \phi  \hat{\chi}_a^{\;\, \mu} ( \bvt^{a} \partial_{\mu}  \vt - \bvt \partial_{\mu}  \vt^a ) -  \sqrt{\chi}\, \hat{\chi}^{\mu\nu} \bvt \partial_{\mu}\partial_{\nu} \vt \, ,
\ee
with $\bvt^{ab} = -\bvt^{ba} $ and $ \vt^{ab} = -\vt^{ba}$ the antisymmetric ghost variables. By making a special gauge transformation, i.e., $\chi_{\mu}^{\;\, a} \to \chi^{1/4} \chi_{\mu}^{\;\, a} $ and $\phi \to \chi^{-1/4}  \phi$, we can rewrite the above Lagrangian in the unitary basis $\chi = \det \chi_{\mu}^{\;\, a} = 1$: 
\be
& & {\cal L}_{Gf} = - \, \frac{1}{2}\lambda_{\omega} \, \hat{\chi}^{\mu\nu} \hat{\chi}^{\mu'\nu'} [\,\partial_{\mu}\Om_{\nu}^{ab}  \partial_{\mu'}\Om_{\nu'ab} 
+ \partial_{\mu}W_{\nu}  \partial_{\mu'}W_{\nu'} \, ] \, , \nonumber \\
& & \qquad \quad + \, \frac{1}{4}\lambda_{\chi} \alpha_w \phi^2 [\,  \hat{\chi}^{\mu\nu} \hat{\chi}^{\mu'\nu'}  d_{\mu} \chi_{\nu}^{\;\, a} d_{\mu'}\chi_{\nu' a}   +  \hat{d}_{\mu}\hat{\chi}^{\;\, \mu}_{a} \hat{d}_{\nu}\hat{\chi}^{a \nu } \, ]  \, , \nonumber \\
& & {\cal L}_{Gs} = - \hat{\chi}^{\mu\nu}  \bvt^{ab}  \partial_{\mu} (\, \eta_{ac}\eta_{bd}  \partial_{\nu}  
+  g_s \eta_{ac} \Om_{\nu bd}  \, ) \vt^{cd} \, \nonumber \\
& & \qquad \quad - \sqrt{\alpha_w}\, \phi \hat{\chi}^{\mu\nu} d_{\mu} \chi_{\nu c} \eta_{ab}  ( \bvt^{b}  \vt^{ac} -\bvt^{ac} \vt^b ) 
+   \sqrt{\alpha_w}\, \phi \tilde{d}_{\mu} \hat{\chi}^{\mu\nu}  \eta_{ab} \bvt^{ac} \chi_{\nu c}  \vt^b \,  \nonumber \\
& & \qquad \quad  -  \sqrt{\alpha_w}\, \phi  \eta_{ab}\hat{\chi}_{c}^{\;\,\mu} ( \bvt^{b} \partial_{\mu} \vt^{ac} - \bvt^{ac} \partial_{\mu} \vt^{b} )  \, \nonumber \\ 
& & \qquad \quad  - \sqrt{\alpha_w}\, \phi  \hat{\chi}_a^{\;\, \mu} ( \bvt^{a} \partial_{\mu}  \vt - \bvt \partial_{\mu}  \vt^a ) -   \hat{\chi}^{\mu\nu} \bvt \partial_{\mu}\partial_{\nu} \vt \, ,
\ee
with the notations 
\be 
d_{\mu} \equiv  \partial_{\mu} + \frac{1}{4} \partial_{\mu} \ln \chi \, , \quad \hat{d}_{\mu} \equiv  \partial_{\mu} - \frac{1}{4} \partial_{\mu} \ln \chi \, , \quad \tilde{d}_{\mu} \equiv  \partial_{\mu} - \frac{1}{2} \partial_{\mu} \ln \chi \, .
\ee
The gauge-invariant Lagrangian for the gauge theory of gravity in the unitary basis reads
\begin{eqnarray}
\label{actionQ}
{\cal L}_G & = & \frac{1}{2} [\, \hat{\chi}^{\mu\nu} ( \bar{\Psi} \chi_{\mu} i {\mathcal D}_{\nu}   \Psi  + \bar{\psi} \chi_{\mu} i \td_{\nu}   \psi  ) + H.c.\, ] - y_s \bar{\psi} \phi \psi   \nonumber \\
& - &  \frac{1}{4}  \hat{\chi}^{\mu\mu'} \hat{\chi}^{\nu\nu'} [\,  {\cal F}^I_{\mu\nu} {\cal F}^{I}_{\mu'\nu'} + {\cal R}_{\mu\nu}^{ab} {\cal R}_{\mu'\nu'ab} + {\cal W}_{\mu\nu} {\cal W}_{\mu'\nu'}  - \alpha_W\,  \phi^2\,  {\cal G}_{\mu\nu}^a {\cal G}_{\mu'\nu' a}  \, ]\nonumber \\
& + &  \frac{1}{2} \hat{\chi}^{\mu\nu} d_{\mu} \phi d_{\nu}\phi -  \alpha_E g_s \phi^2 \hat{\chi}^{\mu\mu'} \hat{\chi}^{\nu\nu'} \chi_{\mu}^{\;\,a } \chi_{\nu}^{\;\,b}  {\cal R}_{\mu'\nu' a b} - \lambda_s \phi^4 \, + {\cal L}'(x)    \, .
\end{eqnarray}
The total effective Lagrangian for the theory of quantum gravity is obtained by putting all parts together:  
\be
{\cal L }_{eff} = {\cal L}_G +  {\cal L}_{Gf}  + {\cal L}_{Gs} \, .
\ee

\subsection{Perturbative expansion and renormalizability of quantized gravity theory} 

It is noticed that the tensor gravifield $\hat{\chi}^{\mu\nu}$ couples to all kinematic terms and gauge interaction terms. Namely, the gravity does interact with all fields and the motion of all fields is surrounded by the gravitational interactions, as $\hat{\chi}^{\mu\nu}$ is determined by the inverse of the gauge-type gravifield $\chi_{\mu}^{\;\, a}$, which causes the nonlinear nature of the theory. To study the quantum theory of gravity, it is useful to work out a practically calculating framework. Based on the conformally flat background gravifield spacetime $\langle \chi_{\mu}^{\;\, a}(x)\rangle = \eta_{\mu}^a $ in the unitary basis, it is not difficult to show that the leading Lagrangian with dimensionless couplings has the following form
\be
& & L_G = \frac{1}{2} [\, \eta^{\mu}_{a} ( \bar{\Psi} \gamma^a i {\mathcal D}_{\mu}   \Psi  + \bar{\psi} \gamma^a i \td_{\mu}   \psi  ) + H.c.\, ] - y_s \bar{\psi} \phi \psi   \nonumber \\
& & \quad \quad -   \frac{1}{4}  \eta^{\mu\mu'} \eta^{\nu\nu'} [\,  {\cal F}^I_{\mu\nu} {\cal F}^{I}_{\mu'\nu'} + {\cal R}_{\mu\nu}^{ab} {\cal R}_{\mu'\nu'ab} + {\cal W}_{\mu\nu} {\cal W}_{\mu'\nu'}  -  a^2\,  {\bf G}_{\mu\nu}^a {\bf G}_{\mu'\nu' a}  \, ]\nonumber \\
& & \quad \quad +  \frac{1}{2} \eta^{\mu\nu} d_{\mu} \phi d_{\nu}\phi -  \alpha_E g_s \phi^2 \eta^{\mu}_{a} \eta^{\nu}_{b}  {\cal R}_{\mu\nu}^{ab} - \lambda_s \phi^4 \, + L'(x)    \, , 
\ee
with the definition
\be
{\bf G}_{\mu\nu}^a = (\nabla_{\mu} h_{\nu}^{\;\; a}  - \nabla_{\nu} h_{\mu}^{\;\; a}) +   g_w M_W (W_{\mu} \eta_{\nu}^{\;\; a} -  W_{\nu} \eta_{\mu}^{\;\; a} ) + g_s M_W (\Om_{\mu b}^a \eta_{\nu}^{\;\; b}  -  \Om_{\nu b}^a \eta_{\mu}^{\;\; b} ) \, .
\ee
For the gauge-fixing and Faddeev-Popov ghost terms, we have
\be
& & L_{Gf} = - \, \frac{1}{2}\lambda_{\omega} \, [\,\partial^{\mu}\Om_{\mu}^{ab}  \partial^{\nu}\Om_{\nu ab} + \partial^{\mu}W_{\mu}  \partial^{\nu}W_{\nu} \, ] \, , \nonumber \\
& & \quad \quad + \, \frac{1}{2}\lambda_{\chi} a^2 (\partial_{\mu} h^{\mu a} \partial_{\nu}h^{\nu}_{\;\, a}  + \frac{1}{16} \partial_{\mu} h\partial^{\mu} h  )   \, , \nonumber \\
& & L_{Gs} = - \bvt^{ab}  \partial^{\mu} (\, \eta_{ac}\eta_{bd}  \partial_{\mu}  
+  g_s \eta_{ac} \Om_{\mu bd}  \, ) \vt^{cd} \, \nonumber \\
& & \quad \quad - a (\partial^{\mu} h_{\mu c} + \frac{1}{4} \eta_{\mu c} \partial^{\mu} h) \eta_{ab}  ( \bvt^{b}  \vt^{ac} -\bvt^{ac} \vt^b ) \, \nonumber \\
& & \quad \quad -  a  (\partial_{\mu} H^{\mu\nu} - \frac{1}{4} \partial^{\nu} H) \eta_{\nu c} \eta_{ab} \bvt^{ac} \vt^b 
\,  \nonumber \\
& & \quad \quad  -  \sqrt{\alpha_w}\, \phi  \eta_{ab}\eta_{c}^{\;\,\mu} ( \bvt^{b} \partial_{\mu} \vt^{ac} - \bvt^{ac} \partial_{\mu} \vt^{b} )  \, \nonumber \\ 
& & \quad \quad  - \sqrt{\alpha_w}\, \phi  \eta_a^{\;\, \mu} ( \bvt^{a} \partial_{\mu}  \vt - \bvt \partial_{\mu}  \vt^a ) -  \bvt \partial_{\mu}\partial^{\mu} \vt \, ,
\ee
with 
\[ H^{\mu\nu} = h^{\mu\nu} + h^{\nu\mu}\, , \quad H = H^{\mu\nu} \eta_{\mu\nu}= 2h \, , \quad h_{\mu\nu} = h_{\mu}^{a}\eta_{\nu a}\, , \quad h = h_{\mu}^a\eta^{\mu}_a = h^{\mu\nu}\eta_{\mu\nu}\, . \] 
Thus the leading effective Lagrangian and effective action are given by 
\be
L_{eff} = L_G + L_{Gf} + L_{Gs} \, ,\quad  S_{eff} = \int d^4x \, L_{eff} \, ,
\ee
and the total effective Lagrangian and effective action are written as
\be
{\cal L}_{eff} = L_{eff} + \hat{L}_{eff}
\ee

It is shown that the identification of the physical quantum degrees of freedom and the definition of a quantum gravity theory are similar to the quantization of Yang-Mills gauge theory.  By decomposing the above leading effective action into a free part and interacting part after considering the gauge symmetry braking, one can write down the Feynman rules in a standard way. 

The question arises from the nonlinear nature of the tensor gravifield $\hat{\chi}^{\mu\nu}$ for the quantized gauge-type gravifield $\chi_{\mu}^{\;\, a} = \eta_{\mu}^a + h_{\mu}^{\;\, a}/M_W$, i.e., 
\be
& &\hat{\chi}_a^{\;\; \mu}(x)  = \eta_a^{\; \mu} - h_a^{\; \mu}/M_W  +  \Nh_a^{\;\; \mu}/M_W^2 \, , \quad 
\Nh_a^{\;\; \mu} = \sum_{n=1}^{\infty}  \frac{(-1)^{n-1}}{M_W^{n-1}} (h^{n+1})_{a}^{\;\; \mu}  \, ,  \nonumber \\
& & \chih^{\mu\nu} = \hat{\chi}_a^{\;\, \mu}\hat{\chi}^{a \nu}=  \eta^{\mu\nu} - H^{\mu\nu} /M_W  + \hat{N}^{\mu\nu}/M_W^2 \, ,\;\; \hat{N}^{\mu\nu} =  \sum_{n=1}^{\infty}\frac{ (-1)^{n-1} }{M_W^{n-1}} ({\cal H}^{n+1})^{\mu\nu}  \, ,
\ee
which leads to the high dimensionful interaction terms
\be
\hat{L}_G & = & -\frac{1}{2M_W} (h_a^{\; \mu} -  \Nh_a^{\;\; \mu}/M_W) [\,  ( \bar{\Psi} \gamma^a i {\mathcal D}_{\mu}   \Psi  + \bar{\psi} \gamma^a i \td_{\mu}   \psi  ) + H.c.\, ]   \nonumber \\
& + & \frac{1}{2M_W}[\, (H^{\mu\mu'} - \hat{N}^{\mu\mu'}/M_W) \eta^{\nu\nu'}  -  \frac{1}{2M_W}(H^{\mu\mu'} - \hat{N}^{\mu\mu'}/M_W)(H^{\nu\nu'} - \hat{N}^{\nu\nu'}/M_W) \, ]  
\nonumber \\ 
& \cdot & [\,  {\cal F}^I_{\mu\nu} {\cal F}^{I}_{\mu'\nu'} + {\cal R}_{\mu\nu}^{ab} {\cal R}_{\mu'\nu'ab} + {\cal W}_{\mu\nu} {\cal W}_{\mu'\nu'}  -  a^2\,  {\bf G}_{\mu\nu}^a {\bf G}_{\mu'\nu' a}  \, ]\nonumber \\
& + & 2\alpha_E g_s \phi^2 \frac{1}{M_W} [\, (h_a^{\; \mu} -  \Nh_a^{\;\; \mu}/M_W) \eta^{\nu}_b 
- \frac{1}{2} (h_a^{\; \mu} -  \Nh_a^{\;\; \mu}/M_W) (h_b^{\; \nu} -  \Nh_b^{\;\; \nu}/M_W)\, ]  {\cal R}_{\mu\nu}^{ a b} \nonumber \\
& - &  \frac{1}{2M_W} (H^{\mu\nu} - \hat{N}^{\mu\nu}/M_W)  d_{\mu} \phi d_{\nu}\phi  + \tilde{L}'(x)     \, .
\ee
Similarly, one can write down the high dimensionful  Lagrangian for the gauge-fixing and Faddeev-Popov ghost terms. 

The high dimensionful interaction terms may cause the theory to be nonrenormalizable in the usual sense that the divergencies cannot be absorbed into the basic parameters of the theory.  On the other hand, it has been shown that the gravifield spacetime is associated with a noncommutative geometry due to the nonvanishing field strength of gravifield. It is conceivable that there exists a fundamental energy scale that characterizes the ultraviolet behavior of the theory. In such a case, there is in principle no divergences appearing in the theory of quantum gravity, so that the quantum contributions from the high dimensionful interaction terms become finite and meaningful. It is particularly interesting to make a detailed investigation for such an expectation. In Refs.\cite{LR,wu2014}, it has been realized that there exists a symmetry-preserving and infinite-free regularization and renormalization method, which allows us to introduce intrinsically two meaningful energy scales to avoid infinities without spoiling gauge symmetries of the original theory. One of the energy scales is the so-called characterizing energy scale that plays the role as the ultraviolet cutoff. The consistency and applicability of the method have been demonstrated in a series of works\cite{Cui:2008uv, Cui:2008bk, Ma:2005md, Ma:2006yc, cui:2011,HW,HLW}. In particular, it has been applied to calculate the gravitational contributions to gauge Green's functions and show the asymptotic free power-law running of gauge coupling\cite{Tang2011}. More recently, it has been initiated to explore the quantum electroweak symmetry breaking mechanism for understanding the hierarchy problem\cite{QEWSB}.

\subsection{Quantum inflation of early Universe}

As has been shown, the background gravifield spacetime describes either an inflationary or a deflationary Universe. In any case, the Universe will undergo an extremely rapid exponential evolution in light of the cosmic proper time,  while it remains unclear how the inflation of the Universe occurs and which epoch the inflationary Universe gets end. It is hard to make a reliable issue on such a question without considering the quantum effect.
 
Let us first check the Lagrangian density and energy-momentum tensor for the background gravifield spacetime after gravitational gauge symmetry breaking. The Lagrangian for the background fields can be shown to have the following form
\be
\bar{{\cal L}}  & = &  - 6 \partial_{\mu} \ob_{\nu} \partial^{\mu} \ob^{\nu} + 12 \alpha_E g_s^2 \ob_{\mu}^2 \pb^2 - \lambda_s \pb^4 \, \nonumber \\
& = & - 6 g_s^2 \ob_{\mu}^2 \ob_{\mu}^2 + 12 \alpha_E g_s^2 \ob_{\mu}^2 \pb^2 - \lambda_s \pb^4  \, .
\ee

The energy-momentum tensor for the background gravifield spacetime is found to be
\be  \label{EMT2}
\bar{{\cal T}}_{\mu\nu} & & = - \eta_{\mu\nu}2[\, d_{\rho}\ob_{\sigma} d^{\rho}\ob^{\sigma} + 2 g_s d_{\rho}\ob^{\rho} \ob_{\sigma}\ob^{\sigma} 
+ 3 g_s^2  (\ob_{\sigma}\ob^{\sigma})^2 - \alpha_W \Ob_{\sigma}\Ob^{\sigma}\pb^2 \, ] \nonumber \\
& & \;\;\; - \eta_{\mu\nu} 2 \alpha_Eg_s (d_{\rho}\ob^{\rho} + 3g_s \ob_{\rho}\ob^{\rho})  \pb^2 
- [\,  \bar{W}_{\mu\sigma} \bar{W}_{\nu}^{\sigma} -  d_{\mu}\pb d_{\nu}\pb  - 2 \alpha_W  \Ob_{\mu} \Ob_{\nu} \pb^2 ] \nonumber \\
& &\;\;\; - 2[\, d_{\mu}\ob_{\sigma}  d_{\nu}\ob^{\sigma}  +  (d_{\mu}\ob_{\nu}  + d_{\nu}\ob_{\mu} ) d_{\sigma}\ob^{\sigma} - d_{\sigma}\ob_{\mu} d^{\sigma}\ob_{\nu} + 2 g_s  (d_{\mu}\ob_{\nu} + d_{\nu}\ob_{\mu}  ) \ob_{\sigma}\ob^{\sigma} \, ] \nonumber \\
& & \;\;\; +  4 \alpha_Eg_s \pb^2 d_{\nu}\ob_{\mu} -\eta_{\mu\nu}\bar{\cal L} \, .
\ee 
Under the ansatz that the conformally covariant derivative of the background scalinon field vanishes, $d_{\mu} \pb = 0 $, which means that the background scalinon field has vanishing conformally covariant kinetic energy, we then arrive at the relation $ g_s \ob_{\mu}(x) = g_w \wb_{\mu}(x) $ and the vanishing covariant kinetic energy for the background gauge fields, i.e., $d_{\mu}\ob_{\nu} = (\partial_{\mu} - g_s \ob_{\mu} ) \ob_{\nu}= (\partial_{\mu} - g_w \wb_{\mu} ) \ob_{\nu} = 0$. It is then not difficult to check that the above energy-momentum tensor is conserved. In fact, we yield 
\be
\bar{{\cal T}}_{\mu\nu} = 0 \, , 
\ee
which shows that the total energy-momentum of the background gravifield spacetime vanishes. Namely, the background gravifield spacetime represents the whole Universe, as it has no energy-momentum exchanging with its exterior.    

Let us examine the solution of the background scalinon field under the coordinate translation in the flat Minkowski spacetime,
\be
& & x_{\mu} \to x^{'\mu} = x^{\mu} + \alpha^{\mu}\, , \nonumber \\
& &  \pb(x) \to \pb_{\kappa\pm}(x') = \frac{\mk}{\alpha_S(1 \mp x^{'\mu}\kappa_{\mu}) } =  \frac{m_{\kappa'\pm} }{\alpha_S(1 \mp x^{\mu}\kappa'_{\mu \pm} )} \equiv \pb_{\kappa'\pm}(x) \, , 
\ee
with
\be
\kappa'_{\mu\pm} = \kappa_{\mu}/(1 \mp \alpha^{\mu} \kappa_{\mu} )\, , \qquad m_{\kappa'\pm} = \sqrt{\kappa'_{\mu}\kappa^{'\mu}}  = \mk/(1 \mp \alpha^{\mu} \kappa_{\mu} )
\ee
which corresponds to the rescaled cosmic vector and cosmological mass scale, respectively.  It indicates that the solutions of the background fields in difference reference frames of coordinates under the coordinate translation can equivalently be characterized by the rescaled cosmic vector. In other words, the background gravifield spacetime can well be described by the cosmic vector $\kappa_{\mu}$.
 
From the above Lagrangian, it is easy to find  the minimal condition for the background scalinon field $\pb(x)$, which leads to the following relation and a simplified Lagrangian
\be 
& & g_s^2 \ob_{\mu}^2 = \frac{\lambda_s}{6\alpha_E} \pb^2 =\alpha_S^2 \pb^2\, , \nonumber \\
& & \quad \bar{{\cal L}}  =  - ( 6\alpha_S^4/g_s^2 - \lambda_s )\, \pb^4 =  - \lambda_s \left(\frac{\alpha_S^2}{\alpha_E g_s^2} -1  \right)\, \pb^4  \, ,
\ee
where the relation between the background vector field and scalinon field is consistent with the solutions obtained from the equations of motion given in Eqs. (\ref{BSL0}) and (\ref{BSL6}). Thus, the minimal of the background scalar potential leads the background scalinon field to approach  
\be 
& & \pb \to 0\, , \; \; \qquad \mbox{for} \quad   6\alpha_S^4/g_s^2  >  \lambda_s = 6 \alpha_E\alpha_S^2  > 6 \alpha_E^2 g_s^2  \, , \nonumber \\
& & \pb \to \pm \infty \, , \; \quad \mbox{for} \quad 6\alpha_S^4/g_s^2 <   \lambda_s = 6 \alpha_E\alpha_S^2  < 6 \alpha_E^2 g_s^2  \, ,
\ee
which implies that the Universe characterized by the background gravifield spacetime evolves spontaneously, without considering quantum effect, to be stabilized at  an infinitely small conformal size $\pb \to 0 $ for $ \alpha_S^2 > \alpha_E g_s^2 $ ($ \alpha_S^4  >  \lambda_s g_s^2/6 $) or to be stabilized at an infinitely large conformal size $\pb \to \pm \infty$ for $ \alpha_S^2  < \alpha_E g_s^2 $ ($ \alpha_S^4  < \lambda_s g_s^2/6 $). 

Let us now discuss the quantum effect on the background gravifield spacetime. For our present purpose, we are going to demonstrate only a one-loop quantum effect. It can be shown that the leading terms of the effective Lagrangian for the background scalinon field get the following form: 
 \be
  {\cal L}_S & \simeq & \mu_U^2\, (\, \pb(x) + \varphi(x)\, )^2 - \lambda_U\, (\, \pb(x) +\varphi(x)\, )^4  \nonumber \\ 
  & + &  2\lambda_s (\, \pb(x) + \varphi(x)\, )^2 \pb^2(x)
  +  \bar{\mu}_U^2\, \pb^2(x) - \bar{\lambda}_U\, \pb^4(x) \,  , 
  \ee
with 
\be
& & \mu_U^2 (M_U/\mu) \simeq \frac{2}{(4\pi)^2} y_S^2 \left(M_U^2-\mu^2 \right) \,  , \nonumber \\
& & \lambda_U(M_U/\mu)  \simeq  \lambda_s + \frac{4}{(4\pi)^2} \lambda_S^2 \, \ln \frac{M_U^2}{\mu^2}  \, , \nonumber \\
& & \bar{\mu}_U^2 (M_U/\mu) \simeq \frac{2}{(4\pi)^2} \bar{y}_S^2 \left(M_U^2-\mu^2 \right)\, , \nonumber \\
& & \bar{\lambda}_U(M_U/\mu)  \simeq   \frac{\alpha_S^2}{g_s^2\alpha_E}  \lambda_s  + \frac{4}{(4\pi)^2} \bar{\lambda}_S^2 \, \ln \frac{M_U^2}{\mu^2}  \, , 
\ee
with
\be
& &  y_S^2 \equiv y^2_s- 6\lambda_s- c_w g_w^2 -c_s g_s^2  \, , \nonumber \\
 & & \lambda_S^2  \equiv  -12 \lambda_s^2  + \left(  2 y_s^2 - \lambda_s \right) y_s^2+ c'_w g_w^4 + c'_s g_s^4  \, , \nonumber \\
 & & \bar{y}_S^2 \equiv \lambda_s +  \bar{c}_s \alpha_S^2  \, , \qquad \bar{\lambda}_S^2 \equiv -\lambda_s^2 +  \bar{c}'_s \alpha_S^4  \, ,
\ee
where $M_U$ defines a basic energy scale of QFT in the ultraviolet (UV) region and $\mu$ is the sliding energy scale reflecting the infrared (IR) property of QFT. The effective parameters $\mu_U^2$ and $\lambda_U$ receive contributions from loops of the singlet fermion field and scalinon field characterized by the terms proportional to the parameters $y_s$ and $\lambda_s$, respectively.  The contributions from the scaling and spin gauge fields as well as the scalinon field are given by the terms proportional to the parameters $g_w$ and $g_s$ as well as $\alpha_S$. The coefficients $c_w$, $c'_w$, $c_s$, $c'_s$, $\bar{c}_s$, and $\bar{c}'_s$ represent the magnitudes of the contributions, which will be discussed in detail elsewhere, as their values do not affect our present general considerations. 

It is seen that the purely quantum-induced effective mass parameters $\mu^2_U$ and $\bar{\mu}^2_U$ arise from the loop quadratic contributions characterized by the UV basic energy scale $M_U$ and  the sliding energy scale $\mu$ as well as the coupling constants at a given energy scale $\mu$. Note that it is such a loop quadratic contribution that  causes the breaking of the global scaling symmetry. It has recently been shown that the quantum loop quadratic contributions can play an important role for understanding the electroweak symmetry breaking and hierarchy problem within the SM of particle physics\cite{QEWSB}. Similarly, for a quantum-induced positive mass parameter $\mu_U^2, \bar{\mu}^2_U  > 0$ with $y_S^2, \bar{y}_S^2 > 0$ and $\mu < M_U$,  it causes an unstable potential and generates the inflation of the early Universe; namely, the Universe begins from an infinitely small conformal size $\pb \sim 0$ and approaches to an infinitely large conformal size $\pb\to \infty$.  Once the effective coupling parameters $\lambda_U$ and $\bar{\lambda}_U$ including the quantum loop contributions gain positive values $\lambda_U, \bar{\lambda}_U > 0$, the effective potential for the background scalinon field is broken down spontaneously to an evolving minimal vacuum, which leads the inflationary Universe to end at a global minimal. Since the couplings $y_s$, $\lambda_s$, $g_s$, $g_w$, and $\alpha_S$ are all free parameters, one can always yield a solution to satisfy the requirements. In general,  the background scalinon field will get an evolving vacuum expectation value (eVEV) 
\be
  V_S^2(M_U/\mu) & = & \langle (\pb + \varphi)^2 \rangle   
  =  \frac{\mu_U^2 + \lambda_s \bar{\mu}_U^2/\bar{\lambda}_U  }{ 2( \lambda_U- \lambda_s^2/\bar{\lambda}_U  ) } \nonumber \\
  \bar{V}_S^2(M_U/\mu) & = & \langle \pb^2 \rangle   
  =  \frac{\bar{\mu}_U^2}{2\bar{\lambda}_U} +  \frac{\lambda_s}{\bar{\lambda}_U}\,  V_S^2(M_U/\mu)  \, , 
  \ee
which makes the inflationary Universe evolve via the minimalizing vacuum state when the sliding energy scale runs down. 

The scalinon field not only acts as a quantum field but also characterizes the conformal size of the Universe. At the beginning of the Universe with an extreme small conformal size, all quantum fields must have a high energy-momentum around the UV basic energy scale $M_U$. On the other hand, the smallness of the scalinon field characterized by the cosmological mass scale $m_{\kappa}$ indicates that a small quantum fluctuation can cause a significant change to the magnitude of the scalinon field. Consider the quantum fluctuation to be at the order of cosmological mass scale $m_{\kappa}$; namely, the sliding energy scale is taken to be $\mu_{\kappa} \sim M_U - m_{\kappa}$, and the quadratic contribution is given by $M_U^2 - (M_U-m_{\kappa})^2 =  2M_Um_{\kappa} + m_{\kappa}^2 \simeq 2M_Um_{\kappa} $. Thus, the eVEV will be stabilized around
\be
V_S & \equiv&  V_S(M_U/\mu_{\kappa} ) \simeq \frac{1}{(4\pi)} \sqrt{ \frac{y_S^2 + \bar{y}_S^2 \lambda_s/\bar{\lambda}_U }{\lambda_U - \lambda_s^2/\bar{\lambda}_U  }   }\,  \sqrt{2M_Um_{\kappa} } \, , \nonumber \\
\bar{V}_S & \equiv &  \bar{V}_S(M_U/\mu_{\kappa} ) \simeq \frac{1}{(4\pi)}   \left[ \frac{\bar{y}_S^2}{ \bar{\lambda}_U  }    +  (\frac{\lambda_s}{\bar{\lambda}_U})\, \frac{y_S^2 + \bar{y}_S^2 \lambda_s/\bar{\lambda}_U }{\lambda_U - \lambda_s^2/\bar{\lambda}_U  }  \right]^{1/2} \, \sqrt{2M_Um_{\kappa} } \, ,
\ee
which determines the epoch when the inflationary Universe ends. The corresponding scale factor of the Universe is given by 
\be
\ab_e = \langle \ab(x) \rangle = \frac{\bar{V}_S}{M_S} =  \frac{\mk }{\alpha_S M_S} \frac{1}{1 - \kh\, c \eta_e } = \ab_0 e^{ \Mk c \tau_e}\, .
\ee
 
With the above considerations, we come to the conclusion that it is the quantum effect with the loop quadratic contribution that causes the breaking of global scaling symmetry and generates the inflation of the Universe; its conformal size is characterized by the eVEV. The inflationary Universe eventually ends when the eVEV of the background scalar potential reaches a stabilized minimal value. Such an inflation of the Universe may be referred as the {\it quantum scalinon inflation}. We shall discuss its phenomena in detail elsewhere.

\section{Spacetime Gauge Field and Quantum Dynamics with Goldstone-like Gravifield and Gravimetric Field }

It has been shown that, when taking the gravifield basis $\{\chi^a=\chi_{\mu}^{\;a}dx^{\mu}\} $ as a one-form with the gauge potential ${\mathsf G}_{\mu}(x) \equiv \chi_{\mu}^{\;\; a}(x)\frac{1}{2} \gamma_a$ in the flat Minkowski spacetime, we have the corresponding gauge covariant field strength ${\cal G}_{\mu\nu}^a(x) =  \nabla_{\mu}\chi_{\nu}^{\;\; a} (x)  - \nabla_{\nu}\chi_{\mu}^{\;\; a}(x) $ which describes the gravitational force in the inertial reference frame of coordinates. Therefore, the locally flat gravifield spacetime characterized by the gravifield basis $\{\chi^a\}$ is regarded as a dynamical spacetime. The quantization of gravitational fields and matter fields leads to the quantum dynamics of the gravifield spacetime. To correlate the quantum dynamics of the gravifield spacetime with the quantum dynamics of coordinate spacetime, we are going to settle a special gauge condition by utilizing the gravifield in such a way that the spin gauge symmetry is transmuted into a hidden gauge symmetry in the bosonic gravitational interactions. Thus, it enables us to compare with the Einstein theory of general relativity.

\subsection{Spacetime gauge field with Goldstone-like gravifield $\&$ gravimetric field }

Let us express the quantum fields in the Einstein-type basis, which is realized by making a scaling gauge transformation 
\[ 
\chi_{\mu}^{\; a} (x) \to \achi(x) \chi_{\mu}^{\; a}(x)\, , \quad  \phi(x) \to  \phi(x)/\achi(x)\, , \quad g_w W_{\mu}(x) \to g_w W_{\mu}(x)  -  \partial_{\mu} \ln \achi(x) \, . \]
 As a consequence, the quantum fields get the following forms in the Einstein-type basis:
\be
& & \chi_{\mu}^{\;\; a} (x) = \achi(x) [\, \eta_{\mu}^{\; a} + h_{\mu}^{\;\; a}(x)/M_W\, ] = ( \ab(x) + \varphi(x)/M_S)  [\, \eta_{\mu}^{\; a} + h_{\mu}^{\;\; a}(x)/M_W\, ]  \, , \nonumber \\
& &  \hat{\chi}_a^{\;\; \mu}(x)  \equiv \achi^{-1}(x) [\, \eta_a^{\; \mu} - \hat{h}_a^{\; \mu}(x)/M_W \, ] = ( \ab(x) + \varphi(x)/M_S)^{-1}  [\, \eta_{\mu}^{\; a} + h_{\mu}^{\;\; a}(x)/M_W\, ]^{-1}  \, , \nonumber \\
& & \chi (x) = \det \chi_{\mu}^{\;\; a}(x) = \achi^4(x) = ( \ab(x) + \varphi(x)/M_S)^4 \, , \quad   \phi(x) =   M_S \, , \nonumber \\
& &  \Om_{\mu}^{ab}(x) =  \tilde{\chi}_{\mu\rho}^{[ab]} \ob^{\rho}(x) +  \Omega_{\mu}^{ab}(x) \, , \quad g_s \ob_{\mu}(x) = \partial_{\mu} \ln \ab(x) \, , \nonumber \\
& & W_{\mu}(x) = \bar{w}_{\mu}(x)  + w_{\mu}(x)\, , \quad  \wb_{\mu}(x) = 0\, ,
\ee
where we have omitted the label $``E''$ in all quantities for convenience.

It is meaningful to construct an alternative gauge field through the gravifield and spin gauge field as follows:
\be
\Gm_{\mu\nu}^{\sigma}(x)  =  \hat{\chi}_a^{\; \sigma} \td_{\mu} \chi_{\nu}^{\; a} = \hat{\chi}_a^{\; \sigma} (\partial_{\mu} \chi_{\nu}^{\; a} + g_s \Omega_{\mu b}^{a} \chi_{\nu}^{\; b}  )
\ee
where the spin gauge symmetry becomes a hidden symmetry and the gravifield appears as a Goldstone-like field that transmutes the local spin gauge symmetry into the global Lorentz symmetry. The gauge field $\Gm_{\mu\nu}^{\sigma}(x) $ is a Lorentz tensor field defined in the flat Minkowski spacetime; we may refer it as a {\it spacetime gauge field} so as to distinguish from the affine connection defined in the curved spacetime.  

Analogous to the background field approach, let us decompose the spin gauge field into two parts, so that they have the following properties under the spin gauge transformation $S(x)= e^{i\alpha_{ab}(x) \Sigma^{ab} /2} \in$ SP(1,3): 
\be
& & \Omega_{\mu}^{ab}(x)  =   \omega_{\mu}^{ab}(x) +  \ot_{\mu}^{ab}(x) \, \nonumber \\
 & & \omega_{\mu}(x) \to \omega'_{\mu}(x) = S \omega_{\mu}(x) S^{-1} + S \partial_{\mu} S^{-1} \, , \quad \omega_{\mu}(x) = \omega_{\mu}^{ab}(x) \frac{1}{2} \Sigma_{ab} \, , \nonumber \\
& & \ot_{\mu}(x) \to \ot'_{\mu}(x) = S \ot_{\mu}(x) S^{-1}\, , \quad \ot_{\mu}(x) = g_s \ot_{\mu}^{ab}(x) \frac{1}{2} \Sigma_{ab}\, .
 \ee
Correspondingly, the {\it spacetime gauge field } $\Gm_{\mu\nu}^{\sigma}(x)$ consists of two gauge-invariant parts under the spin gauge transformation
\be
& & \Gm_{\mu\nu}^{\sigma}(x) \equiv  \Gamma_{\mu\nu}^{\sigma}(x) + \Gmt_{\mu\nu}^{\sigma}(x) \, , \nonumber \\
& &  \Gamma_{\mu\nu}^{\sigma} (x)  \equiv \hat{\chi}_a^{\; \sigma} (\partial_{\mu} \chi_{\nu}^{\; a} + g_s \omega_{\mu b}^{a} \chi_{\nu}^{\; b}  ) \, , \quad \Gmt_{\mu\nu}^{\sigma}(x) \equiv g_s \hat{\chi}_a^{\; \sigma} \ot_{\mu b}^{a} \chi_{\nu}^{\; b}  \, ,
\ee
where the gauge fields $\omega_{\mu}^{ab}(x)$ and $\ot_{\mu}^{ab}(x) $ should not be all independent; half of their degrees of freedom must be reduced so that the independent degrees of the freedom are compatible with the initial ones in $\Omega_{\mu}^{ab}(x) $. From the above definition, it can be checked that $\Gmt_{\mu\nu}^{\sigma}(x)$ reflects an antisymmetric part of the {\it spacetime gauge field }  $\Gm_{\mu\nu}^{\sigma}(x)$ with the following relation:
\be
\Gmt_{\mu \nu}^{\sigma'}(x)\, \chi_{\sigma' \sigma} = - \Gmt_{\mu \sigma}^{ \nu'}(x)\, \chi_{\nu' \nu} \, .
\ee
Thus let us impose a reasonable condition that  $ \Gamma_{\mu\nu}^{\sigma}(x) $ is symmetric, i.e., $  \Gamma_{\mu\nu}^{\sigma}(x) =  \Gamma_{\nu\mu}^{\sigma} (x)$. As a result, the spin gauge field$\omega_{\mu}^{ab}(x) $ is characterized purely by the {\it Goldstone-like gravifield} $ \chi_{\mu}^{\;a} $. Its explicit form is found to be
\be
  g_s \omega_{\mu}^{ab}(x) & = &  \hat{\chi}^{a\nu} \chi_{\mu\nu}^b - \hat{\chi}^{b\nu} \chi_{\mu\nu}^a -  \hat{\chi}^{a\rho}  \hat{\chi}^{b\sigma}  \chi_{\rho\sigma}^c \chi_{\mu c }\, , \quad   \chi_{\mu\nu}^a = \partial_{\mu} \chi_{\nu}^{\; a} -  \partial_{\nu} \chi_{\mu}^{\;a} \, .
\ee
Here $\omega_{\mu}^{ab}(x) $ appears as a pure gauge like field. With such a form of $\omega_{\mu}^{ab}(x) $, it is not difficult to check that $  \Gamma_{\mu\nu}^{\sigma}(x) $ gets the following explicit form:
\be \label{GS}
& &  \Gamma_{\mu\nu}^{\sigma} = \Gamma_{\nu\mu}^{\sigma} = \frac{1}{2} \hat{\chi}^{\sigma\lambda} [\, \partial_{\mu} \chi_{\nu \lambda} + \partial_{\nu} \chi_{\mu \lambda} - \partial_{\lambda}\chi_{\mu\nu} \, ]   \, , \nonumber \\
& &  \chi_{\mu\nu}(x) = \chi_{\mu}^{\; a}(x)\chi_{\nu}^{\; b}(x)\eta_{ab}\, , \qquad \hat{\chi}^{\mu\nu}(x) = \hat{\chi}_a^{\;\mu}(x) \hat{\chi}_b^{\; \nu}(x)\eta^{ab}  \, .
\ee
$  \Gamma_{\mu\nu}^{\sigma}(x) $ is characterized by the tensor field $\chi_{\mu\nu}(x)$ [or its inverse $\hat{\chi}^{\mu\nu}(x)$] and appears as a pure gauge like {\it spacetime gauge field}. $\chi_{\mu\nu}(x)$ behaves as a {\it Goldstone-like gravimetric field} of spacetime and concerns only ten degrees of freedom. This  reflects the fact that the spin gauge symmetry is fixed by the {\it Goldstone-like gravifield} $\chi_{\mu}^{\; a}$ to be as a hidden symmetry.  

From the decomposition of the spin gauge field,  the corresponding field strength of the spin gauge field is also decomposed into two parts:
\be
& & {\mathsf R}_{\mu\nu}^{ab}  = R_{\mu\nu}^{ab}  +  \Rmt_{\mu\nu}^{ab}  \nonumber \\
& & R_{\mu\nu}^{ab} = \partial_{\mu}\omega_{\nu}^{ab} -   \partial_{\nu}\omega_{\mu}^{ab} + g_s( \omega_{\mu c}^{a} \omega_{\nu}^{cb}  - \omega_{\nu c}^{a} \omega_{\mu}^{cb} ) \, , \nonumber \\
& & \Rmt_{\mu\nu}^{ab} = \td_{\mu}\ot_{\nu}^{ab} -   \td_{\nu}\ot_{\mu}^{ab}  + g_s ( \ot_{\mu c}^{a} \ot_{\nu}^{cb}  - \ot_{\nu c}^{a} \ot_{\mu}^{cb} ) \, , \nonumber \\
& & \td_{\mu}\ot_{\nu}^{ab} = \partial_{\mu} \ot_{\nu}^{ab} + g_s \omega_{\mu c}^{a} \ot_{\nu}^{cb} + g_s  \omega_{\mu c}^{b} \ot_{\nu}^{ac}\, .
\ee
By using the relations and identities between the gauge covariant spin field and the spacetime gauge field 
\be
& & g_s\ot_{\mu}^{ab}(x)  =  \chi_{\sigma}^{\; a} \Gmt_{\mu\nu}^{\sigma}(x) \chih^{\nu b} =  \frac{1}{2} \Gmt_{\mu\nu}^{\sigma}(x) 
\tilde{\chi}_{\sigma\rho}^{[ab]} \eta^{\rho\nu}   \, , \nonumber \\
& &  \partial_{\mu} \chi_{\nu}^{\; a} + g_s \omega_{\mu b}^{a} \chi_{\nu}^{\; b}  - \Gamma_{\mu\nu}^{\sigma} (x)  \chi_{\sigma}^{\; a} =0\, , \nonumber \\
& & \partial_{\mu} \chih^{ a \rho}+ g_s \omega_{\mu b}^a \chih^{b\rho} + \Gamma_{\mu\nu}^{\rho} (x)  \chih^{a \nu}=0\, ,
\ee
it is not difficult to demonstrate that the field strength $R_{\mu\nu}^{ab}$ for the spin gauge potential $\omega_{\mu}^{ab}$  is related to the field strength $R_{\mu\nu}^{\rho\sigma} $ which is characterized solely by the {\it spacetime gauge potential} $\Gamma_{\mu\rho}^{\sigma}$. Explicitly, we have
\be
& & g_s R_{\mu\nu}^{ab} = \frac{1}{2} R_{\mu\nu\rho}^{\sigma}\, \tilde{\chi}_{ \sigma\rho'}^{[ab]} \eta^{\rho'\rho}   \, , \nonumber \\
& & R_{\mu\nu\rho}^{\;\sigma}  = \partial_{\mu} \Gamma_{\nu\rho}^{\sigma} - \partial_{\nu} \Gamma_{\mu\rho}^{\sigma} -
\Gamma_{\mu\rho}^{\lambda} \Gamma_{\nu\lambda}^{\sigma} + \Gamma_{\nu\rho}^{\lambda} \Gamma_{\mu\lambda}^{\sigma}\, ,\nonumber \\
& & \tilde{\chi}_{\sigma\rho}^{[ab]}(x) = \chi_{\sigma}^{\;\; a} \hat{\chi}^b_{\;\;\rho}- \chi_{\sigma}^{\;\; b}\hat{\chi}^a_{\;\;\rho}  \, .
\ee
Similarly, we yield the following relation between  $\Rmt_{\mu\nu}^{ab}$  and $\Rmt_{\mu\nu}^{\rho\sigma} $
\be
& & g_s \Rmt_{\mu\nu}^{ab}   =  \frac{1}{2} \Rmt_{\mu\nu\rho}^{\sigma}\, \tilde{\chi}_{ \sigma\rho'}^{[ab]} \eta^{\rho'\rho}\, , \nonumber \\
& & \Rmt_{\mu\nu\rho}^{\;\sigma}  = \nabla_{\mu} \Gmt_{\nu\rho}^{\sigma} - \nabla_{\nu} \Gmt_{\mu\rho}^{\sigma} -
\Gmt_{\mu\rho}^{\lambda} \Gmt_{\nu\lambda}^{\sigma} + \Gmt_{\nu\rho}^{\lambda} \Gmt_{\mu\lambda}^{\sigma} \, , \nonumber \\
& & \nabla_{\mu} \Gmt_{\nu\rho}^{\sigma} = \partial_{\mu} \Gmt_{\nu\rho}^{\sigma} - \Gamma_{\mu\rho}^{\lambda} \Gmt_{\nu\lambda}^{\sigma} + \Gamma_{\mu\lambda}^{\sigma}  \Gmt_{\nu\rho}^{\lambda} \, .
\ee

In terms of the spacetime gauge field, the field strength of the gravifield is found to be 
\be
G_{\mu\nu}^a & = & \btd_{\mu} \chi_{\nu}^{\;\; a}  - \btd_{\nu} \chi_{\mu}^{\;\; a}   =   \partial_{\mu} \chi_{\nu}^{\;\; a}  - \partial_{\nu} \chi_{\mu}^{\;\; a}  +    g_s (\Omega_{\mu b}^a \chi_{\nu}^{\;\; b}  -  \Omega_{\nu b}^a \chi_{\mu}^{\;\; b} ) \nonumber \\ 
& = & (\Gm_{\mu\nu}^{\sigma} - \Gm_{\nu\mu}^{\sigma} ) \chi_{\sigma}^{\; a} =
(\Gmt_{\mu\nu}^{ \sigma} - \Gmt_{\nu\mu}^{ \sigma} ) \chi_{\sigma}^{\; a} \equiv \Gmt_{[\mu\nu]}^{ \sigma}\, \chi_{\sigma}^{\; a}\, ,
\ee
where $ \Gmt_{[\mu\nu]}^{ \sigma}$ defines a totally antisymmetric spacetime gauge field. 

In general, the field strength for the spin gauge field can be represented by the field strength of the spacetime gauge field as follows:
\be
g_s{\mathsf R}_{\mu\nu}^{ab}  & = & \frac{1}{2} \Rm_{\mu\nu\rho}^{\sigma}\, \tilde{\chi}_{ \sigma\rho'}^{[ab]} \eta^{\rho'\rho}
=  \frac{1}{2} (R_{\mu\nu\rho}^{\sigma} +  \Rmt_{\mu\nu\rho}^{\sigma} )\,  \tilde{\chi}_{ \sigma\rho'}^{[ab]} \eta^{\rho'\rho} \, ,
\ee 
which shows a {\it hidden gauge formalism}.
 
 \subsection{Quantum dynamics of spacetime in the hidden gauge formalism}
 
 In the {\it hidden gauge formalism} with the spacetime gauge field and corresponding field strength,  we are able to reformulate the action for the gravitational interactions with spin and scaling gauge symmetry breaking into the following general form in the Einstein-type basis:
\be
\label{action5}
S_{\chi}  & = & \int d^{4}x\, \chi\; \{\, \frac{1}{2}  [\,  \hat{\chi}^{\mu\nu} (\bar{\Psi} \chi_{\mu} i {\mathsf D }_{\nu}   \Psi  + \bar{\psi} \chi_{\mu}  i \td_{\nu}   \psi )  + H.c. \, ]  - y_s M_S \bar{\psi} \psi +  {\cal L}'(x)  \nonumber \\
& - & \frac{1}{4}  \hat{\chi}^{\mu\mu'} \hat{\chi}^{\nu\nu'} [\, {\cal F}^I_{\mu\nu} {\cal F}^{I}_{\mu'\nu'}  + W_{\mu\nu} W_{\mu'\nu'} + g_s^{-2} {\mathsf R}_{\mu\nu}^{\; \rho\sigma}  {\mathsf R}_{\mu'\nu' \rho \sigma}    \, ]    \nonumber \\
& + &   \alpha_W M_S^2 \hat{\chi}^{\mu\mu'}  [  \frac{1}{4}  \hat{\chi}^{\nu\nu'} \chi_{\sigma\sigma'}  \Gmt_{[\mu\nu]}^{ \sigma}  \Gmt_{[\mu'\nu']}^{ \sigma'} + \frac{1}{2}( 3+ \frac{1}{\alpha_W}) g_w^2 w_{\mu}w_{\mu'} \, ] \nonumber \\
& - & \frac{1}{4}  \hat{\chi}^{\mu\mu'} \hat{\chi}^{\nu\nu'}(\, \chi_{\sigma\sigma'} \hat{\chi}^{\rho\rho'}   -  \eta_{\sigma}^{\rho} \eta_{\sigma'}^{\rho'} \, )  \ob_{\rho}\ob_{\rho'} \Gmt_{[\mu\nu]}^{ \sigma}  \Gmt_{[\mu'\nu']}^{ \sigma'} \nonumber \\ 
& + &   \alpha_E M_S^2 (\, 1 - \hat{\chi}^{\rho\sigma} \ob_{\rho}\ob_{\sigma}/\alpha_E M_S^2 \, )   {\mathsf R}   -  \lambda_s M_S^4    \nonumber \\
& + & g_s^{-1} (\td_{\sigma'} \ob_{\sigma}   - g_s \ob_{\sigma'} \ob_{\sigma}  \, )  \chih^{\mu\sigma'} 
(\, \chih^{\nu\rho} {\mathsf R}_{\mu\nu\rho}^{\sigma} + \chih^{\nu\sigma} {\mathsf R}_{\mu\nu} \, )    \nonumber \\
& + & g_s^{-1} \frac{1}{2} ( \chi_{\sigma\sigma'}\chih^{\rho\rho'} - \eta^{\rho}_{\sigma}\eta^{\rho'}_{\sigma'} )   \hat{\chi}^{\mu\mu'} \hat{\chi}^{\nu\nu'}  {\mathsf R}_{\mu\nu\rho}^{\sigma} \Gmt_{[\mu'\nu']}^{ \sigma'}  \ob_{\rho'} -  \hat{\chi}^{\mu\mu'} \hat{\chi}^{\nu\nu'}  \Gmt_{[\mu\nu]}^{ \sigma}  \ob_{\mu'} \td_{\nu'}\ob_{\sigma}  \,    \nonumber \\
& + & 6 \alpha_E g_s M_S^2  (\,  \hat{\chi}^{\mu\nu}  \td_{\mu}\ob_{\nu}  + g_s \hat{\chi}^{\mu\nu} \ob_{\mu}\ob_{\nu}  \, )  -3 g_s^2  (\, \hat{\chi}^{\mu\nu} \ob_{\mu}\ob_{\nu} )^2   \nonumber \\
& - & 2 \hat{\chi}^{\mu\mu'} \hat{\chi}^{\nu\nu'}  (\td_{\mu} \ob_{\nu})( \td_{\mu'}  \ob_{\nu'}) - 
(\chih^{\mu\nu} \td_{\mu} \ob_{\nu} )^2  \nonumber \\
 & + & 2 g_s \hat{\chi}^{\mu\nu}  \partial_{\mu}(\hat{\chi}^{\rho\sigma}\ob_{\rho} \ob_{\sigma}) \ob_{\nu} 
 - 4g_s \hat{\chi}^{\mu\nu} (\td_{\mu}\ob_{\nu} )\, \hat{\chi}^{\rho\sigma}\ob_{\rho} \ob_{\sigma}   \, \}
\ee
with 
\be
& & \td_{\mu} \ob_{\nu} \equiv \partial_{\mu}\ob_{\nu} - \Gm_{\mu\nu}^{\sigma} \ob_{\sigma}\, ,  
\ee
where the {\it spacetime gauge field } $\Gm_{\mu\nu}^{\sigma}(x)$ and its field strength $  \Rm_{\mu\nu\rho}^{\sigma} $ consist of two parts: 
\be
& & \Gm_{\mu\nu}^{\sigma}(x) \equiv  \Gamma_{\mu\nu}^{\sigma}(x) + \Gmt_{\mu\nu}^{\sigma}(x)  \, , \quad \Rm_{\mu\nu\rho}^{\sigma} \equiv  R_{\mu\nu\rho}^{\sigma} +  \Rmt_{\mu\nu\rho}^{\sigma}   \, ,
\ee
with the rank-2 tensor field strength $\Rm_{\mu\nu}$ and the scalar tensor field strength ${\mathsf R}$ defined as follows:
\be
& &  \Rm_{\mu\nu} \equiv - \Rm_{\rho\mu\nu}^{\;\sigma} \eta_{\sigma}^{\rho}  = \Rm_{\mu\sigma\nu}^{\;\sigma}  =  
R_{\mu\nu}  +   \Rmt_{\mu\nu} \, , \quad {\mathsf R} = \hat{\chi}^{\mu\nu}  {\mathsf R}_{\mu\nu}  = R + \Rmt  \, , \nonumber \\
& & R_{\mu\nu}  = \partial_{\mu} \Gamma_{\sigma\nu}^{\sigma} - \partial_{\sigma} \Gamma_{\mu\nu}^{\sigma} -
\Gamma_{\mu\nu}^{\lambda} \Gamma_{\sigma\lambda}^{\sigma} + \Gamma_{\mu\lambda}^{\sigma}\Gamma_{\nu\sigma}^{\lambda} = R_{\nu\mu}\, , \quad \Gamma_{\sigma\nu}^{\sigma}  = \partial_{\nu} \ln \chi \, , \nonumber \\
 & & \Rmt_{\mu\nu}  =  \nabla_{\mu} \Gmt_{\sigma\nu}^{\sigma} - \nabla_{\sigma} \Gmt_{\mu\nu}^{\sigma}  -
\Gmt_{\mu\nu}^{\rho} \Gmt_{\sigma\rho}^{\sigma} + \Gmt_{\sigma\nu}^{\rho} \Gmt_{\mu\rho}^{\sigma} \, , \; \quad \Gmt_{\sigma\nu}^{\sigma}  \equiv \Gmt_{\nu} = \frac{1}{2} \chi_{\nu a} \chih_b^{\;\sigma} g_s \ot_{\sigma }^{ab}  \, , 
\ee
where $R_{\mu\nu}$ is a symmetric rank-2 tensor field strength characterized purely by the symmetric Goldstone-like gravimetric field $\chi_{\mu\nu}$. The above action possesses a global Lorentz and scaling invariance except for the fermionic interactions. It is manifest that after the gauge symmetry breaking  $\langle \chi_{\mu\nu}\rangle =  \eta_{\mu\nu}$, the totally antisymmetric spacetime gauge field $\Gmt_{[\mu\nu]}^{\sigma'} $ and the scaling gauge field $w_{\mu}(x)$ as well as the singlet fermion $\psi(x)$ become massive in the Einstein-type basis. 

The gravitational  interactions can be described by the antisymmetric {\it spacetime gauge field }  $\Gm_{\mu\nu}^{\sigma}(x)$ with $\Gmt_{\mu \nu}^{\sigma'}(x)\, \chi_{\sigma' \sigma} = - \Gmt_{\mu \sigma}^{ \nu'}(x)\, \chi_{\nu' \nu}$ and the symmetric Goldstone-like {\it gravimetric field} $ \chi_{\mu\nu}$ except for the fermionic interactions that always couple with the gravifield (where we have used the notation $\chi_{\mu} \equiv \chi_{\mu}^{\;\, a}\gamma_a$). Though the gravitational gauge symmetries become hidden ones in the bosonic gravitational interactions, the independent physical quantum degrees of freedom in terms of the  quantized fields should be the same as the gauge theory of quantum gravity analyzed in the previous section. 

Therefore, in terms of the hidden gauge formalism, we come to the conclusion that all the bosonic gravitational interactions can be characterized by the symmetric gravimetric field $\chi_{\mu\nu}$ with ten independent field components and the antisymmetric spacetime gauge field that has the same independent field components as the spin gauge field. It is manifest that the six independent field components of the gauge-type gravifield $\chi_{\mu}^{\;\, a}$ are reduced in the hidden gauge formalism for the bosonic gravitational interactions, which enables us to compare the bosonic part of the above action with the Einstein theory of general relativity. It is noticed from the above action that there is an interaction term given by the scalar tensor field strength ${\mathsf R} = \hat{\chi}^{\mu\nu}  {\mathsf R}_{\mu\nu}  = R + \Rmt $. Here the first part of the scalar tensor field strength $R$ does characterize the Einstein theory of general relativity as it is governed purely by the gravimetric field $\chi_{\mu\nu}$. Nevertheless, the fermionic interactions always involve the gravifield $\chi_{\mu}^{\;\, a}$ or $\hat{\chi}_a^{\;\,\mu}$, so that the basic gravitational field should be the gravifield $\chi_{\mu}^{\;\, a}$ rather than the gravimetric field $\chi_{\mu\nu}$ when considering the basic fermionic interactions. It indicates that the Einstein theory of general relativity must result as an effective low-energy theory by integrating out the fermion fields and also integrating the high-energy contributions into the renormalized coupling constants and quantum fields.

\section{Gravity Equation Beyond and Extension to Einstein's Equation and Hidden General Coordinate Invariance}
 
To be more explicit in comparison with the Einstein theory of general relativity, we shall reformulate, in terms of the hidden gauge formalism, the equation of motion for the gravifield given in Eqs.(\ref{GEM1}) and (\ref{GEM2}) in the Einstein-type basis.  For that, let us first rewrite the gauge-invariant action of Eq.(\ref{action3}) into the following expression in the hidden gauge formalism:   
\be
\label{action6}
S_{\chi}  & = & \int d^{4}x\, \chi\,  {\cal L}  \, , \nonumber \\
{\cal L} & = & \frac{1}{2}  [\,  \hat{\chi}^{\mu\nu} (\bar{\Psi} \chi_{\mu} i {\mathsf D }_{\nu}   \Psi  + \bar{\psi} \chi_{\mu}  i \td_{\nu}   \psi )  + H.c. \, ]  - y_s M_S \bar{\psi} \psi +  {\cal L}'(x)  \nonumber \\
& - & \frac{1}{4}  \hat{\chi}^{\mu\mu'} \hat{\chi}^{\nu\nu'} [\, {\cal F}^I_{\mu\nu} {\cal F}^{I}_{\mu'\nu'}  
+ {\cal W}_{\mu\nu} {\cal W}_{\mu'\nu'} + g_s^{-2} {\cal R}_{\mu\nu}^{\; \rho\sigma}  {\cal R}_{\mu'\nu' \rho \sigma}   \, ]    \nonumber \\
& + &   \alpha_W M_S^2  \frac{1}{4}  \hat{\chi}^{\mu\mu'}  \hat{\chi}^{\nu\nu'} \chi_{\sigma\sigma'} (\, \Gmct_{[\mu\nu]}^{ \sigma}  + g_w W_{[\mu\nu]}^{\sigma} \, ) (\, \Gmct_{[\mu'\nu']}^{ \sigma'} +  g_w W_{[\mu'\nu']}^{\sigma'} \, ) \nonumber \\
& + &   \alpha_E M_S^2 \,  {\cal R}  + M_S^2 g_w^2 \hat{\chi}^{\mu\mu'}  W_{\mu}W_{\mu'} -   \lambda_s M_S^4    \, ,
\ee
with the spacetime gauge field 
\be
& & \Gmc_{\mu\nu}^{\sigma}(x)  =  \hat{\chi}_a^{\; \sigma} \td_{\mu} \chi_{\nu}^{\; a} = \hat{\chi}_a^{\; \sigma} (\partial_{\mu} \chi_{\nu}^{\; a} + g_s \Om_{\mu b}^{a} \chi_{\nu}^{\; b}  )  \, , \quad  \Om_{\mu b}^{a} = \omega_{\mu b}^{a} + \Omt_{\mu b}^{a}\, , \nonumber \\
& & \Gmc_{\rho\nu}^{\sigma}  =  \Gamma_{\rho\nu}^{\sigma}  + \Gmct_{\rho\nu}^{\sigma} \, , \quad  \Gamma_{\mu\nu}^{\sigma} (x)  \equiv \hat{\chi}_a^{\; \sigma} (\partial_{\mu} \chi_{\nu}^{\; a} + g_s \omega_{\mu b}^{a} \chi_{\nu}^{\; b}  )\, , \quad \Gmct_{\mu\nu}^{\sigma}(x) \equiv g_s \hat{\chi}_a^{\; \sigma} \Omt_{\mu b}^{a} \chi_{\nu}^{\; b} \, , \nonumber \\
& & \Gmct_{[\mu\nu]}^{ \sigma} =  \Gmc_{\mu\nu}^{ \sigma} - \Gmc_{\nu\mu}^{ \sigma}  = \Gmct_{\mu\nu}^{ \sigma} - \Gmct_{\nu\mu}^{ \sigma}  \, , \quad W_{[\mu\nu]}^{\sigma} = W_{\mu}\eta_{\nu}^{\sigma}- W_{\nu}\eta_{\mu}^{\sigma}\, .
 \ee 
 and the field strength tensors
 \be
 & & {\cal R}_{\mu\nu}^{\; \rho\sigma}  = {\cal R}_{\mu\nu \rho'}^{\sigma} \hat{\chi}^{\rho\rho'}\, , \quad  {\cal R}_{\mu\nu \rho\sigma}  = {\cal R}_{\mu\nu \rho}^{\sigma'} \chi_{\sigma'\sigma}\, , \nonumber \\
& & \Rc =  \Rc_{\mu\nu} \hat{\chi}^{\mu\nu} = \Rc_{\nu}^{\;\,\mu} \eta^{\nu}_{\mu} \, , \quad  \Rc_{\mu\nu} \equiv - \Rc_{\rho\mu\nu}^{\;\sigma} \eta_{\sigma}^{\rho}  = \Rc_{\mu\sigma\nu}^{\;\sigma}  \, , \nonumber \\
& &  \Rc_{\mu}^{\;\,\nu} \equiv \Rc_{\rho\mu\sigma}^{\;\nu} \hat{\chi}^{\rho\sigma}   \, , \quad  \Rc_{\mu\nu} = \Rc_{\mu}^{\;\, \rho} \chi_{\rho\nu} \, ,
\ee
where the symmetric spacetime gauge field $\Gamma_{\mu\nu}^{\sigma}$ is governed purely by the gravimetric field $\chi_{\mu\nu}$ as shown in Eq.(\ref{GS}) and $\Gmct_{\mu\nu}^{\sigma}(x)$ is an antisymmetric part of the spacetime gauge field  with the antisymmetric relation $\Gmct_{\mu\nu}^{\sigma}(x) \chi_{\sigma\rho} = - \Gmct_{\mu\rho}^{\sigma}(x) \chi_{\sigma\nu}$. Note that here we have used different notations for the gauge fields so as to distinguish two cases: with and without the spin and scaling gauge symmetry breaking.

Once the spacetime gauge field  is decomposed into symmetric and antisymmetric parts $ \Gmc_{\mu\nu}^{\sigma}(x) \equiv  \Gamma_{\mu\nu}^{\sigma}(x) + \Gmct_{\mu\nu}^{\sigma}(x)$,  its field strength $  \Rc_{\mu\nu\rho}^{\sigma} $ can be written into two parts, respectively, 
\be
& &  \Rc_{\mu\nu\rho}^{\sigma} \equiv  R_{\mu\nu\rho}^{\sigma} +  \Rct_{\mu\nu\rho}^{\sigma}   \, ,
\ee
where $R_{\mu\nu\rho}^{\sigma} $ is determined solely by the symmetric spacetime gauge field $\Gamma_{\mu\nu}^{\sigma}$. The rank-2 tensor field strength $\Rc_{\mu\nu}$ ($\Rc_{\nu}^{\;\,\mu}$) and the scalar tensor field strength $\Rc$ are also decomposed into two parts
\be
& &  \Rc_{\mu\nu} \equiv  R_{\mu\nu}  +   \Rct_{\mu\nu} \, , \quad \Rc_{\mu}^{\;\,\nu} \equiv R_{\nu}^{\;\,\mu}  +  \Rct_{\nu}^{\;\,\mu} \, , \nonumber \\
& &  \Rc \equiv R + \Rct \, , \quad R=R_{\mu\nu}\hat{\chi}^{\mu\nu}=R_{\nu}^{\;\,\mu}\eta_{\mu}^{\nu}\, , 
\ee
where the first part $R_{\mu\nu} $ is the symmetric tensor field strength that is governed solely by the symmetric gravimetric field. The corresponding scalar tensor field strength $R$ in the action describes the Einstein theory of general relativity. 

The gauge-invariant {\it gravifield tensor} ${\cal G}^{\; \mu\rho}_{ \nu}$ and {\it gravifield tensor current}  $ {\cal G}_{\nu}^{\;\; \mu} $ defined in Eqs.(\ref{GEM1}) and (\ref{GEM2}) get the following expressions in the hidden gauge formalism: 
\be \label{GE1}
& & {\cal G}^{\; \mu\rho}_{ \nu}  \equiv \alpha_W M_S^2  \chi [\,  \hat{\chi}^{\mu \mu'} \hat{\chi}^{\rho\nu'} \Gmct_{[\mu'\nu']}^{\sigma} \chi_{\sigma\nu}  + g_w ( \hat{\chi}^{\mu \mu'} \eta^{\rho}_{\nu} - \hat{\chi}^{\rho\mu'} \eta^{\mu}_{\nu} ) W_{\mu'} \, ] = - {\cal G}^{\; \rho\mu}_{ \nu}  \, , \nonumber \\
& &   {\cal G}_{\nu}^{\;\; \mu} \equiv  \Gmc_{\rho\nu}^{\sigma}  {\cal G}^{\; \mu\rho}_{ \sigma}  + g_w W_{\rho}  {\cal G}^{\; \mu\rho}_{ \nu}   \, .
\ee

In terms of the hidden gauge formalism, the gravity equation can simply be expressed as
\be \label{GE2}
 \td_{\rho} {\cal G}^{\;\mu\rho}_{ \nu}  &  = &   {\cal T}^{\;\, \mu}_{\nu} \,  , 
 \ee
 with a definition for the gauge-invariant covariant derivative
 \be
 \td_{\rho} {\cal G}^{\;\mu\rho}_{ \nu}  & \equiv & \partial_{\rho}  {\cal G}^{\; \mu\rho}_{ \nu}  
 - \Gmc_{\rho\nu}^{\sigma}  {\cal G}^{\; \mu\rho}_{ \sigma}  - g_w W_{\rho}  {\cal G}^{\; \mu\rho}_{ \nu} \, .
\ee

The energy-momentum tensor is rewritten as 
\be \label{EMT}
{\cal T}^{\;\, \mu}_{\nu}& = & - \eta^{\mu}_{\; \nu}\chi {\cal L}  + \frac{1}{2}\chi  [\, 
i\hat{\chi}_a^{\;\,\mu} \bar{\Psi} \gamma^a {\mathcal D}_{\nu}\Psi + i \hat{\chi}_a^{\;\,\mu}\bar{\psi} \gamma^{a} \td_{\nu}\psi  + H.c.\,  ] \nonumber \\
& - & \chi \hat{\chi}^{\mu\mu'} \hat{\chi}^{\rho\sigma}  [\,  {\cal F}^I_{\mu'\rho} {\cal F}^{I}_{\nu \sigma} 
+ {\cal W}_{\mu'\rho} {\cal W}_{\nu \sigma} 
+ g_s^{-2} {\cal R}_{\mu'\rho}^{\; \alpha\beta}  {\cal R}_{\nu\sigma \alpha \beta}  \, ]\, \nonumber \\
& + & \chi \alpha_W M_S^2  \hat{\chi}^{\mu\mu'}  [\,  \hat{\chi}^{\rho\sigma} \chi_{\lambda\lambda'}  \Gmct_{[\mu'\rho]}^{ \lambda}  \Gmct_{[\nu\sigma]}^{ \lambda'}  + ( 2 + 1/\alpha_W)g_w^2 W_{\mu'}W_{\nu}  + g_w^2 \chi_{\mu'\nu} \hat{\chi}^{\rho\sigma} W_{\rho}W_{\sigma}   \, ]    \nonumber \\
& + &   2 \chi  \alpha_E M_S^2 \, \Rc_{\nu}^{\;\,\mu} 
\ee
with the Lagrangian density given in Eq.(\ref{action6}).

In general, the above gravity equation holds for 16 components, as the energy-momentum tensor is not symmetric.  Let us decompose the energy-momentum tensor into symmetric and antisymmetric parts through a redefinition 
\be
{\cal T}_{\mu\nu} =  {\cal T}^{\;\, \sigma}_{\mu}  \chi_{\sigma\nu} \equiv  {\bf G}_{\mu\nu} + {\bf T}_{\mu\nu} + {\cal T}_{[\mu\nu]}  \, ,
\ee
where ${\bf G}_{\mu\nu}$ and  ${\bf T}_{\mu\nu}$ are the symmetric parts and ${\cal T}_{[\mu\nu]} $ is the antisymmetric part.  They are explicitly given by
\be \label{EMTS}
{\bf G}_{\mu\nu}  & = &  2\alpha_E M_S^2 \chi \, [\,  R_{\mu\nu} -   \frac{1}{2} \chi_{\mu\nu} \, R \, ] \, , \nonumber \\
{\bf T}_{\mu\nu} & = & \chi_{\mu\nu} T^{S} + T_{\mu\nu}^{F} - \chi_{\mu\nu} T^F +  T_{\mu\nu}^{R} - \frac{1}{2}\chi_{\mu\nu} T^R
+ T_{\mu\nu}^{G} - \frac{1}{4} \chi_{\mu\nu} T^G \, ,
\ee
for the symmetric part with the definitions
\be
T^S & = & \chi y_s M_S \bar{\psi} \psi  + \chi \lambda_s M_S^4  \, , \nonumber \\
T_{\mu\nu}^{F} & = & \frac{1}{4}\chi  [\, i\bar{\Psi} ( \chi_{\mu}{\mathcal D}_{\nu} + \chi_{\nu}{\mathcal D}_{\mu} ) \Psi 
+ i\bar{\psi} (\chi_{\mu} \td_{\nu} + \chi_{\nu} \td_{\mu} ) \psi  + H.c.\,  ] \, , \nonumber \\
T_{\mu\nu}^{R} & = &  \chi  \alpha_E M_S^2 \, ( \Rct_{\mu\nu} + \Rct_{\nu\mu} )  + \chi M_S^2 g_w^2 W_{\mu}W_{\nu}  \, , \nonumber \\
T_{\mu\nu}^{G} & = &  -  \chi \hat{\chi}^{\rho\sigma}  [\,  {\cal F}^I_{\mu\rho} {\cal F}^{I}_{\nu \sigma} 
+ {\cal W}_{\mu\rho} {\cal W}_{\nu \sigma} + g_s^{-2} {\cal R}_{\mu\rho }^{\; \alpha \beta}  {\cal R}_{\nu\sigma \alpha \beta} \, ]\, \nonumber \\
& + & \chi \alpha_W M_S^2    [\,  \hat{\chi}^{\rho\sigma}  \Gmct_{[\mu\rho]}^{ \alpha}  \Gmct_{[\nu\sigma\alpha]}  + 2 g_w^2 W_{\mu}W_{\nu}   \, ]  \, , \nonumber \\
T^F & = & T_{\mu\nu}^{F} \hat{\chi}^{\mu\nu} \, , \quad T^R  =  T_{\mu\nu}^{R} \hat{\chi}^{\mu\nu} \, , \quad  T^G  =  T_{\mu\nu}^{G} \hat{\chi}^{\mu\nu} \, , \quad \Gmct_{[\nu\sigma\alpha]}  \equiv \Gmct_{[\nu\sigma]}^{\beta} \chi_{\beta\alpha}  \, ,
\ee
and
\be \label{EMTA}
{\cal T}_{[\mu\nu]} & = & {\bf G}_{[\mu\nu]}  + {\bf T}_{[\mu\nu]} \, , \nonumber \\
{\bf G}_{[\mu\nu]} & = &  \chi  \alpha_E M_S^2 \, ( \Rct_{\mu\nu} - \Rct_{\nu\mu} ) \, , \nonumber \\
{\bf T}_{[\mu\nu]}  & = & \frac{1}{4}\chi  [\, i\bar{\Psi} ( \chi_{\mu}{\mathcal D}_{\nu} - \chi_{\nu}{\mathcal D}_{\mu} ) \Psi 
+ i\bar{\psi} (\chi_{\mu} \td_{\nu} - \chi_{\nu} \td_{\mu} ) \psi  + H.c.\,  ]  \, 
\ee
for the antisymmetric part.

Let us also decompose the gravifield tensor into symmetric and antisymmetric parts:
\be
{\cal G}^{\;\rho}_{ \mu\nu}   = {\cal G}^{\; \sigma\rho}_{ \mu} \chi_{\sigma\nu} = G_{\mu\nu}^{\; \rho} +   {\cal G}^{\;\rho}_{ [\mu\nu]}
\ee
with
\be \label{GSA}
G_{\mu\nu}^{\; \rho} & = & \alpha_W M_S^2 g_w \chi   [\,  \frac{1}{2} ( \eta_{\mu}^{\rho} W_{\nu} + \eta_{\nu}^{\rho} W_{\mu} ) - \chi_{\mu\nu} \hat{\chi}^{\rho\sigma} W_{\sigma} \, ] \, , \nonumber \\
 {\cal G}^{\;\rho}_{ [\mu\nu]} & = & \alpha_W M_S^2  \chi [\,  \Gmct_{[\nu\sigma\mu]}\hat{\chi}^{\sigma\rho}  + g_w \frac{1}{2}  ( \eta_{\mu}^{\rho} W_{\nu} - \eta_{\nu}^{\rho} W_{\mu} )   \, ] \, . 
\ee
The covariant derivative of the gravifield tensor ${\cal G}^{\;\rho}_{ \mu\nu} $ can also be expressed into the corresponding symmetric and antisymmetric parts:
\be
& & {\cal T}^{{\cal G}}_{\mu\nu} \equiv  \td_{\rho} {\cal G}_{\mu\nu}^{\; \rho} =  T^{{\cal G}}_{\mu\nu} + {\cal T}^{{\cal G}}_{[\mu\nu]} \, , \nonumber \\
& & \td_{\rho} {\cal G}_{\mu\nu}^{\; \rho}  = ( \partial_{\rho} - g_w W_{\rho} ) {\cal G}_{\mu\nu}^{\; \rho} 
- \Gamma_{\rho\mu}^{\sigma} {\cal G}_{\sigma\nu}^{\; \rho} - \Gamma_{\rho\nu}^{\sigma} {\cal G}_{\mu\sigma}^{\; \rho} 
 - \Gmct_{\rho\mu}^{\sigma} {\cal G}_{\sigma\nu}^{\; \rho}    \, , 
\ee
with  $T^{{\cal G}}_{\mu\nu} $ the symmetric part and ${\cal T}^{{\cal G}}_{[\mu\nu]} $ the antisymmetric part. Their explicit forms read
\be
 T^{{\cal G}}_{\mu\nu} & = & ( \partial_{\rho} - g_w W_{\rho} ) G_{\mu\nu}^{\; \rho} 
- \Gamma_{\rho\mu}^{\sigma} G_{\sigma\nu}^{\; \rho} - \Gamma_{\rho\nu}^{\sigma} G_{\sigma\mu}^{\; \rho} 
 - \frac{1}{2} [\, \Gmct_{\rho\mu}^{\sigma} {\cal G}_{\sigma\nu}^{\; \rho} + \Gmct_{\rho\nu}^{\sigma} {\cal G}_{\sigma\mu}^{\; \rho} \, ] \, \nonumber \\
 & = & T_{\mu\nu}^W - \chi_{\mu\nu} T^W + \alpha_W M_S^2 \chi \frac{1}{2} [\,   \hat{\chi}^{\rho\sigma}
 (\Gmct_{\rho\mu}^{\; \alpha} \Gmct_{[\sigma\nu\alpha]} + \Gmct_{\rho\nu}^{\; \alpha} \Gmct_{[\sigma\mu\alpha]}) 
 - g_w( \Gmct_{\sigma\mu}^{\; \sigma} W_{\nu} + \Gmct_{\sigma\nu}^{\; \sigma} W_{\mu})  \, ] \, \nonumber \\
 {\cal T}^{{\cal G}}_{[\mu\nu]} & = & ( \partial_{\rho} - g_w W_{\rho} ) {\cal G}_{[\mu\nu]}^{\; \rho} 
- \Gamma_{\rho\mu}^{\sigma} {\cal G}_{[\sigma\nu]}^{\; \rho} + \Gamma_{\rho\nu}^{\sigma} {\cal G}_{[\sigma\mu]}^{\; \rho} 
- \frac{1}{2} [\, \Gmct_{\rho\mu}^{\sigma} {\cal G}_{\sigma\nu}^{\; \rho} - \Gmct_{\rho\nu}^{\sigma} {\cal G}_{\sigma\mu}^{\; \rho} \, ] \, \nonumber \\
& = & \alpha_W M_S^2 \chi  \frac{1}{2} [\,  \hat{\chi}^{\rho\sigma}  ( \td_{\rho} \Gmct_{[\nu\sigma\mu]} - \td_{\rho} \Gmct_{[\mu\sigma\nu]} )  - g_w W_{\mu\nu} - g_w ( \Gmct_{\sigma\mu}^{\; \sigma} W_{\nu} - \Gmct_{\sigma\nu}^{\; \sigma} W_{\mu}) \, ]
\ee
with definitions
\be
& & T_{\mu\nu}^W  =  \alpha_W M_S^2 g_w \chi  \frac{1}{2} ( \td_{\mu} W_{\nu} + \td_{\nu} W_{\mu} ) \, , \quad T^W = \hat{\chi}^{\mu\nu} T_{\mu\nu}^W \, ,\nonumber \\
& & \td_{\mu} W_{\nu} =  ( \partial_{\mu} - g_w W_{\mu} ) W_{\nu} - \Gamma_{\mu\nu}^{\rho} W_{\rho} \, , \quad W_{\mu\nu} = \partial_{\mu}W_{\nu} - \partial_{\nu}W_{\mu}\, , \nonumber \\
& & \td_{\rho} \Gmct_{[\nu\sigma\mu]}  =  (\partial_{\rho} - g_wW_{\rho}) \Gmct_{[\nu\sigma\mu]}  
- \Gamma_{\rho\nu}^{\alpha} \Gmct_{[\alpha\sigma\mu]} - \Gamma_{\rho\sigma}^{\alpha} \Gmct_{[\nu\alpha\mu]}
- \Gamma_{\rho\mu}^{\alpha} \Gmct_{[\nu\sigma\alpha]}   - \Gmct_{\rho\mu}^{\; \alpha} \Gmct_{[\nu\sigma\alpha]} 
\ee
From the above analyses, we arrive at two types of gravity equation by comparing the symmetric part and antisymmetric part of  equation $\td_{\rho} {\cal G}_{\mu\nu}^{\; \rho}  =  {\cal T}_{\mu\nu}$, namely
\be
& & {\bf G}_{\mu\nu} =   - {\bf T}_{\mu\nu} + T^{{\cal G}}_{\mu\nu} \, , \\
& &  {\bf G}_{[\mu\nu]} =   - {\bf T}_{[\mu\nu]} + T^{{\cal G}}_{[\mu\nu]} \, , 
\ee
where the first equation is the extension to Einstein's equation of general relativity, i.e., 
\be
& & 2\alpha_E M_S^2 \chi \, [\,  R_{\mu\nu} -   \frac{1}{2} \chi_{\mu\nu} \, R \, ]  \nonumber \\
& & \qquad  =  -[ \chi_{\mu\nu} T^{S} + T_{\mu\nu}^{F} - \chi_{\mu\nu} T^F +  T_{\mu\nu}^{R} - \frac{1}{2}\chi_{\mu\nu} T^R
+ T_{\mu\nu}^{G} - \frac{1}{4} \chi_{\mu\nu} T^G ] +  T_{\mu\nu}^W - \chi_{\mu\nu} T^W \nonumber \\
& & \qquad  + \alpha_W M_S^2 \chi \frac{1}{2} [\,   \hat{\chi}^{\rho\sigma}
 (\Gmct_{\rho\mu}^{\; \alpha} \Gmct_{[\sigma\nu\alpha]} + \Gmct_{\rho\nu}^{\; \alpha} \Gmct_{[\sigma\mu\alpha]}) 
 -g_w ( \Gmct_{\sigma\mu}^{\; \sigma} W_{\nu} + \Gmct_{\sigma\nu}^{\; \sigma} W_{\mu})  \, ] \, ,
\ee
while the second equation is a new type of gravitational equation 
\be
& & 2  \alpha_E M_S^2\chi \, ( \Rct_{\mu\nu} - \Rct_{\nu\mu} ) + \alpha_W M_S^2 \chi  [\,  \hat{\chi}^{\rho\sigma}  ( \td_{\rho} \Gmct_{[\mu\sigma\nu]} - \td_{\rho} \Gmct_{[\nu\sigma\mu]}  )   + g_w ( \Gmct_{\sigma\mu}^{\; \sigma} W_{\nu} - \Gmct_{\sigma\nu}^{\; \sigma} W_{\mu}) \, ]   \nonumber \\
& &  = - \frac{1}{2}\chi  [\, i\bar{\Psi} ( \chi_{\mu}{\mathcal D}_{\nu} - \chi_{\nu}{\mathcal D}_{\mu} ) \Psi 
+ i\bar{\psi} (\chi_{\mu} \td_{\nu} - \chi_{\nu} \td_{\mu} ) \psi  + H.c.\,  ] - \chi \alpha_W M_S^2  g_w W_{\mu\nu} \, ,
\ee
which characterizes the twisting and torsional effects. 

It is natural to ask how the first gravity equation in the hidden gauge formalism becomes a natural extension to Einstein's equation of general relativity though we start from the principles and postulates alternative to the ones of general relativity. In the general relativity, the main postulate made by Einstein is that {\it the physical laws of nature are to be expressed by equations which hold good for all systems of coordinates}, which indicates that the physical laws should be invariant under the general linear transformations of local GL($4,R$) symmetry.  In fact, it is interesting to note that via a linear group transformation the action of Eq.(\ref{action6}) is indeed invariant under a general coordinate transformation
\be
& & dx^{'\mu} = \frac{\partial x^{'\mu}}{\partial x^{\nu} }\, dx^{\nu}\, , \quad \partial'_{\mu} = \frac{\partial x^{\nu}}{\partial x^{'\mu} }\,  \partial_{\nu} \, \, \nonumber \\ 
& & V^{'\mu}_k =  \frac{\partial x^{'\mu}}{\partial x^{\nu} }\,  V^{\nu}_k \, , \quad  A_{\mu}^{'k} =  \frac{\partial x^{\mu}}{\partial x^{'\nu} }\,  A_{\nu}^k 
\ee
with $V^{\mu}_k$ and $A_{\mu}^k$ representing vector fields, such as $V^{\mu}_k = \hat{\chi}_a^{\;\, \mu} $ and
 $A_{\mu}^k = ( {\cal A}_{\mu}^I\, , \Om_{\mu}^{ab}\, ,  \chi_{\mu}^{\; \, a} \, , W_{\mu} ) $.  Although the gauge theory of gravity is built based on the global Lorentz symmetry of coordinates in the flat Minkowski spacetime, either the gauge- and Lorentz-invariant action of Eq.(\ref{action3}) or the Lorentz-invariant action of Eq.(\ref{action6}) in the hidden gauge formalism, they all possess a hidden local symmetry of linear group GL($4,R$) under the general coordinate transformation. Such a {\it hidden general coordinate invariance} is a natural consequence of the postulate that the action must be expressed to be coordinate-independent in the gravifield spacetime as shown in Eq.(\ref{action2}). Therefore, the postulate of the {\it gauge invariance and coordinate independence } is thought to be a more general one. 

Before ending this section, we would like to emphasize again that the gravitational interactions with boson fields  can be characterized by the symmetric gravimetric fields $\chi_{\mu\nu}$ in terms of the hidden gauge formalism, while the gravitational interactions with fermion fields have to be described by the gravifield field $\chi_{\mu}^{\;\, a}$. In the gravity equations, it is explicitly seen that the fermionic energy-momentum tensors $T_{\mu\nu}^F$ and ${\bf T}_{[\mu\nu]}$ always couple to the gravifield $\chi_{\mu} \equiv \chi_{\mu}^{\;\, a} \gamma_a$.

By comparing to the Einstein theory of general relativity, we are able to fix the combination of the coupling parameter and scaling mass scale as
\be  \label{relation}
2\alpha_E M_S^2 = \frac{1}{8\pi G} = \frac{1}{8\pi } M_P^2 \simeq 4\alpha_W M_S^2 \; ,
\ee
with $M_P$ the Planck mass and $\alpha_E$ and $\alpha_W$ the effective couplings at low energy. Such a relation results from the requirement that the free part of the gravifield in quantizing the gauge theory of gravity Eq.(\ref{action3}) is equivalent to the free part of the gravimetric field in quantizing the general theory of relativity as an effective field theory at the low energy. Namely, the high-order derivative terms of the gravimetric field in Eq.(\ref{action6}) is considered to be negligible in a good approximation when the energy scale is much below the Planck scale $M_P$. 

Furthermore, we would like to point out that the hidden gauge formalism Eq.(\ref{action6}) for the gauge theory of gravity [or Eq.(\ref{action5}) with gravitational gauge symmetry breaking] is presented in order to compare with the general theory of relativity formulated by Einstein based on geometrical considerations in the curved spacetime. In general, for a quantum theory of gravity, it is demonstrated in the present paper that the gauge theory of gravity Eq.(\ref{action3}) based on gauge invariance and coordinate independence should provide a better approach than the usual straightforward methods of perturbative quantum gravity based on the general theory of gravity. In principle, only the gravifield is essentially thought to be a basic gravitational field that interacts with the basic building blocks of fermionic matter fields, and also only the gravifield is really treated as a gauge-type field that is associated with the gauge field in the coset $T^{1,3}_G$ = PG(1,3)/SP(1,3) of internal Poincar$\acute{\mbox{e}}$  gauge group PG(1,3) in the spinor representation of gravifield spacetime. Thus, the quantization of the gravifield can be carried out straightforwardly in a standard approach within the framework of relativistic QFT in the perturbative expansion. In contrast,  as shown in the hidden gauge formalism Eq.(\ref{action6}), the Goldstone-like gravimetric field in the bosonic interactions concerns high derivative gravitational interactions, i.e., $g_s^{-2} \hat{\chi}^{\mu\nu} \hat{\chi}^{\rho\sigma} R_{\mu\rho }^{\; \alpha \beta}  R_{\nu\sigma \alpha \beta}$, which makes the quantization of gravity unusual. Only in the low-energy limit with an energy scale much below the Planck scale $M_P$ does it enable us to take appropriately the general relativity of gravity as an effective field theory to make a perturbative quantization of gravity\cite{JFD}.  

Moreover, it is interesting to notice that the relation $\alpha_E \simeq 2 \alpha_W$ given in Eq.(\ref{relation}) indicates that the bosonic gravitational interactions in the gauge theory of gravity Eq.(\ref{action3}) possess only a local gauge symmetry of spin group SP(1,3) rather than a local symmetry of internal Poincar$\acute{\mbox{e}}$  gauge group PG(1,3). This is due to the fact that only in the case with $\alpha_E = \alpha_W$ does the bosonic part of the action in Eq.(\ref{action3}) get an enlarged symmetry of gauge group SP(1,4) that can be regarded as a gauged internal Poincar$\acute{\mbox{e}}$  group PG(1,3) in the spinor representation of gravifield spacetime. In fact, for the gravitational interactions with fermionic fields, there possesses in any case only a local gauge symmetry of spin group SP(1,3). Thus, the gravitational interactions for both bosonic and fermionic parts can only have a local gauge symmetry of spin group SP(1,3) as a subgroup of internal Poincar$\acute{\mbox{e}}$  gauge group PG(1,3). With such an observation, the hidden gauge formalism of the gauge theory of gravity Eq.(\ref{action6})  is, in general, not expected to be obtained as an extension of the standing way of gauging the global Poincar$\acute{\mbox{e}}$  group P(1,3) in flat Minkowski spacetime.

In conclusion, within the framework of relativistic QFT, the postulate of gauge invariance and coordinate independence should be a key principle for establishing a gauge theory of quantum gravity shown in Eqs.(\ref{action2})  and (\ref{action3}) with the gravifield as a basic gauge-type gravitational field.

\section{Conclusions and Remarks}

By treating the gravitational force on the same footing as the electroweak and strong forces within the framework of QFT in flat Minkowski spacetime, we have described the QFT of gravity based on the spin and scaling gauge symmetries. A biframe spacetime has been initiated to describe such a QFT of gravity. One frame spacetime is the globally flat coordinate Minkowski spacetime that acts as an inertial reference frame for the motions of fields, and the other is the locally flat noncoordinate gravifield spacetime that functions as an interaction representation frame for the degrees of freedom of fields. The gauge-invariance and coordinate-independence in the gravifield spacetime have been shown to be a more general postulate for establishing the QFT of gravity. Such a quantum gravity theory has been demonstrated to have the following properties: i) The basic gravitational interaction is characterized by a bicovariant vector field $\chi_{\mu}^{\;\; a}(x)$ sided on both a locally flat noncoordinate spacetime and a globally flat Minkowski spacetime. $\chi_{\mu}^{\;\; a}(x)$  is an essential ingredient for gauging the spin symmetry SP(1,3) and scaling symmetry of the basic fermion fields; it couples to all kinetic terms and interaction terms of the gauge field and is referred as gravifield for short; ii) The locally flat noncoordinate spacetime is spanned by the gravifield to form a locally flat {\it gravifield spacetime} characterized by the dual gravifield bases $\{\chi^{\, a}= \chi_{\mu}^{\;\; a}(x)dx^{\mu} \}$ and $\{\hat{\chi}_{\, a}= \hat{\chi}^{\;\;\mu}_{a}(x)\partial_{\mu} \}$; such a gravifield spacetime shows a property of non-commutative geometry. The vector field $\chi^{\, a}(x)= \chi_{\mu}^{\;\; a}(x)dx^{\mu} $ has been regarded as the one-form gauge-type potential in the Minkowski spacetime; its field strength describes the gravitational interaction. It has enabled us to construct a gauge invariant and coordinate independent action for the QFT of gravitational interaction. iii) The globally flat Minkowski spacetime in QFT sets an inertial frame for a reference to describe motions of quantum fields and to make a meaningful definition for the momentum and energy; we are able to derive equations of motion for all quantum fields. iv) Such a QFT of gravity has allowed us to obtain basic conservation laws with respect to all symmetries. An alternative equation of motion for the gravifield tensor has been deduced in connection with the energy-momentum tensor. v) The spin and scaling gauge symmetries have been assumed to be broken down to a background structure that has global Lorentz and scaling symmetries.  A unitary basis defined by fixing a special scaling gauge condition has been adopted to yield an exact solution for equations of motion for the background fields. vi)  The geometric property of the gravifield spacetime is, in general, characterized by the gravimetric field and the scalinon field. The resulting background gravifield spacetime has been found to coincide with a Lorentz-invariant and conformally flat Minkowski spacetime of coordinates. The background gravifield spacetime has been shown to be characterized by the cosmic vector with a nonzero cosmological mass scale. Such a background gravifield spacetime is, in general, not isotropic in terms of the conformal proper time except in a special comoving reference frame. The conformal size of the Universe has been shown to be singular when the light travels close to the cosmological horizon in terms of the conformal proper time or in the comoving frame. It has also been demonstrated that the Universe appears inflationary or deflationary in light of the cosmic proper time.

The action of the quantized gravitational interactions has explicitly been constructed in the unitary basis. The gauge-fixing contributions to the quantization of gravity gauge theory have been presented by using the Faddeev-Popov formalism in the path integral approach. It has been shown that it is, in general, no more difficult to make the identification of the physical quantum degrees of freedom and the definition of a quantum gravity theory analogous to the quantization of Yang-Mills gauge theory. Based on the background gravifield spacetime, we have explicitly written down the leading effective action which enables us to read the Feynman rules and investigate the quantum effects. The main issue is the universal coupling of the inverse gauge-type gravifield, which causes the nonlinear nature of the theory and may result in the nonrenormalizability of the theory.  On the other hand, the gravitational interaction with the scaling gauge invariance indicates the existence of a fundamental energy scale that characterizes the ultraviolet behavior of the theory, so that the theory appears to be infinite-free and meaningful by appropriately treating the divergence integrals. When the spin and scaling gauge symmetries are broken down to a background structure that possesses the global Lorentz and scaling symmetries, there exist rich gravitational interactions with the background fields. The quantum effect on the inflation of the Universe has been discussed based on the effective background scalar potential of the scalinon field at the one-loop level. We have concluded that it is the quantum loop quadratic contribution that causes the breaking of global scaling symmetry and generates the inflation of the early Universe, and the end of the inflationary Universe occurs when the evolving vacuum expectation value of the background scalar potential approaches a minimal condition.

When taking the gravifield as a Goldstone-like field that transmutes the local spin gauge symmetry into the global Lorentz symmetry, an alternative spacetime gauge field has been constructed from the spin gauge field, i.e., $\Gmc_{\mu\nu}^{\sigma}(x) = \hat{\chi}_a^{\; \sigma} (\partial_{\mu} \chi_{\nu}^{\; a} + g_s \Om_{\mu b}^{a} \chi_{\nu}^{\; b} ) $, so that the spin gauge symmetry becomes a hidden gauge symmetry. Such a spacetime gauge field is a Lorentz tensor field defined and valued in flat Minkowski spacetime. With this consideration, we have presented an alternative action of quantum gravity which shows that the bosonic gravitational interactions can be described by the Goldstone-like gravimetric field $\chi_{\mu\nu}(x)$ and the spacetime gauge field except for the gravitational interactions with the fermion fields, which enables us to compare the present gauge theory of gravity with the Einstein theory of general relativity. It is manifest that the Einstein theory of general relativity results as an effective low-energy theory. Two types of gravity equation have been obtained; one is as the extension to Einstein's equation of general relativity, and the other is a new type of gravitational equation that reflects the twisting effect.  It is interesting to see that  both the gauge-invariant action and the Lorentz-invariant action in the hidden gauge formalism possess a hidden symmetry of local linear group GL($4,R$). Namely, the resulting gauge theory of gravity possesses a {\it hidden general coordinate invariance}, which indicates that the {\it gauge invariance and coordinate independence } in the gravifield spacetime become a more general postulate for establishing a gauge theory of gravity.

Finally, we would like to remark that in this paper we have described mainly a general framework for the quantum field theory of gravity based on the spin and scaling gauge symmetries and paid special attention to the mechanism of quantum scalinon inflation of the Universe. We shall leave other important issues elsewhere for further investigations in detail.

\centerline{{\bf Acknowledgement}}

The author thanks R.G. Cai for useful discussions. He is also grateful to many colleagues for valuable conversations during the International Conference on Gravitation and Cosmology/the fourth Galileo-Xu Guangqi Meeting, held at KITPC/ITP-CAS and UCAS, 2015, for celebrating the 100th anniversary of Albert Einstein's presentation of the theory of general relativity. This work was supported in part by the National Science Foundation of China (NSFC) under Grants \#No. 11475237, No.~11121064, and No.~10821504 and also by the CAS Center for Excellence in Particle Physics (CCEPP).
   

\end{document}